\documentclass[12pt]{article}
\usepackage{a4,graphicx,amssymb}
\usepackage[small,bf]{caption}

\newcommand{\note}[1]{}                    

\newcommand{\sixth}{\mbox{\small $\frac{1}{6}$}}         
\newcommand{\half}{\mbox{\small $\frac{1}{2}$}}          
\newcommand{\third}{\mbox{\small $\frac{1}{3}$}}         
\newcommand{\twothird}{\mbox{\small $\frac{2}{3}$}}      
\newcommand{\threehalf}{\mbox{\small $\frac{3}{2}$}}     
\newcommand{\eighth}{\mbox{\small $\frac{1}{8}$}}        
\newcommand{\tenth}{\mbox{\small $\frac{1}{10}$}}        
\newcommand{\fortieth}{\mbox{\small $\frac{1}{40}$}}      
\newcommand{\R}{\mbox{\tiny $R$}}                        
\newcommand{\Si}{\mbox{\tiny $S$}}                       
\newcommand{\NS}{\mbox{\tiny $N\!S$}}                    
\newcommand{\PQ}{\mbox{\tiny $P\!Q$}}                    
\newcommand{\full}{\mbox{\tiny $full$}}                  
\newcommand{\plaquette}{\mbox{\tiny $Plaquette$}}        
\newcommand{\rectangle}{\mbox{\tiny $Rectangle$}}        

\def\lsim{\mathrel{\rlap{\lower4pt\hbox{\hskip1pt$\sim$}}
    \raise1pt\hbox{$<$}}}                
\def\gsim{\mathrel{\rlap{\lower4pt\hbox{\hskip1pt$\sim$}}
    \raise1pt\hbox{$>$}}}                

\begin{document}

\title{
\vspace{-3.0cm}
\flushright{\normalsize DESY 11-030} \\
\vspace{-0.35cm}
{\normalsize Edinburgh 2011/09} \\
\vspace{-0.35cm}
{\normalsize LTH 909} \\
\vspace{-0.35cm}
{\normalsize September 22, 2011} \\
\vspace{0.5cm}
\centering{\Large \bf Flavour blindness and patterns of flavour symmetry
           breaking in lattice simulations of
           \newline 
           up, down and strange quarks}}

\author{\large
        W. Bietenholz$^a$, V. Bornyakov$^b$,
        M. G\"ockeler$^c$, R. Horsley$^d$, \\
        W.~G. Lockhart$^e$, Y. Nakamura$^f$,
        H. Perlt$^g$, D. Pleiter$^h$, \\
        P.~E.~L. Rakow$^e$, G. Schierholz$^{ci}$,
        A. Schiller$^g$, T. Streuer$^{c}$, \\
        H. St\"uben$^j$, F. Winter$^d$
        and J.~M. Zanotti$^d$ \\[1em]
         -- QCDSF-UKQCD Collaboration -- \\[1em]
        \small $^a$ Instituto de Ciencias Nucleares,
               Universidad Aut\'{o}noma de M\'{e}xico,\\[-0.5em]
        \small A.P. 70-543, C.P. 04510 Distrito Federal, Mexico \\[0.25em]
        \small $^b$ Institute for High Energy Physics,
               142281 Protovino, Russia and \\[-0.5em]
        \small Institute of Theoretical and
               Experimental Physics,
               117259 Moscow, Russia \\[0.25em]
        \small $^c$ Institut f\"ur Theoretische Physik,
               Universit\"at Regensburg, \\[-0.5em]
        \small 93040 Regensburg, Germany \\[0.25em]
        \small $^d$ School of Physics and Astronomy,
               University of Edinburgh, \\[-0.5em]
        \small Edinburgh EH9 3JZ, UK \\[0.25em]
        \small $^e$ Theoretical Physics Division,
               Department of Mathematical Sciences, \\[-0.5em]
        \small University of Liverpool,
               Liverpool L69 3BX, UK \\[0.25em]
        \small $^f$ RIKEN Advanced Institute for
               Computational Science, \\[-0.5em]
        \small Kobe, Hyogo 650-0047, Japan \\[0.25em]
        \small $^g$ Institut f\"ur Theoretische Physik,
               Universit\"at Leipzig, \\[-0.5em]
        \small 04109 Leipzig, Germany \\[0.25em]
        \small $^h$ Deutsches Elektronen-Synchrotron DESY, \\[-0.5em] 
        \small 15738 Zeuthen, Germany \\[0.25em]
        \small $^i$ Deutsches Elektronen-Synchrotron DESY, \\[-0.5em]
        \small 22603 Hamburg, Germany \\[0.25em]
        \small $^j$ Konrad-Zuse-Zentrum
               f\"ur Informationstechnik Berlin, \\[-0.5em]
        \small 14195 Berlin, Germany }

\date{}

\maketitle


\clearpage

\begin{abstract}
   QCD lattice simulations with 2+1 flavours (when two quark flavours
   are mass degenerate) typically start at rather large up-down and
   strange quark masses and extrapolate first the strange quark mass
   and then the up-down quark mass to its respective physical value.
   Here we discuss an alternative method of tuning the quark masses,
   in which the singlet quark mass is kept fixed. Using group theory
   the possible quark mass polynomials for a Taylor expansion about
   the flavour symmetric line are found, first for the general $1+1+1$
   flavour case and then for the $2+1$ flavour case. This ensures
   that the kaon always has mass less than the physical kaon mass.
   This method of tuning quark masses then enables highly constrained
   polynomial fits to be used in the extrapolation of hadron masses
   to their physical values. Numerical results for the  $2+1$ flavour case
   confirm the usefulness of this expansion and an extrapolation
   to the physical pion mass gives hadron mass values to within a
   few percent of their experimental values. Singlet quantities
   remain constant which allows the lattice spacing to be determined
   from hadron masses (without necessarily being at the physical point).
   Furthermore an extension of this programme to include partially
   quenched results is given.
\end{abstract}

\clearpage


\tableofcontents 

\clearpage


\section{Introduction} 
\label{introduction}


The QCD interaction is flavour blind. Neglecting electromagnetic
and weak interactions, the only difference between quark flavours
comes from the quark mass matrix, which originates from the coupling
to the Higgs field. We investigate here how flavour blindness
constrains hadron masses after flavour $SU(3)$ symmetry is broken by
the mass difference between the strange and light quarks.
The flavour structure illuminates the pattern of symmetry breaking
in the hadron spectrum and helps us extrapolate $2+1$ flavour
lattice data to the physical point. (By $2+1$ we mean that the
$u$ and $d$ quarks are mass degenerate.) 

\begin{figure}[htb]
   \vspace*{0.15in}
   \begin{center}
      \includegraphics[width=7.0cm]{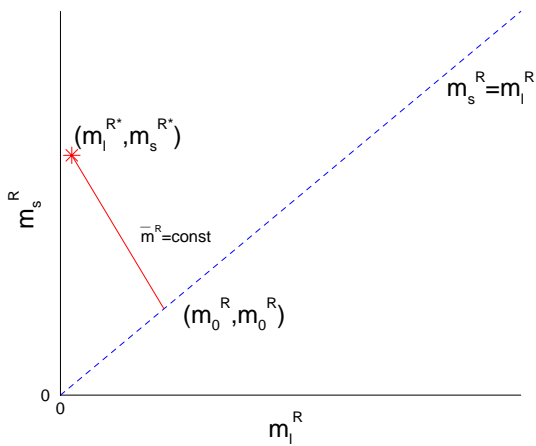}
   \end{center} 
   \caption{Sketch of the path (red, solid line) in the
            $m_l^{\R}$--$ m_s^{\R}$ plane to the physical point
            denoted by $(m_l^{\R\,*}, m_s^{\R\,*})$. The dashed
            diagonal line is the $SU(3)$-symmetric line.}
   \label{sketch_mlR_msR+path} 
\end{figure}

We have our best theoretical understanding when all $3$ quark flavours
have the same masses (because we can use the full power of flavour
$SU(3)$ symmetry); nature presents us with just one instance of the
theory, with $ m_s^{\R}/m_l^{\R} \approx 25$ (where the superscript $^{\R}$
denotes the renormalised mass). We are interested in interpolating
between these two cases.  
We consider possible behaviours near the symmetric point, and find
that flavour blindness is particularly helpful if we approach 
the physical point, denoted by $(m_l^{\R\,*}, m_s^{\R\,*})$,
along a path in the $m_l^{\R}$--$m_s^{\R}$ plane starting at
a point on the $SU(3)$ flavour symmetric line
($m_l^{\R} = m_s^{\R} = m_0^{\R}$) and holding the sum of the quark masses
$\overline{m}^{\R} = \third(m_u^{\R} + m_d^{\R} + m_s^{\R})
\equiv \third(2m_l^{\R} + m_s^{\R})$ constant \cite{bietenholz10a},
at the value $m_0^{\R}$ as sketched in Fig.~\ref{sketch_mlR_msR+path}.
The usual procedure (path) is to estimate the physical strange quark mass
and then try to keep it fixed, i.e.\ $m^{\R}_s = \mbox{constant}$,
as the light quark mass is reduced towards its physical value.
However on that path the problem is that the kaon mass%
\footnote{In this article quark masses will be denoted by $m$, and
hadron masses by $M$.}
is always larger than its physical value. Choosing instead a path
such that the singlet quark mass is kept fixed has the advantage
that we can vary both quark masses over a wide range,
with the kaon mass always being lighter than its physical value
along the entire trajectory. Starting from the symmetric point
when masses are degenerate is particularly useful for strange quark
physics as we can track the development of the strange quark mass.
Also if we extend our measurements beyond the symmetric point
we can investigate a world with heavy up-down quarks and a
lighter strange quark.

The plan of this article is as follows. Before considering the $2+1$
quark flavour case, we consider the more general $1+1+1$ case in
section~\ref{theory_1p1p1}. This also includes a discussion of the
renormalisation of quark masses for non-chiral fermions.
Keeping the singlet quark mass constant constrains the extrapolation
and in particular it is shown in this section that flavour singlet
quantities remain constant to leading order when extrapolating
from a flavour symmetric point. This motivates investigating
possible quark mass polynomials -- we are able to classify them here
to third order in the quark masses under the $SU(3)$ and $S_3$
(flavour) groups. In section~\ref{theory_2p1} we specialise to
$2+1$ flavours and give quark mass expansions to second order
for the pseudoscalar and vector meson octets
and baryon octet and decuplet. (The relation of this expansion
to chiral perturbation theory is discussed later in
section~\ref{applications_chipt}.) In section~\ref{partial_quenching}
we extend the formalism to the partially quenched case
(when the valence quarks of a hadron do not have to have
the same mass as the sea quarks). This is potentially useful
as the same expansion coefficients occur, which could allow
a cheaper determination of them. We then turn to more specific
lattice considerations in sections~\ref{path}, \ref{lat_generalities}
with emphasis on clover fermions (i.e.\ non-chiral fermions) used here.
This is followed by section~\ref{lat_results}, which first gives
numerical results for the constant singlet quark mass results
used here. Flavour singlet quantities prove to be a good
way of defining the scale and the consistency of some 
choices is discussed. We also investigate possible
finite size effects. Finally in section~\ref{hadron_masses}
the numerical results for the hadron mass spectrum are presented
in the form of a series of `fan' plots where the various masses
fan out from their common value at the symmetric point.
Our conclusions are given in section~\ref{conclusions}.
Several Appendices provide some group theory background for this
article, discuss the action used here and give tables of the hadron
masses found.

Mostly we restrict ourselves to the constant surface.
However, in a few sections, we also consider
variations in $\overline{m}^{\R}$ (for example in the derivation
of the quark mass expansion polynomials, section~\ref{taylor_expansion},
the discussion of $O(a)$ improvement in section~\ref{improvement},
and in section~\ref{mbar_varies} where we generalise a constant
$\overline{m}^{\R}$ formula).


\section{Theory for $\mathbf{1 + 1 + 1}$ flavours}
\label{theory_1p1p1}


Our strategy is to start from a point with all three sea quark
masses equal, 
\begin{equation}
    m_u^{\R} = m_d^{\R} = m_s^{\R} \equiv m_0^{\R} \,,
\end{equation}
and extrapolate towards the physical point,
$(m_u^{\R *}, m_d^{\R *}, m_s^{\R *})$, keeping the average sea quark mass
\begin{equation}
   \overline{m}^{\R} = \third (m_u^{\R} + m_d^{\R} + m_s^{\R}) \,,
\end{equation}
constant at the value $m_0^{\R}$. For this trajectory to reach the
physical point we have to start at a point where
$m_0^{\R} \approx \third m_s^{\R*}$. As we approach the physical point,
the $u$ and $d$ quarks become lighter, but the $s$ quark becomes heavier.
Pions are decreasing in mass, but $K$ and $\eta$ increase in mass
as we approach the physical point.


\subsection{Singlet and non-singlet renormalisation}
\label{S_and_NS_renorm}


Before developing the theory, we first briefly comment on the
renormalisation of the quark mass. While for chiral fermions
the renormalised quark mass is directly proportional to the bare
quark mass, $m_q^{\R} = Z_m m_q$, the problem,
at least for Wilson-like fermions which have
no chiral symmetry, is that singlet and non-singlet quark mass
can renormalise differently \cite{gockeler04a,rakow04a}%
\footnote{Perturbative computations showing this effect,
which starts at the two-loop order, are given in
\cite{skouroupathis07a,skouroupathis08a}.}
\begin{equation}
   m^{\R}_q = Z_m^{\NS}(m_q - \overline{m}) + Z_m^{\Si}\overline{m} \,,
   \qquad q = u, d, s \,,
\label{mr2mbare_noalphaZ}
\end{equation}
where $m_q$ are the bare quark masses,
\begin{equation}
   \overline{m} = \third (m_u + m_d + m_s) \,,
\end{equation}
$Z_m^{\NS}$ is the non-singlet renormalisation constant,
and $Z_m^{\Si}$ is the singlet renormalisation constant
(both in scheme $R$). It is often convenient to re-write
eq.~(\ref{mr2mbare_noalphaZ}) as
\begin{equation}
   m_q^{\R} = Z_m^{\NS}(m_q + \alpha_Z\overline{m}) \,,
\label{mr2mbare}
\end{equation}
where
\begin{equation}
   \alpha_Z =  r_m - 1 \,, \quad r_m = { Z_m^{\Si} \over Z_m^{\NS} } \,,
\label{alphaZ}
\end{equation}
represents the fractional difference between the renormalisation constants.
(Numerically we will later see that this factor $\alpha_Z$ is $\sim O(1)$,
and is thus non-negligible at our coupling.) This then gives
\begin{equation}
   \overline{m}^{\R} = Z_m^{\NS}(1 + \alpha_Z)\overline{m} \,.
\label{mbarR}
\end{equation}
This means that even for Wilson-type actions it does not matter
whether we keep the bare or renormalised average sea quark mass constant.
Obviously eq.~(\ref{mbarR}) also holds for a reference point
$(m_0, m_0, m_0)$ on the flavour symmetric line, i.e.\
\begin{equation}
   m_0^{\R} = Z_m^{\Si}m_0 = Z_m^{\NS}(1 + \alpha_Z)m_0 \,.
\end{equation}
Furthermore introducing the notation 
\begin{equation}
   \delta m_q^{\R} \equiv m_q^{\R} - \overline{m}^{\R} \,, \qquad
   \delta m_q \equiv m_q - \overline{m} \,,  \qquad q = u, d, s \,,
\label{delta_mq}
\end{equation}
for both renormalised and bare quark masses, we find that
\begin{equation}
   \delta m_q^{\R} = Z_m^{\NS}\, \delta m_q \,.
\end{equation}
So by keeping the singlet mass constant we avoid the need
to use two different $Z$s and as we will be considering 
expansions about a flavour symmetric point, they will be similar
using either the renormalised or bare quark masses. (Of course the
value of the expansion parameters will be different, but the
structure of the expansion will be the same.) We shall discuss
this point a little further in section~\ref{improvement}.

So in the following we need not usually distinguish between bare
and renormalised quark masses.

Note that it follows from the definition that
\begin{equation}
   \delta m_u + \delta m_d + \delta m_s = 0 \,,
\label{zerosum}
\end{equation}
so we could eliminate one of these symbols. However we shall keep
all three symbols as we can then write some expressions in a more
obviously symmetrical form.


\subsection{General strategy}
\label{gen_strat}


With this notation, the quark mass matrix is 
\begin{eqnarray}
   \cal M &=&  \left( \begin{array}{ccc}
                         m_u     & 0       & 0        \\
                         0       & m_d     & 0        \\
                         0       & 0       & m_s      \\ 
                      \end{array}
               \right)
                                                           \nonumber \\ 
          &=& \overline{m}
              \left(  \begin{array}{ccc}
                         1 & 0 & 0 \\
                         0 & 1 & 0 \\
                         0 & 0 & 1 \\
                      \end{array}
              \right)
                                                           \nonumber \\ 
          & & + \half (\delta m_u - \delta m_d)
              \left(  \begin{array}{ccc}
                         1 & 0  & 0  \\
                         0 & -1 & 0  \\
                         0 & 0  & 0  \\
                      \end{array}
              \right)
              + \half \delta m_s
              \left(  \begin{array}{ccc}
                        -1 & 0 & 0 \\
                         0 &-1 & 0 \\
                         0 & 0 & 2 \\
                      \end{array}
              \right) \,.
\label{massmat} 
\end{eqnarray}  
The mass matrix ${\cal M}$  has a singlet part (proportional to $I$)
and an octet part, proportional to $\lambda_3$, $\lambda_8$. 
We argue here that the theoretically cleanest way to
approach the physical point is to keep the singlet part
of ${\cal M}$ constant, and vary only the non-singlet parts.

An important advantage of our strategy is that it strongly constrains
the possible mass dependence of physical quantities, and so simplifies
the extrapolation towards the physical point. 
Consider a flavour singlet quantity, which we shall denote by $X_S$,
at a symmetric point $(m_0, m_0, m_0)$. Examples are the scale%
\footnote{There is no significance here to using $r_0$ or $r_0^{-1}$;
however defining $X_r = r_0^{-1}$ is more consistent with later
definitions.}
$X_r = r_0^{-1}$, or the plaquette $P$ (this will soon be generalised
to other singlet quantities). If we make small changes in the quark masses,
symmetry requires that the derivatives at the symmetric point are equal
\begin{eqnarray}
   {\partial X_S \over \partial m_u}
      = {\partial X_S \over \partial m_d}
      = {\partial X_S \over \partial m_s} \,.
\end{eqnarray}
So if we keep $m_u + m_d + m_s$ constant, then any arbitrary small changes
in the quark masses mean that $\Delta m_s + \Delta m_u + \Delta m_d = 0$ so
\begin{equation}
   \Delta X_S = {\partial X_S \over \partial m_u}\Delta m_u
                + {\partial X_S \over \partial m_d}\Delta m_d
                + {\partial X_S \over \partial m_s}\Delta m_s = 0 \,.
\label{symarg} 
\end{equation}
The effect of making the strange quark heavier exactly cancels
the effect of making the light quarks lighter, so we know that
$X_S$ must be stationary at the symmetrical point.
This makes extrapolations towards the physical point much easier,
especially since we find that in practice quadratic terms in the
quark mass expansion are very small. Any permutation of the quarks,
such as an interchange $ u \leftrightarrow s$, or a cyclic permutation
$ u \to d \to s \to u $ does not change the physics,
it just renames the quarks. Any quantity unchanged by all
permutations will be flat at the symmetric point, like $X_r$. 

\begin{figure}[htb]
   \begin{center}
   \begin{tabular}{cc}
      \includegraphics[width=6.00cm]{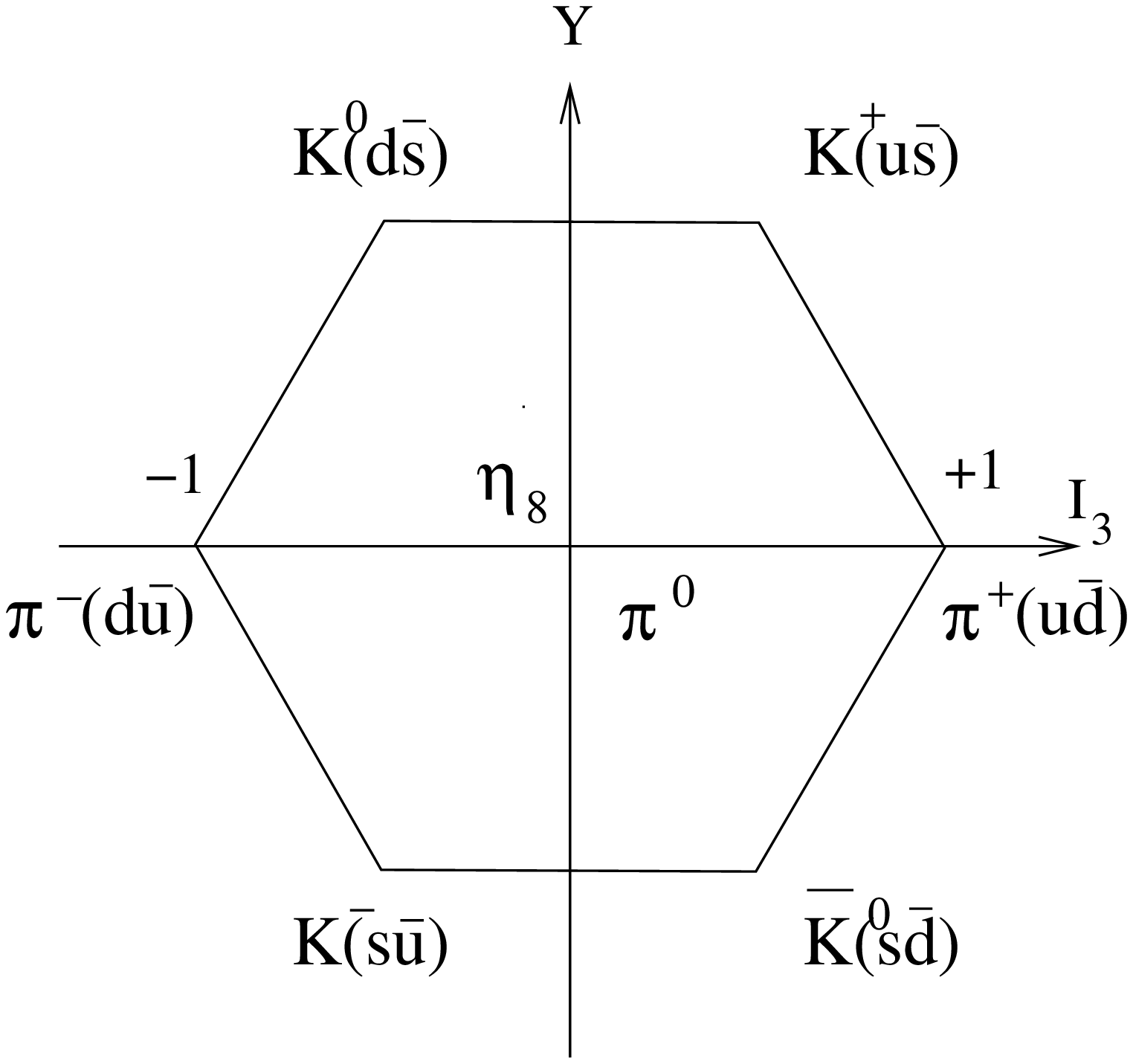}   &
      \includegraphics[width=6.00cm]{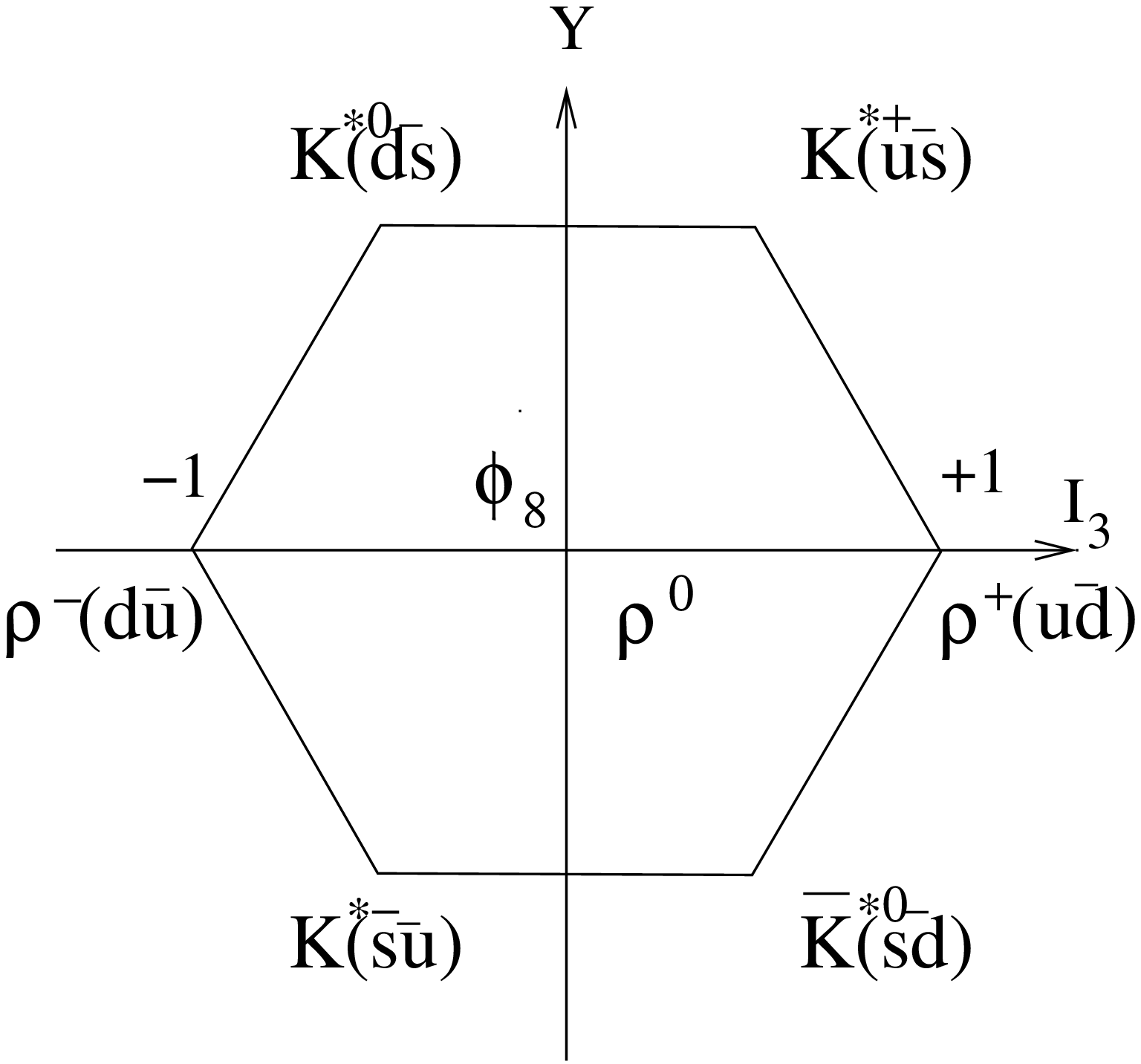} 
   \end{tabular}
   \end{center}
   \caption{The octets for spin 0 (pseudoscalar) and
            spin 1 (vector) mesons (plotted in the $I_3$--$Y$ plane).
            $\eta_8$ and $\phi_8$ are pure octet states,
            ignoring any mixing with the singlet mesons.}
\label{meson_mults}
\end{figure}

\begin{figure}[htb]
   \begin{center}
   \begin{tabular}{cc}
      \includegraphics[width=6.00cm]{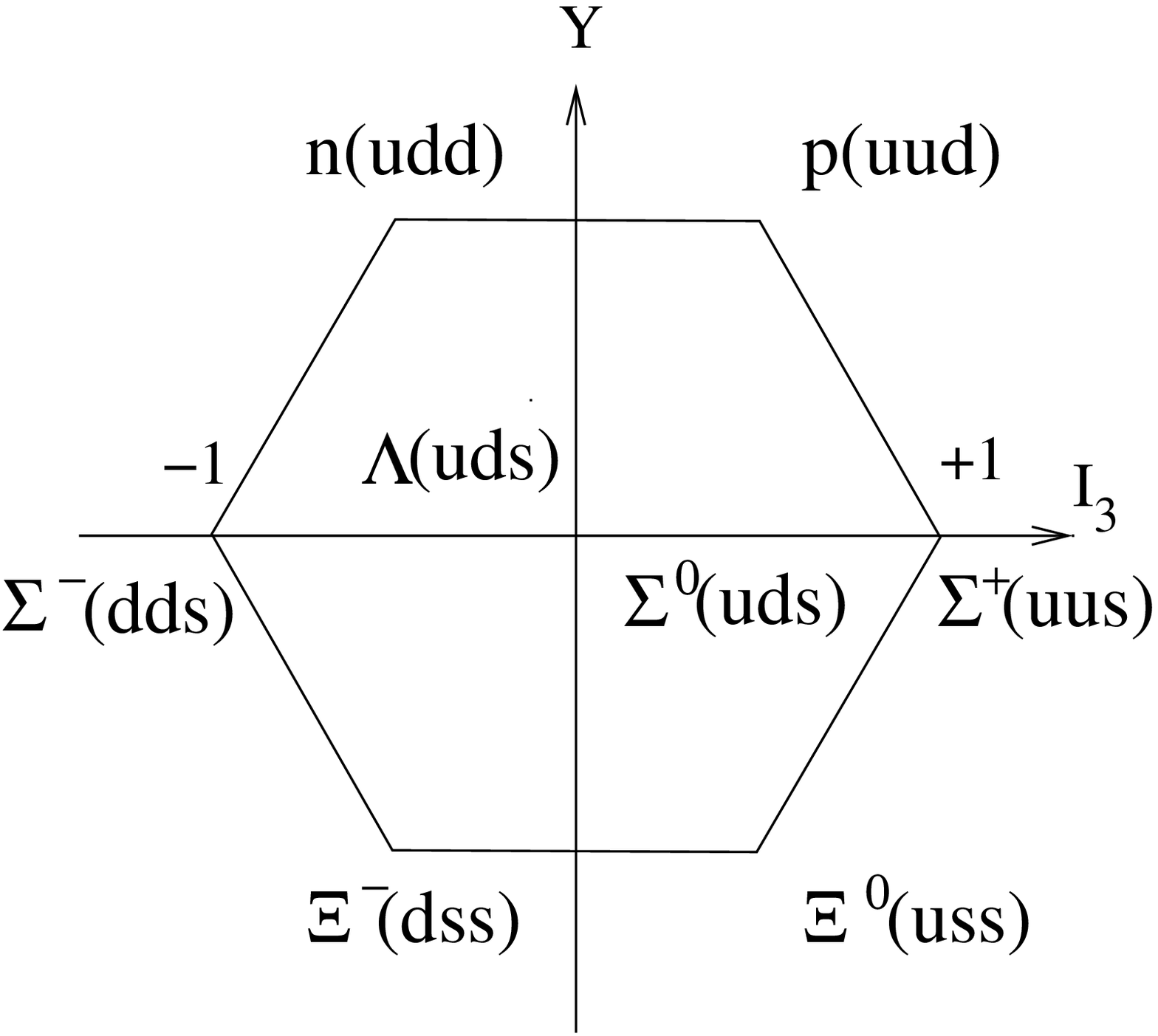}       &
      \includegraphics[width=5.00cm]{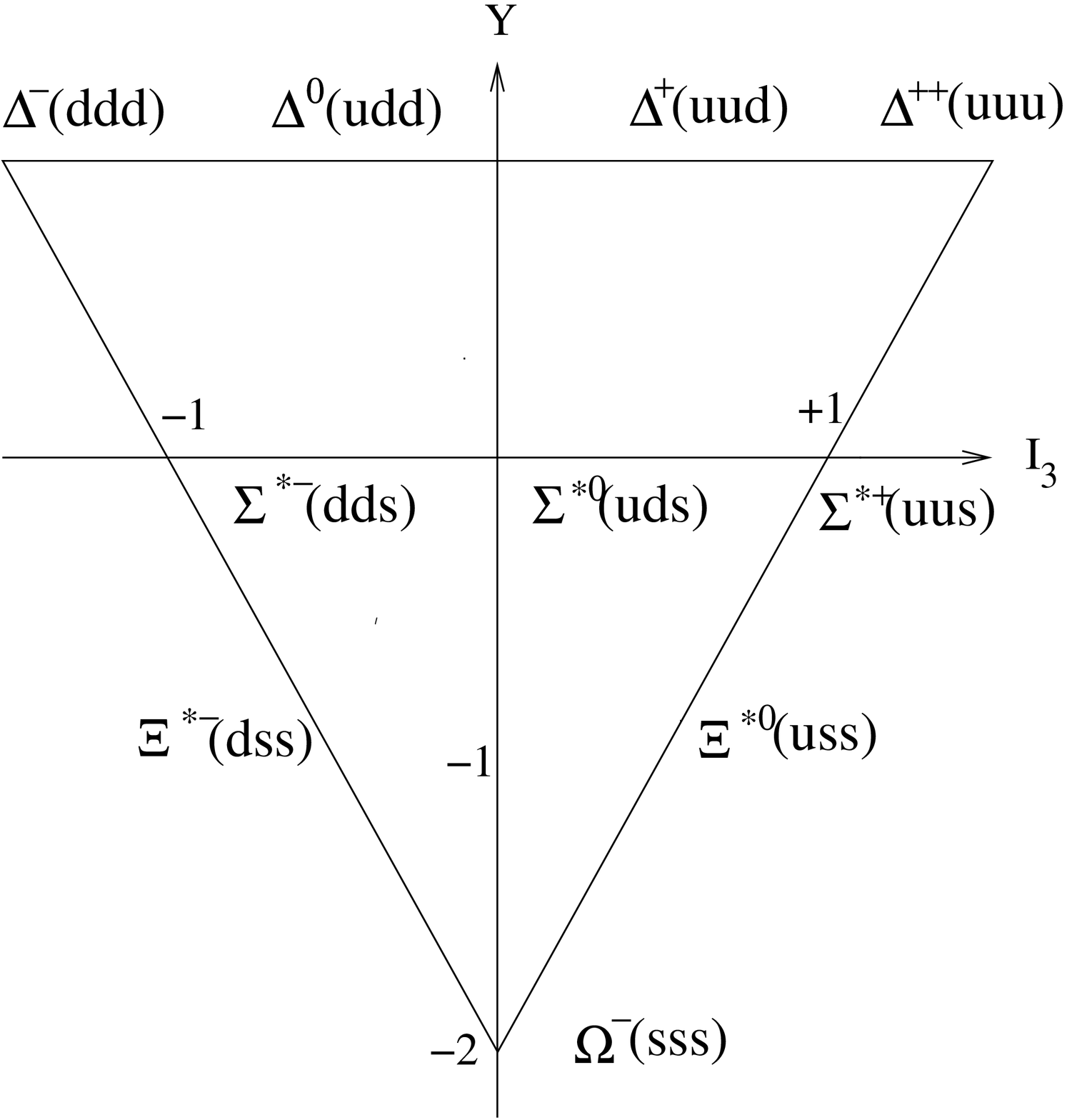} 
   \end{tabular}
   \end{center}
   \caption{The lowest octet and decuplet for the spin $\half$
            and for the spin $\threehalf$ baryons.}
\label{baryon_mults}
\end{figure}

We can also construct permutation-symmetric combinations of hadrons.
For orientation in Fig.~\ref{meson_mults} we give the octet multiplets
for spin $0$ (pseudoscalar) and spin $1$ (vector) mesons and in
Fig.~\ref{baryon_mults} the lowest octet and decuplet multiplets
for the spin $\half$ and for the spin $\threehalf$ baryons
(all plotted in the $I_3$--$Y$ plane).

\begin{figure}[htb]
   \vspace*{0.15in}
   \begin{center}
      \includegraphics[width=4cm,angle=270]{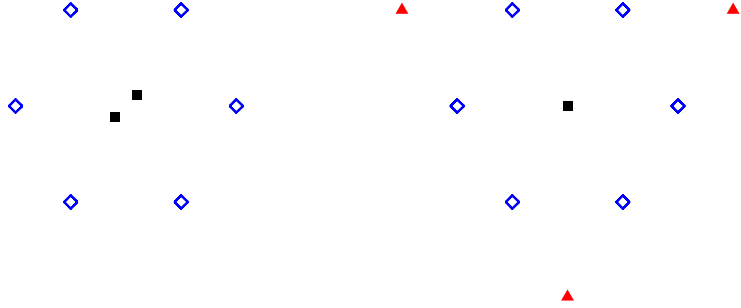}
    \end{center} 
\caption{The behaviour of the octet and decuplet under the
         permutation group $S_3$. The colours denote sets
         of particles which are invariant under permutations
         of the quark flavours (red or filled triangles,
         blue or open diamonds and black or filled squares).}
\label{permset}
\end{figure}

For example, for the decuplet, any permutation of the quark labels
will leave the $\Sigma^{*0}(uds)$ unchanged, so the $\Sigma^{*0}$
is shown by a single black (square) point in Fig.~\ref{permset}.
On the other hand, a permutation (such as $u \to d \to s$)
can change a $\Delta^{++}(uuu)$ into a $\Delta^-(ddd)$
or (if repeated) into an $\Omega^-(sss)$, so these three
particles form a set of baryons which is closed under
quark permutations, and are all given the same colour
red (triangle) in Fig.~\ref{permset}. Finally the $6$
baryons containing two quarks of one flavour,
and one quark of a different flavour, form an invariant set,
shown in blue (diamond) in Fig.~\ref{permset}.

If we sum the masses in any of these sets, we get a flavour symmetric 
quantity, which will obey the same argument we gave in 
eq.~(\ref{symarg}) for the quark mass (in)dependence of
the scale $r_0$. We therefore expect that the $\Sigma^{*0}$
mass must be flat at the symmetric point, and furthermore
that the combinations $(M_{\Delta^{++}} + M_{\Delta^-} + M_{\Omega^{-}})$ 
and $(M_{\Delta^{+}} + M_{\Delta^0} + M_{\Sigma^{*+}}+  M_{\Sigma^{*-}} 
+ M_{\Xi^{*0}} + M_{\Xi^{*-}} )$ will also be flat. 
Technically these symmetrical combinations are in the $A_1$
singlet representation of the permutation group $S_3$.
This is the symmetry group of an equilateral triangle, $C_{3v}$.
This group has $3$ irreducible representations,
\cite{atkins70a}, two different singlets, $A_1$ and $A_2$ and a
doublet $E$, with elements $E^+$ and $E^-$. Some details of this group
and its representations are given in Appendix~\ref{perm_group},
while Table~\ref{S3_simplified} gives a summary of the transformations.
\begin{table}[htb]  
   \begin{center}
   \begin{tabular}{ccccc}
         & $A_1 $ & \multicolumn{2}{c}{$E$} & $A_2$ \\
      \cline{3-4} Operation & & $E^+$ & $E^-$ & \\
      \hline 
      Identity & $+$ & $+$ & $+$ & $+$ \\
      $u \leftrightarrow d$ & $+$ & $+$ & $-$ & $-$ \\
      $u \leftrightarrow s$  & $+$ & \multicolumn{2}{c}{mix} & $-$ \\
      $d \leftrightarrow s$  & $+$ & \multicolumn{2}{c}{mix} & $-$ \\
      $ u \to d \to s \to u$ & $+$ & \multicolumn{2}{c}{mix} & $+$ \\
      $ u \to s \to d  \to u $  & $+$ & \multicolumn{2}{c}{mix} & $+$ \\
      \hline
   \end{tabular} 
   \end{center} 
\caption{A simplified table showing how the group operations of
         $S_3$ act in the different representations: $+$ refers
         to unchanged; $-$ refers to states that are odd under the
         group operation.} 
\label{S3_simplified}
\end{table} 

We list some of these invariant mass combinations in Table~{\ref{perminv}}.
\begin{table}[htb] 
   \begin{center} 
      \begin{tabular}{clc}
         \hline
         Pseudoscalar
                  & $X_\pi^2 = \sixth( M_{K^+}^2 + M_{K^0}^2 
                                     + M_{\pi^+}^2 + M_{\pi^-}^2
                                     + M_{\overline{K}^0}^2 + M_{K^-}^2)$
                                                                  & blue  \\
         mesons   & $X_{\eta_8}^2 
                            = \half(M^2_{\pi^0} + M^2_{\eta_8})$    & black  \\
      \hline 
         Vector   & $X_\rho = \sixth( M_{K^{*+}} + M_{K^{*0}}
                                     + M_{\rho^+} +  M_{\rho^-}
                                     + M_{\overline{K}^{*0}} + M_{K^{*-}})$
                                                                  & blue   \\
         mesons   & $X_{\phi_8} = \half(M_{\rho^0} + M_{\phi_8})$    & black  \\
                  & $X_{\phi_s} = \third(2M_{\rho^0} + M_{\phi_s})$  &        \\
      \hline 
      \hline 
         Octet    & $X_N = \sixth( M_p + M_n
                                 + M_{\Sigma^+} +  M_{\Sigma^-}
                                 + M_{\Xi^0} + M_{\Xi^-} )$         & blue  \\
         baryons  & $X_\Lambda = \half( M_{\Lambda} + M_{\Sigma^0})$
                                                                  & black \\
      \hline
         Decuplet & $X_\Delta = \third( M_{\Delta^{++}} + M_{\Delta^-}
                                     + M_{\Omega^-})$              & red   \\
         baryons  & $X_{\Xi^*} = \sixth( M_{\Delta^+} + M_{\Delta^0}
                                       + M_{\Sigma^{*+}} + M_{\Sigma^{*-}} 
                                       + M_{\Xi^{*0}} + M_{\Xi^{*-}})$
                                                                  & blue   \\
                  & $X_{\Sigma^*} = M_{\Sigma^{*0}}$                 & black \\
      \hline 
      \end{tabular}
   \end{center}
\caption{Permutation invariant mass combinations,
         see Fig.~\protect\ref{permset}. $\phi_s$ is a
         fictitious $s \overline{s}$ particle; $\eta_8$
         and $\phi_8$ are pure octet mesons.
         The colours in the third column correspond to
         Fig.~\protect\ref{permset}.}
\label{perminv}
\end{table}
The permutation group $S_3$ yields a lot of useful relations, but
cannot capture the entire structure. For example, there is no way to
make a connection between the $\Delta^{++}(uuu)$
and the $\Delta^+(uud)$ by permuting quarks. 
To go further, we need to classify physical quantities by $SU(3)$
(containing the permutation group $S_3$ as a subgroup), which we
shall consider now.


\subsection{Taylor expansion}
\label{taylor_expansion}


We want to describe how physical quantities depend on the quark masses.
To do this we will Taylor expand about a symmetric reference point 
\begin{eqnarray}
   (m_u, m_d, m_s) = ( m_0, m_0, m_0 ) \,.
\label{sympoint} 
\end{eqnarray}
Our results will be polynomials in the quark masses, we will 
express them in terms of $\overline{m}$ and $\delta m_q$ of
eq.~(\ref{delta_mq}). The main idea is to classify all possible
mass polynomials by their transformation properties under the
permutation  group $S_3$ and under the full flavour group $SU(3)$,
and classify hadronic observables in the same way. $\overline{m}$
and $\delta m_q$ are a natural basis to choose as $\overline{m}$
is purely singlet and $\delta m_q$ is non-singlet.
The alternative $m_q - m_0$ would be less useful as it contains
a mixture of singlet and non-singlet quantities.

The Taylor expansion of a given observable can only include
the polynomials of the same symmetry as the observable.
The Taylor expansions of hadronic quantities in the same
$SU(3)$ multiplet but in different $S_3$ representations
will have related expansion coefficients. (We will show examples
of the latter, e.g.\ in 
eqs.~(\ref{decup_mass_rel1})--(\ref{decup_mass_rel3}).)

While we can always arrange polynomials to be in definite permutation
group states, when we get to polynomials of $O(\delta m_q^2)$ we find
that a polynomial may be a mixture of several $SU(3)$ representations, 
but the classification is still useful. In Table~\ref{cubic}
\begin{table}[htb] 
   \begin{center} 
   \begin{tabular} {ccccccccc}
   Polynomial &  & $S_3$ & \multicolumn{6}{c}{$SU(3)$}                      \\
   \hline 
   $1$ & \checkmark & $A_1$ & $1$ &  &  &  &  &                             \\
   \hline
   \hline 
   $(\overline{m} - m_0)$ &  & $A_1$ & $1$ &  &  &  &  &                    \\
   \hline
   $\delta m_s$ & \checkmark & $E^+$ &  & $8$ &  &  &  &                   \\
   $(\delta m_u - \delta m_d)$ & \checkmark & $E^-$ &  & $8$ &  &  &  &   \\
   \hline
   \hline 
   $(\overline{m} - m_0)^2$ & & $A_1$ & 1 &  &  &  &  &                     \\
   $(\overline{m} - m_0) \delta m_s$ &  & $E^+$ &  & $8$ &  &  &  &        \\
   $(\overline{m} - m_0)(\delta m_u - \delta m_d)$ &  & $E^-$ &  & $8$ &  
                                                                   &  &  &  \\
   \hline
   $\delta m_u^2 + \delta m_d^2 +\delta m_s^2$ & \checkmark & $A_1$ & 1 
                                                         &  &  &  & $27$ &  \\
   $3 \delta m_s^2 - (\delta m_u - \delta m_d)^2$ & \checkmark & $E^+$ &
                                                      & $8$ &  &  & $27$ &  \\
   $\delta m_s (\delta m_d - \delta m_u)$ & \checkmark & $E^-$ &  & $8$
                                                           &  &  & $27$ &   \\
   \hline
   \hline
   $(\overline{m} - m_0)^3$ &  & $A_1$ & $1$ &  &  &  &  &                  \\
   $(\overline{m} - m_0)^2 \delta m_s$ &  & $E^+$ &  & $8$ &  &  &  &      \\
   $(\overline{m} - m_0)^2 (\delta m_u - \delta m_d)$ &  & $E^-$ &  &
                                                            $8$ &  &  &  &  \\ 
   $(\overline{m} - m_0)(\delta m_u^2 + \delta m_d^2 +\delta m_s^2)$ & 
                                           & $A_1$ & $1$ &  &  &  & $27$ &  \\
   $(\overline{m} - m_0)\left[3 \delta m_s^2 - (\delta m_u - \delta m_d)^2
                        \right]$  &  & $E^+$ &  & $8$ &  &  & $27$ &        \\
   $(\overline{m} - m_0)\delta m_s(\delta m_d - \delta m_u)$ &  & $E^-$
                                                   &  & $8$ &  &  & $27$ &  \\
   \hline 
   $\delta m_u \delta m_d \delta m_s $ & \checkmark &$A_1$& $1$ &  &  &
                                                             & $27$ & $64$  \\
   $\delta m_s (\delta m_u^2 + \delta m_d^2 +\delta m_s^2 )$
                         & \checkmark & $E^+$ &  & $8$ &  &  & $27$ & $64$  \\
   $(\delta m_u - \delta m_d) (\delta m_u^2 
                    + \delta m_d^2 +\delta m_s^2 )$ & \checkmark & $E^-$
                                              &  & $8$ &  &  & $27$ & $64$  \\
   $(\delta m_s - \delta m_u)(\delta m_s-\delta m_d)
                      (\delta m_u-\delta m_d)$ & \checkmark & $A_2$ &  & 
                                         & $10$ & $\overline{10}$ &  &$64$  \\
   \hline 
\end{tabular}
\caption{All the quark-mass polynomials up to $O(m_q^3)$, classified by
         symmetry properties. A tick (\checkmark) marks the polynomials
         relevant on a constant $\overline{m}$ surface. These polynomials
         are plotted in Fig.~\ref{triplot}. If we want to make an expansion
         valid when $\overline{m}$ varies, then all the polynomials
         in the table (with and without ticks) are needed.}
\label{cubic}
\end{center} 
\end{table} 
we classify all the polynomials which could occur in a Taylor
expansion about the symmetric point, eq.~(\ref{sympoint}),
up to $O(\delta m_q^3)$.

Many of the polynomials in the table have factors of
$(\overline{m}-m_0)$. These polynomials drop out
if we restrict ourselves to the surface of constant
$\overline{m} = m_0$, leaving only the polynomials marked with
a tick (\checkmark) in Table~\ref{cubic}. At $O(m_q^k)$ there
are $k+1$ independent polynomials needed to describe functions
on the constant $\overline{m}$ surface (the polynomials with the ticks),
but ${1 \over 2}(k+1)(k+2)$ polynomials needed if the constraint
$\overline{m} = \mbox{constant}$ is dropped (all polynomials, with
and without ticks). Thus the advantage of working in the constant
$\overline{m}$ surface increases as we proceed to higher order
in $m_q$. 

Since we are keeping $\overline{m}$ constant, we are only changing
the octet part of the mass matrix in eq.~(\ref{massmat}).
Therefore, to first order in the mass change, only octet
quantities can be affected. $SU(3)$ singlets have no linear
dependence on the quark mass, as we have already seen by
the symmetry argument eq.~(\ref{symarg}), but we now see that
all quantities in $SU(3)$ multiplets higher than the octet cannot
have linear terms. This provides a constraint on the hadron masses
within a multiplet and leads (as we shall see) to the Gell-Mann--Okubo
mass relations \cite{gell-mann62a,okubo62a}.

When we proceed to quadratic polynomials we can construct 
polynomials which transform like mixtures of the $1$, $8$ and $27$ 
multiplets of $SU(3)$, see Table~\ref{cubic}. Further representations,
namely the $10$, ${\overline{10}}$ and $64$, first occur when
we look at cubic polynomials in the quark masses,
see again Table~\ref{cubic}. 

In a little more detail, constructing polynomials with a definite
$S_3$ classification is fairly straightforward, we have
to see what happens to each polynomial under simple interchanges
(e.g.\ $u \leftrightarrow d$) and cyclic permutations
(e.g.\ $u \to s$, $s \to d$, $d \to u$). The $S_3$ column
of Table~\ref{cubic} is easy to check by hand.
The $SU(3)$ assignment of polynomials is less straightforward.
Only the simplest polynomials belong purely to a single
$SU(3)$ multiplet; most polynomials contain mixtures of
several multiplets. The non-singlet mass is an octet of $SU(3)$, 
so quadratic polynomials in $\delta m_q$ can contain representations 
which occur in $8 \otimes 8$, cubic polynomials representations
which occur in $8 \otimes 8 \otimes 8$. We can find out which
representations are present in a given polynomial by using the 
Casimir operators of $SU(3)$~\cite{pfeifer03a,greiner89}.
That operator was programmed in Mathematica, and used to analyse
our polynomial basis. Some more details are presented in
Appendix~\ref{group_theory} (in section~\ref{group_class}).
The results of the calculation are recorded in the $SU(3)$
section of Table~\ref{cubic}.

\begin{figure}[p]
   \begin{center} 
      \includegraphics[width=6.00cm,angle=270]{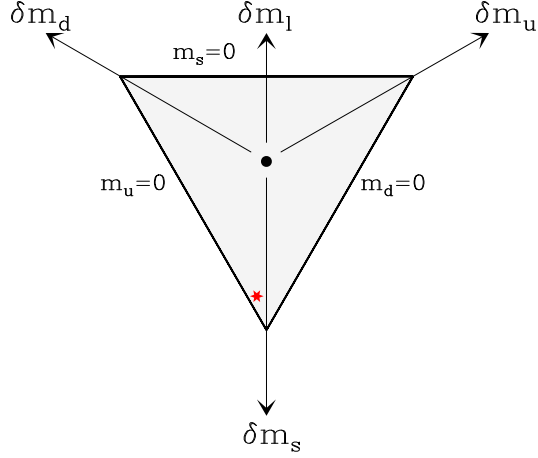}
   \end{center} 
\caption{The allowed quark mass region on the
         $\overline{m} = \mbox{constant}$ surface is an
         equilateral triangle. The black point at the centre
         is the symmetric point, the red star is the physical point.
         $2+1$ simulations lie on the vertical symmetry axis.
         The physical point is slightly off the $2+1$ axis because
         $m_d > m_u$.}
\label{trisk}
\end{figure}

\begin{figure}[p]
   \begin{center} 
      \includegraphics[width=11.00cm]{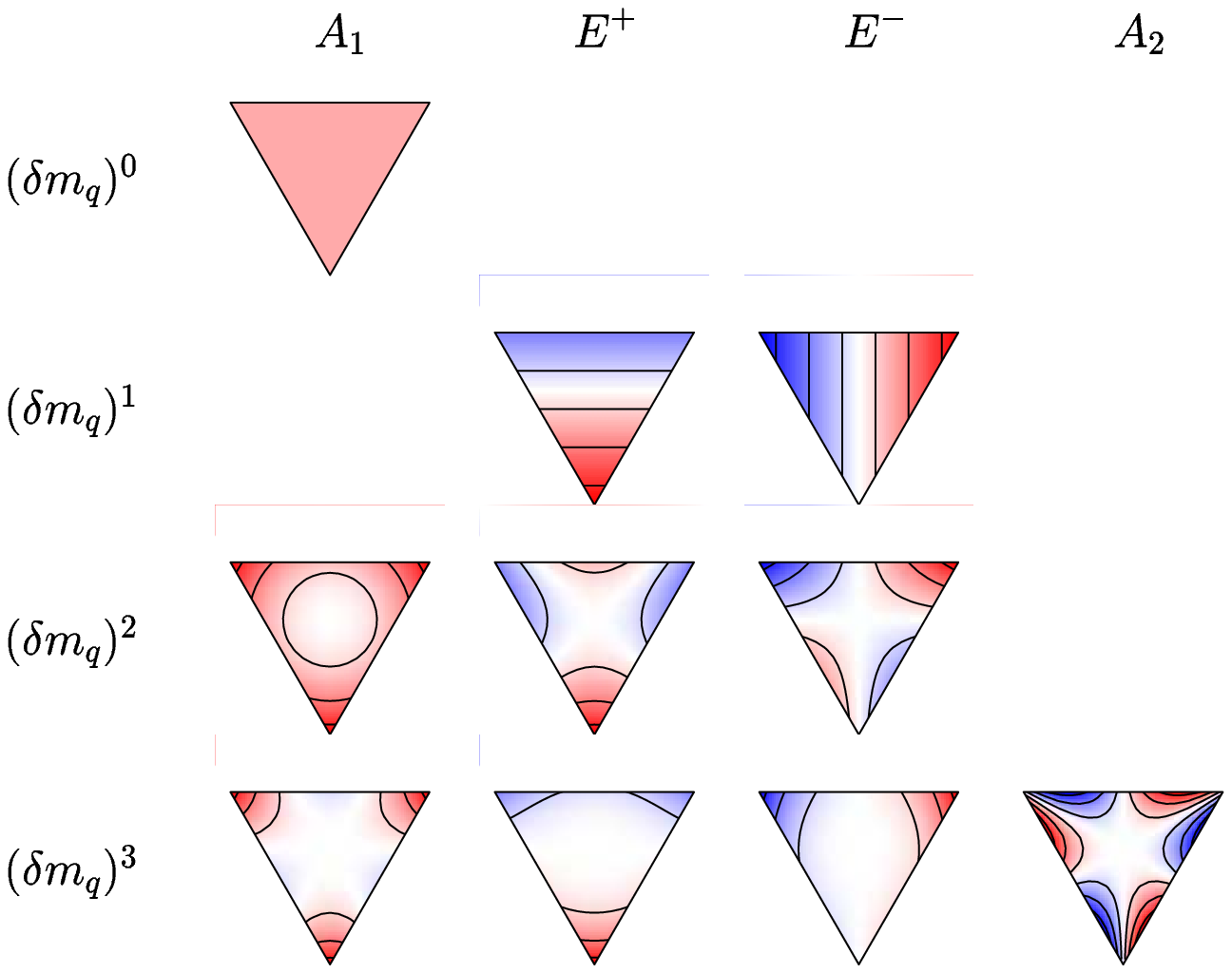}
  \end{center} 
\caption{Contour plots of the polynomials relevant for the
         constant $\overline{m}$ Taylor expansion,
         see Table~\protect\ref{cubic}.
         A red(dish) colour denotes a positive number
         while a blue(ish) colour indicates a negative number.
         If $m_u=m_d$ (the $2+1$ case), only the polynomials in the
         $A_1$ and $E^+$ columns contribute. Each triangle in this
         figure uses the coordinate system explained in 
         Fig.~\protect\ref{trisk}.}
\label{triplot}
\end{figure}

The allowed quark mass region on the $\overline{m} = \mbox{constant}$
surface is an equilateral triangle, as shown in Fig.~\ref{trisk}.
Plotting the polynomials of Table~\ref{cubic} across this triangular
region then gives the plots in Fig.~\ref{triplot}, where the colour
coding indicates whether the polynomial is positive (red) or negative (blue).

As a first example of the use of these tables, consider the 
Taylor expansion for the scale $r_0/a$ up to cubic order in the
quark masses. As discussed previously, this is a gluonic quantity,
blind to flavour, so it has symmetry $A_1$ under the $S_3$
permutation group. Therefore its Taylor expansion only contains
polynomials of symmetry $A_1$. If we keep $\overline{m}$,
the average quark mass, fixed, the expansion of $r_0/a$
must take the form 
\begin{eqnarray}
   {r_0 \over a} = \alpha + \beta (\delta m_u^2 + \delta m_d^2 +\delta m_s^2 )  
                   + \gamma \, \delta m_u \delta m_d \delta m_s \,,
\label{r0_expansion}
\end{eqnarray}
with just $3$ coefficients. Interestingly, we could find all $3$ 
coefficients from $2+1$ data, so we would be able to predict
$1+1+1$ flavour results from fits to $2+1$ data. This is common.
If we allow $\overline{m}$ to vary too, we would need $7$ coefficients
to give a cubic fit for $r_0$ (all the $A_1$ polynomials
in Table~\ref{cubic} both ticked and unticked).
This point is further discussed in section~\ref{mbar_varies}.
If we did not have any information on the flavour symmetry of $r_0$
we would need all the polynomials in Table~\ref{cubic}, which would
require $20$ coefficients.


\subsection{$\mathbf{O(a)}$ improvement of quark masses}
\label{improvement}


Before classifying the hadron mass matrix, we pause and consider
the $O(a)$ improvement of quark masses. (If we are considering chiral
fermions, we have `automatic $O(a)$ improvement', see e.g.\ \cite{capitani99a}
for a discussion.) In writing down expressions for bare and improved
quark masses, it is natural to expand about the chiral point,
all three quarks massless, which means setting $m_0 = 0$
in the expressions in Table~\ref{cubic}. Later, when we consider
lattice results, we want to expand around a point where we can run
simulations, so we will normally have a non-zero $m_0$.  
 
Improving the quark masses requires us to add improvement terms 
of the type $a m_q^2$ to the bare mass. We can add $SU(3)$-singlet
improvement terms to the singlet quark mass, $SU(3)$-octet improvement
terms to the non-singlet quark mass. We are led to the following
expressions for the improved and renormalised quark masses
\begin{eqnarray}
   \overline{m}^{\R} 
      &=& Z_m^{\Si}\left[
          \overline{m} + a\left\{b_1\overline{m}^2 
                         + b_2(\delta m_s^2 + \delta m_u^2 + \delta m_d^2 )
                          \right\}
                   \right]
                                                           \nonumber \\ 
   \delta m_s^{\R} 
      &=& Z_m^{\NS}\left[
          \delta m_s + a\left\{b_3\overline{m}\delta m_s 
                         + b_4(3\delta m_s^2 - (\delta m_u - \delta m_d)^2 )
                        \right\}
                   \right] \,,
\label{strimp} 
\end{eqnarray}
together with $Z_m^{\Si} = Z_m^{\NS} r_m$,
eqs.~(\ref{mr2mbare}), (\ref{alphaZ}).
We have improved $\overline{m}^{\R}$ by adding the two possible singlet 
terms from the quadratic section of Table~\ref{cubic}, and improved
$\delta m_s^{\R}$ by adding the two possible $E^+$ octet polynomials.
Note that if we keep $\overline{m}$ constant, we only need to consider
the improvement terms $b_2$ and $b_4$. The $b_1$ and $b_3$ terms
could be absorbed into the $Z$ factors. Simplifications of this
sort are very common if $\overline{m}$ is kept fixed.

We get expressions for the $u$ and $d$ quark mass improvement
by flavour-permuting eq.~(\ref{strimp})
\begin{eqnarray}
   \delta m_u^{\R} 
      &=& Z_m^{\NS}\left[
          \delta m_u + a\left\{ b_3\overline{m}\delta m_u 
                         + b_4(3\delta m_u^2 - (\delta m_s - \delta m_d)^2 )
                        \right\}
                   \right]
                                                               \nonumber \\
   \delta m_d^{\R} 
      &=& Z_m^{\NS}\left[
          \delta m_d + a\left\{ b_3\overline{m}\delta m_d 
                         + b_4(3\delta m_d^2 - (\delta m_s - \delta m_u)^2 )
                        \right\}
                   \right]
                                                               \nonumber \\
   \delta m_u^{\R} - \delta m_d^{\R} 
      &=& Z_m^{\NS}\left[
          \delta m_u - \delta m_d  \right.
                                                                         \\
      & & \hspace*{0.75in} \left.
               + a\left\{ b_3\overline{m}(\delta m_u - \delta m_d)
               + 6b_4 \delta m_s (\delta m_d - \delta m_u)
             \right\}
                   \right] \,.
                                                               \nonumber
\end{eqnarray} 
The improvement terms for $\delta m_u - \delta m_d$ are proportional
to the two $E^-$, $SU(3)$-octet, quadratic polynomials.
(We have to use the identity, eq.~(\ref{zerosum}), 
to bring the result to the desired form -- which will often be
the case in what follows.)

Table~\ref{cubic} is based purely on flavour arguments,
we would hope that all the results are true whether we use bare
or renormalised quantities, and also independently of whether
we work with a naive bare mass, or a bare mass with $O(a)$ 
improvement terms. Let us check if this is true. The first thing
we need to know is whether the zero-sum identity eq.~(\ref{zerosum})
survives renormalisation and improvement. Using
the previous equations we find
\begin{eqnarray} 
   \delta m_u^{\R} + \delta m_d^{\R} + \delta m_s^{\R}
      &=& Z_m^{\NS} \left[
          (\delta m_u + \delta m_d + \delta m_s) \right.
                                                               \nonumber \\
      & & \hspace*{0.50in} + a\left\{
            b_3  \overline{m}(\delta m_u + \delta m_d + \delta m_s) \right.
                                                               \nonumber \\
      & & \hspace*{0.75in} \left. \left. +
            b_4 (\delta m_u + \delta m_d + \delta m_s)^2 \right\}
                    \right]
                                                               \nonumber \\
      &=& 0 \,,
 \end{eqnarray} 
showing that eq.~(\ref{zerosum}) is not violated by improvement or
renormalisation. 

The next point we want to check is if the symmetry of a polynomial
depends on whether we expand in terms of improved or unimproved masses. 
As an example, let us look at the quadratic polynomial
\begin{eqnarray}
   \delta m_s^{\R} ( \delta m_d^{\R} - \delta m_u^{\R}) \,,
\end{eqnarray}
which has permutation symmetry $E^-$, and $SU(3)$ content octet 
and $27$-plet. Expanding to first order in the lattice spacing $a$ 
we find 
\begin{eqnarray}
   \delta m_s^{\R} ( \delta m_d^{\R} - \delta m_u^{\R})
      &=& (Z_m^{\NS})^2 \left[ 
          \delta m_s ( \delta m_d - \delta m_u ) \right.
                                                               \nonumber \\
      & & \hspace*{0.50in} + a\left\{
             2 b_3 \overline{m} \delta m_s ( \delta m_d - \delta m_u)
             \right.
                                                                         \\
      & & \hspace*{0.85in} + \left.      \left.
             2 b_4 (\delta m_u - \delta m_d)
                    (\delta m_u^2 + \delta m_d^2 + \delta m_s^2)
            \right\}
                        \right] \,.
                                                               \nonumber
\end{eqnarray} 
The mass improvement terms have generated two extra cubic polynomials, 
but they are both polynomials with the same symmetry as the initial
polynomial. The same holds for the other quadratic terms.
This shows that Table~\ref{cubic} applies both to improved and
unimproved masses.

Thus our conclusion is that the flavour expansion results are true
whether we use bare or renormalised quantities, and also independently
of whether we work with a naive bare mass, or a bare mass with $O(a)$ 
improvement terms.

Finally we compare these results with those obtained in \cite{bhattacharya05a},
to see whether we can match the $4$ improvement terms found
in eq.~(\ref{strimp}) to the $4$ terms introduced there, namely
\begin{eqnarray}
   m_q^{\R} &=& Z_m^{\NS} \, \left[ 
               m_q + (r_m-1)\overline{m}
                  \phantom{\left\{ \overline{m^2} \right\}} \right.
                                                   \label{bhatt_iqm} \\
         & & \hspace*{0.35in} \left.
             + a\left\{ b_mm_q^2 + 3\overline{b}_mm_q\overline{m}
                        + (r_md_m - b_m) \overline{m^2}
                        + 3(r_m\overline{d}_m - \overline{b}_m) \overline{m}^2
                    \right\}
                        \right]  \,,
                                                              \nonumber
\end{eqnarray}
where $\overline{m^2} = {1\over 3}(m_u^2 + m_d^2 + m_s^2)$.
At first this looks different from eq.~(\ref{strimp}), but this is
just due to a different choice of basis polynomials. The quadratic
polynomials in eq.~(\ref{bhatt_iqm}) are simple linear combinations
of those in eq.~(\ref{strimp}).

From eq.~(\ref{strimp}) we have
\begin{eqnarray}
   m_s^{\R}
      &=& \delta m_s^{\R} + \overline{m}^{\R} 
                                                                         \\
      &=& Z_m^{\NS} \left[
          m_s + (r_m-1)\overline{m}
              + a\left\{ b_3\overline{m}\delta m_s 
                        + b_4(3\delta m_s^2 - (\delta m_u - \delta m_d)^2 )
                 \right. \right.
                                                               \nonumber \\
       & & \hspace*{1.85in} \left. \left.
                    + r_mb_1\overline{m}^2
                    + r_mb_2(\delta m_s^2 + \delta m_u^2 + \delta m_d^2)
                        \right\}
                   \right] \,,
                                                               \nonumber
\end{eqnarray}
so we now equate the terms to those in eq.~(\ref{bhatt_iqm}).
We must first re-write
\begin{eqnarray}
   3\delta m_s^2 - (\delta m_u - \delta m_d)^2
      = 6\left[m_s^2 - 2\overline{m}m_s - \overline{m^2} + 2\overline{m}^2
         \right] \,,
\end{eqnarray}
so
\begin{eqnarray}
   m_s^{\R} 
      &=& Z_m^{\NS} \, \left[
            m_s + (r_m-1)\overline{m}
                      \right.
                                                               \nonumber \\
      & & \hspace*{0.50in}
          + a \left\{ b_3\overline{m}( m_s - \overline{m})
                         + 6b_4(m_s^2 - 2\overline{m}m_s - \overline{m^2}
                               + 2\overline{m}^2)
              \right.
                                                               \nonumber \\
      & & \hspace*{0.75in} \left. \left.
                      + r_mb_1\overline{m}^2
                      + 3r_mb_2(\overline{m^2} - \overline{m}^2)
              \right\}
                      \right] \,,
\end{eqnarray}
which gives the results
\begin{eqnarray}
   \begin{array}{ccl}
      b_m &=& 6b_4 
                                                              \nonumber \\
      \overline{b}_m
          &=& \third b_3 - 4b_4
                                                              \nonumber \\
      d_m &=& 3b_2
                                                              \nonumber \\
      \overline{d}_m 
          &=& \third b_1 - b_2
                                                              \nonumber \\
   \end{array}
   \,, \qquad \mbox{or} \qquad
   \begin{array}{ccl}
      b_1 &=& 3\overline{d}_m + d_m
                                                              \nonumber \\
      b_2 &=& \third d_m
                                                              \nonumber \\
      b_3 &=& 3\overline{b}_m +2b_m
                                                              \nonumber \\
      b_4 &=& \sixth b_m
                                                              \nonumber \\
   \end{array} \,.
                                                                        \\
\end{eqnarray}


\subsection{$\mathbf{SU(3)}$ and $\mathbf{S_3}$ classification of hadron
            mass matrices} 
\label{class_mass_matrix}


In eq.~(\ref{massmat}) we split the quark mass matrix into a singlet
part and two octet parts. We want to make a similar decomposition 
of the hadron mass matrices. We start with the decuplet mass matrix
because it is simpler than the octet mass matrix.


\subsubsection{The decuplet mass matrix}

\begin{figure}[htb]
   \vspace*{0.15in}
   \begin{center}
      \includegraphics[width=4.50cm,angle=270]{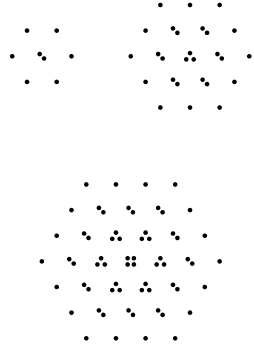}
   \end{center} 
   \caption{An illustration of the octet, $27$-plet and $64$-plet
            representations of $SU(3)$. The number of spots in
            the central location gives the number of
            flavour-conserving operators in each multiplet. 
            In the octet, the $2$ operators form an $E$ doublet of the
            permutation group. In the $27$-plet the $3$ operators are
            an $A_1$ singlet and an $E$ doublet. In the $64$-plet
            the centre operators are an $A_1$ singlet, an $E$ doublet
            and an $A_2$ singlet.}
   \label{decup_weight}
\end{figure}

The decuplet mass matrix is a $10 \times 10$ diagonal matrix. 
From $SU(3)$ group algebra we know
\begin{equation}
   10 \otimes \overline{10} = 1 \oplus 8 \oplus 27 \oplus 64 \,.
\end{equation}
The singlet matrix is the identity matrix, the octet representation
contains $2$ diagonal matrices ($\lambda_3$ and $\lambda_8$),
the $27$-plet has $3$ diagonal matrices, and the $64$-plet includes
$4$ diagonal matrices, see Fig.~\ref{decup_weight}.
This gives us a basis of $10$ diagonal matrices, into which we can
decompose the decuplet mass matrix.

We can use the Casimir operator to project out the diagonal
matrices in a particular $SU(3)$ representation
(see Appendix~\ref{grp_anal_mass} for a fuller discussion).
As an example of a matrix with pure octet symmetry,
we can take the operator $2 I_3$. (We have multiplied $I_3$
by $2$ simply to avoid having fractions in the matrix.)
Since we know the isospins of all the decuplet baryons,
we can write down

\begin{small}
\begin{eqnarray} 
   {\Delta^- \ \, \Delta^0  \,\ \Delta^+ \,\ \Delta^{++}
   \,   \Sigma^{*-} \ \ \Sigma^{*0} \,\ \Sigma^{*+}
   \,\  \Xi^{*-} \,\ \Xi^{*0} \,\ \Omega^- \ } & &
                                                \label{decnot}   \\
   \left( \begin{array}{ccccccccccc}
             -3 & 0 & 0\; & 0\; & 0 & 0 & 0 & 0 & 0 & 0  \\
              0 &-1 & 0\; & 0\; & 0 & 0 & 0 & 0 & 0 & 0  \\
              0 & 0 & 1\; & 0\; & 0 & 0 & 0 & 0 & 0 & 0  \\
              0 & 0 & 0\; & 3\; & 0 & 0 & 0 & 0 & 0 & 0  \\
              0 & 0 & 0\; & 0\; &-2 & 0 & 0 & 0 & 0 & 0  \\
              0 & 0 & 0\; & 0\; & 0 & 0 & 0 & 0 & 0 & 0  \\
              0 & 0 & 0\; & 0\; & 0 & 0 & 2 & 0 & 0 & 0  \\
              0 & 0 & 0\; & 0\; & 0 & 0 & 0 &-1 & 0 & 0  \\
              0 & 0 & 0\; & 0\; & 0 & 0 & 0 & 0 & 1 & 0  \\
              0 & 0 & 0\; & 0\; & 0 & 0 & 0 & 0 & 0 & 0  \\
          \end{array}
   \right)
   &\equiv& \begin{array}{ccccccc}
               -3\!\!\! &        & -1 &   & 1 &   & 3 \\
                        &        &    &   &   &   &   \\
                        & -2\!\! &    & 0 &   & 2 &   \\
                        &        &    &   &   &   &   \\
                        &        & -1 &   & 1 &   &   \\
                        &        &    &   &   &   &   \\
                        &        &    & 0 &   &   &   \\
            \end{array}
                                                        \nonumber
\end{eqnarray}
\end{small}

\noindent
where we have used a more compact notation to record the
diagonal elements on the right-hand side. The entry in the 
$\Delta^-$ column of the matrix is $-3$, so on the right-hand side
we put a $-3$ in the position of the $\Delta^-$ in the usual 
decuplet diagram, and so on. By considering the reflection and
rotation symmetries of the right-hand side of eq.~(\ref{decnot})
we can see that this matrix corresponds to the basis element 
$E^-$ of the doublet representation of $S_3$.

\begin{figure}[htb]
   \begin{center} 
      \includegraphics[width=11cm,angle=270]{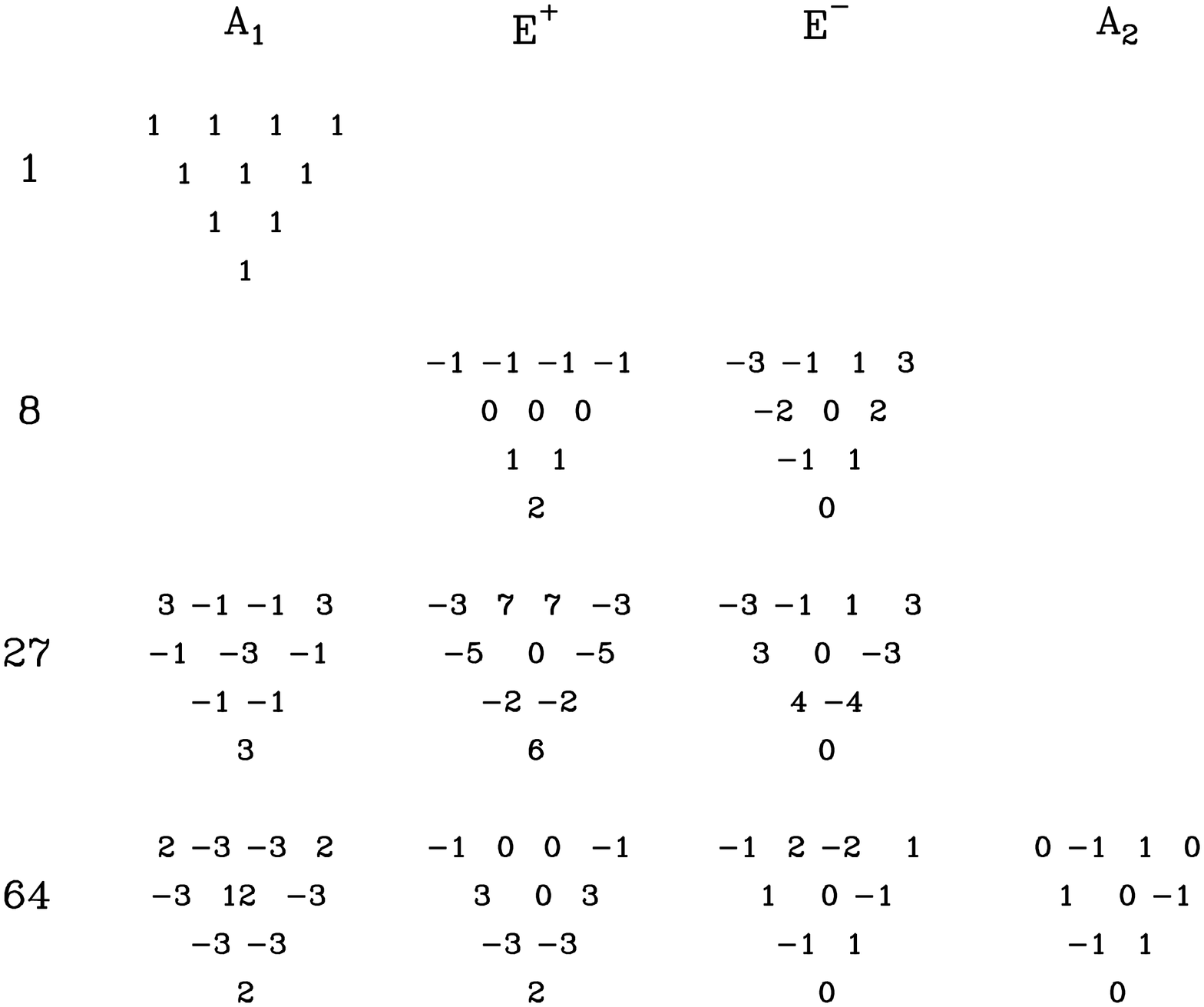}
   \end{center}
\caption{The matrices for projecting out decuplet mass contributions
         of known symmetry -- see eq.~(\protect\ref{decnot})
        for an explanation of the notation.}
\label{decumat}
\end{figure}

In Fig.~\ref{decumat} we 
show all $10$ diagonal matrices, in this compact notation. 
These matrices are orthogonal, in the sense
\begin{equation}
   \mbox{Tr}[ \tau_a \tau_b ] 
      = 0 \quad \mbox{if} \quad a \ne b \,,
\end{equation}
(where $\tau_a$ is any of the matrices of Fig.~\ref{decumat})
so they can be used to project out mass combinations 
which have simple quark mass dependencies,
see Fig.~\ref{decumat}, and Table~\ref{mat10}. 
\begin{table}[htb]
   \begin{center}
   \begin{tabular}{rrrrrrrrrrcc}
      $\Delta^- $   & $\Delta^0 $  & $\Delta^+ $   & $\Delta^{++} $ &
      $\Sigma^{*-}$ & $\Sigma^{*0}$ & $\Sigma^{*+}$ & $\Xi^{*-}$     &
      $\Xi^{*0}$    & $\Omega^-$   & $S_3$          &
      $SU(3)$                                                         \\
      \hline
      1 & 1 & 1 & 1 & 1 & 1 & 1 & 1 & 1 & 1 & $A_1$ & 1               \\
      \hline
      $-1$ & $-1$ & $-1$ & $-1$ & 0 & 0 & 0 & 1 & 1 & 2 & $E^+$ & 8   \\
      $-3$ & $-1$ & 1 & 3 & $-2$ & 0 & 2 & $-1$ & 1 & 0 & $E^-$ & 8   \\
      \hline
      $3$ & $-1$ & $-1$ & 3 & $-1$ & $-3$ & $-1$ & $-1$ & $-1$ & 3 & 
                                                   $A_1$ & 27         \\
      $-3$ & $ 7$ & $ 7$ & $-3$ &$-5$ & $0$ & $-5$ &$-2$ & $-2$ & 6 & 
                                                     $E^+$ & 27       \\
      $-3$ & $-1$ & $ 1$ & $ 3$ &$ 3$ & $0$ & $-3$ & $ 4$ & $-4$ & 0 & 
                                                     $E^-$ & 27       \\
      \hline
      $2$ & $-3$ & $-3$ & 2 & $-3$ & $12$ & $-3$ & $-3$ & $-3$ & 2 & 
                                                   $A_1$ & 64         \\
      $-1$ & $ 0$ & $ 0$ & $-1$ &$ 3$ & $0$ & $ 3$ & $-3$ & $-3$ & 2 & 
                                                     $E^+$ & 64       \\
      $-1$ & $ 2$ & $-2$ & 1 &$ 1$ & $ 0$ & $-1$ & $-1$ & $ 1$ & 0 & 
                                                   $E^-$ & 64         \\
      $ 0$ & $-1$ & $ 1$ & 0 &$ 1$ & $ 0$ & $-1$ & $-1$ & $ 1$ & 0 & 
                                                   $A_2$ & 64         \\
      \hline
   \end{tabular}
   \end{center} 
\caption{Decuplet mass matrix contributions, classified by permutation
         and $SU(3)$ symmetry, see Fig.~\protect\ref{decumat}.}
\label{mat10}
\end{table}

Let us now give some examples of mass formulae.
First we look at the singlet of the decuplet mass matrix.
Because we are keeping $\overline{m} = \mbox{constant}$ only
the terms with ticks in Table~\ref{cubic} contribute.
This gives from Table~\ref{mat10},
\begin{eqnarray}
   \lefteqn{M_{\Delta^-} + M_{\Delta^0} + M_{\Delta^+} + M_{\Delta^{++}}}
   & &                                                    \nonumber      \\
   \lefteqn{
   + M_{\Sigma^{*-}} +  M_{\Sigma^{*0}} + M_{\Sigma^{*+}}
   + M_{\Xi^{*-}} + M_{\Xi^{*0}} + M_{\Omega^{-}}}
   & &                                                    \nonumber      \\
   &=& 10M_0\ + B_1\; (\delta m_u^2+\delta m_d^2+\delta m_s^2)
             + C_1\; \delta m_u \delta m_d \delta m_s  \,.\
\label{decup_sing_mass_rel}
\end{eqnarray}
This equation being a singlet has the same form as for $r_0/a$,
eq.~(\ref{r0_expansion}).

As a further example for the $27$-plet component of the decuplet
mass matrix, we see from Table~\ref{mat10} that there are three
mass combinations which transform as $27$-plets, giving three
related mass relations
\begin{eqnarray}
   \lefteqn{
   3 M_{\Delta^-} - M_{\Delta^0} - M_{\Delta^+} +3 M_{\Delta^{++}}}
   & &                                                      \nonumber \\
   \lefteqn{
   - M_{\Sigma^{*-}} - 3 M_{\Sigma^{*0}} - M_{\Sigma^{*+}}
   - M_{\Xi^{*-}} - M_{\Xi^{*0}} + 3 M_{\Omega^{-}}}
   & &                                                      \nonumber \\
   & & = b_{27} \left[ \delta m_u^2 + \delta m_d^2 + \delta m_s^2 \right]  
       + 9 c_{27}  \delta m_u \delta m_d \delta m_s
\label{decup_mass_rel1}                                               \\
   \lefteqn{
   - 3 M_{\Delta^-} + 7 M_{\Delta^0} + 7 M_{\Delta^+} - 3 M_{\Delta^{++}}}
                                                            \nonumber \\
   \lefteqn{
   - 5 M_{\Sigma^{*-}} - 5 M_{\Sigma^{*+}}
   - 2 M_{\Xi^{*-}} - 2 M_{\Xi^{*0}} + 6 M_{\Omega^{-}}}
   & &                                                      \nonumber \\
   & & = b_{27} \left[ 3\delta m_s^2 -(\delta m_u- \delta m_d)^2 \right]
         + 3 c_{27} \delta m_s \left( \delta m_u^2 + \delta m_d^2 
                                      + \delta m_s^2 \right)
\label{decup_mass_rel2}                                               \\
   \lefteqn{
   - 3 M_{\Delta^-} - M_{\Delta^0} + M_{\Delta^+} + 3 M_{\Delta^{++}} }
   & &                                                      \nonumber \\
   \lefteqn{
   + 3 M_{\Sigma^{*-}} - 3 M_{\Sigma^{*+}}
   + 4 M_{\Xi^{*-}} - 4 M_{\Xi^{*0}}}
   & &                                                      \nonumber \\
   & & = 2 b_{27} (\delta m_d-\delta m_u) \delta m_s
         + c_{27} (\delta m_u - \delta m_d) \left( \delta m_u^2 + \delta m_d^2
                                         + \delta m_s^2  \right) \,.
\label{decup_mass_rel3}
\end{eqnarray}
The coefficients in eqs.~(\ref{decup_mass_rel1})--(\ref{decup_mass_rel3})
are connected, they all involve just one quadratic parameter, $b_{27}$,
and one cubic parameter, $c_{27}$. We now want to explain the
different numerical coefficients in front of these parameters.
These can be checked by considering some simple symmetry limits.
First consider the isospin limit, equal masses for the $u$ and
$d$ quarks, $\delta m_u \to \delta m_l$, $\delta m_d \to \delta m_l$,
$\delta m_s \to -2 \delta m_l$ (from eq.~(\ref{zerosum})).
In this limit, eq.~(\ref{decup_mass_rel3}) reduces to $0=0$, while
eqs.~(\ref{decup_mass_rel1}), (\ref{decup_mass_rel2}) both become
\begin{equation}
   4 M_\Delta - 5 M_{\Sigma^*} - 2 M_{\Xi^*} + 3 M_\Omega
     = 6 b_{27} \delta m_l^2 - 18 c_{27} \delta m_l^3 \,.
\end{equation}
To include eq.~(\ref{decup_mass_rel3}) in our checks,
we can take the $U$-spin limit,
$m_s \to m_d$, i.e.\  $\delta m_s \to \delta m_d$,
$\delta m_u \to -2 \delta m_d$. In this limit all decuplet
baryons with the same electric charge would have equal mass,
because they would be in the same $U$-spin multiplet,
so $M_{\Omega^-} \to M_{\Delta^-}, M_{\Xi^{*-}} \to M_{\Delta^-},
M_{\Sigma^{*-}} \to M_{\Delta^-}$ and similarly for the other charges.
Now, all three equations become identical,
\begin{equation}
   4 M_{\Delta^-}- 5 M_{\Delta^0 }- 2 M_{\Delta^+} + 3 M_{\Delta^{++}}
      = 6 b_{27} \delta m_d^2 - 18 c_{27} \delta m_d^3 \,,
\end{equation}
which again confirms that the numerical coefficients in
eqs.~(\ref{decup_mass_rel1})--(\ref{decup_mass_rel3}) are correct.

Finally note that we can find all the coefficients in these equations
from a $2+1$ simulation, and use them to (fully) predict the 
results of a $1+1+1$ simulation.


\subsubsection{The octet mass matrix}


We can analyse the possible terms in the octet mass matrix in the same way as
we did for the decuplet. We first consider the baryon octet.
Using the same technique as for the decuplet mass matrix we find
the results given in Table~\ref{mat8}.
\begin{table}[htb]
   \begin{center}
   \begin{tabular}{rrrrrrrrcc}
      $n $ & $p $ & $\Sigma^-$ & $\Sigma^0$ & $\Lambda$ & $\Sigma^+$
                               & $\Xi^-$ & $\Xi^0$ & $S_3$  & $SU(3)$   \\
      \hline
      1 & 1 & 1 & 1 & 1 & 1 & 1 & 1 & $A_1$ & 1                         \\
      \hline
      $-1$ & $-1$ & 0 & 0 & 0 & 0 & 1 & 1 & $E^+$ & $8_a$               \\
      $-1$ & 1 & $-2$ & 0 & 0 & 2 & $-1$ & 1 &$ E^-$ &$ 8_a$            \\
      \hline
      1 &  1 & $-2$& $-2$& 2 & $-2$& 1 & 1 & $E^+$ & $8_b$              \\
      $-1$ & 1 & 0 & \multicolumn{2}{c}{\ mix} & 0 & 1 & $-1$ 
                                                       & $E^-$ & $8_b$  \\
      \hline 
      1 & 1 & 1 &$-3$&$-3$& 1 & 1 & 1 & $A_1$ & 27                      \\
      1 & 1 &$-2$&  3 &$-3$&$ -2$& 1 & 1 & $E^+$ & 27                   \\
      $-1$ & 1 & 0 & \multicolumn{2}{c}{\ mix} & 0 & 1 & $-1$ 
                                                       & $E^-$ & $27$   \\
      \hline 
      1 & $-1$ & $-1$ & 0 & 0 & 1 & 1 & $-1$ 
        & $\phantom{^{I^X}} A_2\phantom{^{I^X}}$ & 10,${\overline{10}}$  \\
      0 & 0 & 0 & \multicolumn{2}{c}{\ mix} & 0 & 0 & 0 & $A_2$ 
                                                & 10,${\overline{10}}$  \\
      \hline
   \end{tabular}
   \end{center} 
\caption{Mass matrix contributions for octet baryons, classified by 
         permutation and $SU(3)$ symmetry. Note that the first two octet 
         quantities (the $8_a$) are proportional to the hypercharge $Y$
         and to isospin $I_3$, respectively.}
\label{mat8}
\end{table}
However there is a complication in the octet case which we do not
have in the decuplet, caused by the fact that we have two
particles (the $\Lambda$ and $\Sigma^0$) with the same $Y$ and $I_3$ 
quantum numbers. If $m_u \ne m_d $ these states mix. There are interesting 
connections between the elements of the $\Lambda/\Sigma^0$ mixing matrix and 
the splittings of the other baryons, but since in this article we are 
concerned with $2+1$ simulations, where this mixing does not arise, we will
not discuss this further here. We can however pick out several useful
mass relations which are unaffected by  $\Lambda/\Sigma^0$ mixing
\begin{eqnarray}
   \lefteqn{M_n + M_p + M_\Lambda 
            + M_{\Sigma^{-}} +  M_{\Sigma^{0}} + M_{\Sigma^{+}}
            + M_{\Xi^{-}} + M_{\Xi^{0}} }
   & &                                                    \nonumber      \\
   &=& 8 M_0 + b_1(\delta m_u^2+\delta m_d^2+\delta m_s^2)
             + c_1\delta m_u \delta m_d \delta m_s
                                                          \nonumber      \\
   \lefteqn{M_n + M_p -3  M_\Lambda 
            + M_{\Sigma^{-}} -3  M_{\Sigma^{0}} + M_{\Sigma^{+}}
            + M_{\Xi^{-}} + M_{\Xi^{0}} }
   & &                                                    \nonumber      \\
   &=&  b_{27} (\delta m_u^2+\delta m_d^2+\delta m_s^2) \,.
\label{unaff_lamsig_mix}
\end{eqnarray}
At order $\delta m_q^3$ we meet some quantities in the baryon octet masses
(the $10$ and $\overline{10}$ combinations) which can not be deduced from
$2+1$ flavour measurements -- though valence $1+1+1$ on a $2+1$ background
would be a possible method of estimating these quantities.

One early prediction concerning hyperon masses was the Coleman-Glashow 
relation \cite{coleman61a}
\begin{equation}
   M_n - M_p - M_{\Sigma^-} + M_{\Sigma^+} + M_{\Xi^-} - M_{\Xi^0} 
      \approx 0 \,.
\label{CGdef}
\end{equation} 
Deviations from this relation are barely detectable, using a recent
precision measurement of the $\Xi^0$ mass~\cite{fanti99a}
gives the value \cite{jenkins00a},
$M_n - M_p - M_{\Sigma^-} + M_{\Sigma^+} + M_{\Xi^-} - M_{\Xi^0}
= -0.29 \pm 0.26 \, \mbox{MeV}$. The original Coleman-Glashow argument
showed why the leading electromagnetic contribution to this quantity
vanishes (in modern terms, the leading electromagnetic mass contributions
are unchanged by the operation $d \leftrightarrow s$ because the $s$
and $d$ quarks have the same charge, but the quantity in eq.~(\ref{CGdef})
is odd under this operation). To understand the smallness of the
Coleman-Glashow quantity we also need to explain why the contribution
from flavour $SU(3)$ breaking due to quark mass differences is small.
The mass combination appears in Table~\ref{mat8} as one of the $A_2$ 
quantities. We can understand the success of the Coleman-Glashow 
relation by noting that the only polynomial in Table~\ref{cubic}
with $A_2$ symmetry is $O(\delta m_q^3)$, so that the predicted 
violation of the Coleman-Glashow relation is 
\begin{eqnarray}
   \lefteqn{M_n - M_p - M_{\Sigma^-} + M_{\Sigma^+} + M_{\Xi^-} - M_{\Xi^0} }
   & &                                                    \nonumber      \\
   &=& c_{10}(\delta m_s - \delta m_u)(\delta m_s-\delta m_d)
                  (\delta m_u-\delta m_d) \,.
\end{eqnarray}
The polynomial is zero if any pair of quarks have the same mass, 
so we would need to measure the masses of baryons in a $1+1+1$ 
setting to determine $c_{10}$ and predict the violation of
the Coleman-Glashow relation. 

Turning now to the mesons, both the pseudoscalar and vector meson
octet have a similar mass matrix, so they do not have to be considered
separately. In Table~\ref{mat8mes}
\begin{table}[htb]
   \begin{center}
   \begin{tabular}{rrrrrrrrcc}
      \hline
      $K^0$ & $K^+$ & $\pi^-$ & $\pi^0$ & $\eta_8$ & $\pi^+$
                        & $K^-$ & $\overline{K}^0$ &        &           \\
      \cline{1-8}
      $K^{*0}$ & $K^{*+}$ & $\rho^-$ & $\rho^0$ & $\phi_8$ & $\rho^+$
                   & $K^{*-}$ & $\overline{K}^{*0}$ & $S_3$  & $SU(3)$   \\
      \hline
      \hline
      1 & 1 & 1 & 1 & 1 & 1 & 1 & 1 & $A_1$ & 1                         \\
      \hline
      1 &  1 & $-2$& $-2$& 2 & $-2$& 1 & 1 & $E^+$ & $8_b$              \\
      $-1$ & 1 & 0 & \multicolumn{2}{c}{\ mix} & 0 & 1 & $-1$ 
                                                       & $E^-$ & $8_b$  \\
      \hline 
      1 & 1 & 1 &$-3$&$-3$& 1 & 1 & 1 & $A_1$ & 27                      \\
      1 & 1 &$-2$&  3 &$-3$&$ -2$& 1 & 1 & $E^+$ & 27                   \\
      $-1$ & 1 & 0 & \multicolumn{2}{c}{\ mix} & 0 & 1 & $-1$ 
                                                       & $E^-$ & $27$   \\
      \hline
      \hline
   \end{tabular}
   \end{center} 
\caption{Mass matrix contributions for octet mesons, classified by 
         permutation and $SU(3)$ symmetry.} 
\label{mat8mes}
\end{table}
we give the mass matrix contributions for the octet mesons,
classified by permutation and $SU(3)$ symmetry.

Some contributions allowed for baryons, Table~\ref{mat8},
are absent for mesons because they would violate charge conjugation,
giving (for example) different masses to the $K^+$ and $K^-$.
In particular, there are two octets, $8_a$ and $8_b$, in the baryon
table, Table~\ref{mat8}, but only the $8_b$ octet is permitted in
Table~\ref{mat8mes}. When we write down the mass formulae,
this will mean that the  baryon mass formula will have two
independent terms linear in the quark mass, but the meson mass formula
will only have a single linear term.


\section{Theory for $\mathbf{2 + 1}$ flavours}
\label{theory_2p1}


If we take any mass relation from the previous $1+1+1$ section, and 
put $m_u = m_d = m_l$ we will get a valid mass relation for the
$2+1$ case. In the $2+1$ case only one variable is needed to parametrise
the symmetry breaking, since from eq.~(\ref{zerosum}), 
\begin{eqnarray}
   \delta m_s = - 2 \delta m_l \,,
\end{eqnarray}
where
\begin{eqnarray}
   \delta m_l = m_l - \overline{m} \,.
\end{eqnarray}

In the $2+1$ case we can simplify the mass matrix
Tables~\ref{mat10}, \ref{mat8}, \ref{mat8mes}. The $E^-$ and $A_2$ 
matrices are not needed, their coefficients are always 
proportional to $m_u - m_d$, which we are now setting to zero.  
In the higher representations ($27$-plet and $64$-plet) only 
one linear combination of the $A_1$ and $E^+$ matrices appears
in the $2+1$ case (it is the linear combination which does not 
split particles within an isospin multiplet). Therefore, in this 
section we just need the simplified matrix Tables~\ref{mat10_2p1}, 
\ref{mat8_2p1} and \ref{mat8mes_2p1}.  
\begin{table}[htb]
   \begin{center}
   \begin{tabular}{rrrrrrrrrrc}
      $\Delta^- $   & $\Delta^0 $   & $\Delta^+ $   & $\Delta^{++} $ &
      $\Sigma^{*-}$ & $\Sigma^{*0}$ & $\Sigma^{*+}$ & $\Xi^{*-}$      &
      $\Xi^{*0}$    & $\Omega^-$    & \quad $SU(3)$                     \\
      \hline
      1 & 1 & 1 & 1 & 1 & 1 & 1 & 1 & 1 & 1 &  1                        \\
      \hline
      $-1$ & $-1$ & $-1$ & $-1$ & 0 & 0 & 0 & 1 & 1 & 2 &  8            \\
      \hline
      $3$ & $ 3$ & $ 3$ & 3 & $-5$ & $-5$ & $-5$ & $-3$ & $-3$ & 9 & 27 \\
      \hline
      $-1$ & $-1 $ & $-1$ & -1& $ 4$ & $ 4$ & $ 4$ & $-6$ & $-6$ & 4 & 64  \\
      \hline
   \end{tabular}
   \end{center}
\caption{Decuplet mass matrix contributions for the $2+1$ case,
         classified by $SU(3)$ symmetry. Compare with
         Table~\protect\ref{mat10}.}
\label{mat10_2p1}
\end{table}
\begin{table}[htb]
   \begin{center}
   \begin{tabular}{rrrrrrrrc}
      $n $ & $p $ & $\Sigma^-$ & $\Sigma^0$ & $\Lambda$ & $\Sigma^+$
                               & $\Xi^-$ & $\Xi^0$ &  \quad $SU(3)$     \\
      \hline
      1 & 1 & 1 & 1 & 1 & 1 & 1 & 1 & 1                                 \\
      \hline
      $-1$ & $-1$ & 0 & 0 & 0 & 0 & 1 & 1 & $8_a$                       \\
      \hline
      1 &  1 & $-2$& $-2$& 2 & $-2$& 1 & 1 & $8_b$                      \\
      \hline
      3 & 3 & $-1$ &$-1$&$-9$& $-1$ & 3 & 3 & $27$                      \\
      \hline
   \end{tabular}
   \end{center}
\caption{Mass matrix contributions for octet baryons for the $2+1$ case,
         classified by $SU(3)$ symmetry. Compare with 
         Table~\protect\ref{mat8}. }
\label{mat8_2p1}
\end{table} 
\begin{table}[htb]
   \begin{center}
   \begin{tabular}{rrrrrrrrc}
      \hline
      $K^0$ & $K^+$ & $\pi^-$ & $\pi^0$ & $\eta_8$ & $\pi^+$
                        & $K^-$ & $\overline{K}^0$ &                    \\
      \cline{1-8}
      $K^{*0}$ & $K^{*+}$ & $\rho^-$ & $\rho^0$ & $\phi_8$ & $\rho^+$
                   & $K^{*-}$ & $\overline{K}^{*0}$ &  \quad   $SU(3)$   \\
      \hline
      \hline
      1 & 1 & 1 & 1 & 1 & 1 & 1 & 1         & 1                         \\
      \hline
      1 &  1 & $-2$& $-2$& 2 & $-2$& 1 & 1 &         $8_b$              \\
      \hline
      3 & 3 & $-1$ &$-1$&$-9$& $-1$ & 3 & 3 &        $27$              \\
      \hline
   \end{tabular}
   \end{center}
\caption{Mass matrix contributions for octet mesons for the $2+1$ case,
         classified by $SU(3)$ symmetry. Compare with
         Table~\protect\ref{mat8mes}.}
\label{mat8mes_2p1}
\end{table}

In the $2+1$ limit the decuplet baryons have $4$ different
masses (for the $\Delta$, $\Sigma^*$, $\Xi^*$, and $\Omega$).
Similarly, for the octet baryons there are also $4$ distinct
masses, $(N,\Lambda, \Sigma, \Xi)$; and for octet mesons,
$3$ masses. In the meson octet the $K$ and $\overline{K}$
must have the same mass, but there is no reason why the $N$
and $\Xi$ (which occupy the corresponding places in the baryon
octet, see Figs.~\ref{meson_mults} and \ref{baryon_mults}),
should have equal masses once flavour $SU(3)$ is broken. 

Again we have the singlet quantities $X_S$ which are stationary
at the symmetry point as given in Table~\ref{perminv}, but which
now simplify to give Table~{\ref{perminv_2p1}}.
\begin{table}[htb] 
   \begin{center} 
      \begin{tabular}{cl}
      \hline
         Pseudoscalar
                  & $X_\pi^2 = \third(2 M_K^2 + M_\pi^2)$         \\
         mesons   & $X_{\eta_8}^2 = \half(M^2_\pi + M^2_{\eta_8})$  \\
      \hline 
         Vector   & $X_\rho = \third(2 M_{K^*} + M_\rho)$          \\
         mesons   & $X_{\phi_8} = \half(M_\rho + M_{\phi_8})$       \\
                  & $X_{\phi_s} = \third(2M_\rho + M_{\phi_s})$     \\
      \hline 
      \hline 
         Octet    & $X_N = \third(M_N + M_\Sigma + M_\Xi)$         \\
         baryons  & $X_\Lambda = \half(M_\Sigma + M_\Lambda)$       \\
      \hline
         Decuplet & $X_\Delta = \third(2 M_\Delta + M_\Omega)$      \\
         baryons  & $X_{\Xi^*} = \third(M_\Delta + M_{\Sigma^*} + M_{\Xi^*})$
                                                                  \\
                  & $X_{\Sigma^*} = M_{\Sigma^*}$                    \\
      \hline 
      \end{tabular}
   \end{center}
\caption{Permutation invariant mass combinations,
         see Fig.~\protect\ref{permset}. $\phi_s$ is a
         fictitious $s \overline{s}$ particle; $\eta_8$
         a pure octet meson.}
\label{perminv_2p1}
\end{table} 
In the notation we have now assumed isospin invariance, so that for example
$M_\Delta \equiv M_{\Delta^{++}} = M_{\Delta^+} = M_{\Delta^0} = M_{\Delta^-}$.
(The corresponding mass values we use in this article
are given in section~\ref{hadron_masses}.)

Furthermore this can obviously be generalised. Let us first
define the quark mass combinations $m_\eta = (m_l + 2m_s)/3$
and $m_K = (m_l + m_s)/2$. Then $m_l + m_\eta = 2\overline{m}$
and $m_l + 2m_K = 3\overline{m}$ are constants on our trajectory
and so $\delta m_l + \delta m_\eta = 0$ and also
$\delta m_l + 2\delta m_K = 0$. For example, this means that
any functions of the form
\begin{eqnarray}
   2f(m_K) + f(m_l) \quad \mbox{or} \quad
   g(m_s) + 2g(m_l) \quad \mbox{or} \quad
   h(m_\eta) + h(m_l) \,,
\label{stationary_fun}
\end{eqnarray}
will also have zero derivative at the symmetric point. These equations
generalise the meson sector of Table~\ref{perminv_2p1}.

In Table~\ref{bary2p1} we present the  $2+1$ baryon
\begin{table}[p]
\begin{center} 
   \begin{tabular}{crcccc}
      $SU(3)$ & Mass Combination & \multicolumn{4}{c}{Expansion}         \\
      \hline 
        1  & $ 2 M_N + 3 M_\Sigma + M_\Lambda + 2 M_\Xi $
           & \quad 1, 
           &               
           & $ \delta m_l^2,$ 
           & $ \delta m_l^3 $                                            \\
        8  & $ M_\Xi - M_N $ 
           &          
           & $ \delta m_l, $ 
           & $ \delta m_l^2, $ 
           & $ \delta m_l^3 $                                            \\
        8  & $ -M_N + 3 M_\Sigma - M_\Lambda - M_\Xi $
           &   
           & $ \delta m_l,$ 
           & $ \delta m_l^2,$ 
           & $ \delta m_l^3 $                                            \\
        27 & $ 2 M_N - M_\Sigma - 3  M_\Lambda + 2 M_\Xi $
           &   
           & 
           & $\delta m_l^2, $ 
           & $\delta m_l^3 $                                             \\
      \hline
      \hline
        1  & $ 4 M_\Delta + 3 M_{\Sigma^*} + 2 M_{\Xi^*} + M_{\Omega} $
           & \quad 1,&
           & $ \delta m_l^2, $ 
           & $\delta m_l^3 $                                             \\
        8  & $ -2 M_\Delta  + M_{\Xi^*} + M_{\Omega} $
           &   
           & $ \delta m_l, $ 
           & $ \delta m_l^2, $ 
           & $ \delta m_l^3 $                                            \\
        27 &  $ 4 M_\Delta - 5 M_{\Sigma^*} - 2 M_{\Xi^*} + 3 M_{\Omega} $
           &   
           & 
           & $ \delta m_l^2, $ 
           & $ \delta m_l^3 $                                            \\
        64 & $ - M_\Delta + 3 M_{\Sigma^*} - 3 M_{\Xi^*} +  M_{\Omega} $
           &
           & 
           &
           & $\delta m_l^3 $                                             \\
      \hline 
   \end{tabular}
\end{center}
\caption{Baryon mass combinations classified by $SU(3)$ representation, 
         in the $2+1$ case.}
\label{bary2p1}
\end{table} 
decuplet results corresponding to Table~\ref{mat10} or 
eqs.~(\ref{decup_sing_mass_rel})--(\ref{decup_mass_rel3})
and the baryon octet from Table~\ref{mat8} or
eq.~(\ref{unaff_lamsig_mix}). Similarly the mesons are given
in Table~\ref{meso2p1}. Particular combinations chosen to
\begin{table}[p]
\begin{center} 
   \begin{tabular}{crcccc}
      $SU(3)$ & Mass Combination & \multicolumn{4}{c}{Expansion}         \\
      \hline 
        1  & $ 3 M^2_\pi + 4 M^2_K + M^2_{\eta_8} $
           & \quad 1,
           &
           & $\delta m_l^2, $
           & $\delta m_l^3 $                                             \\
        8  & $ -3 M^2_\pi + 2 M^2_K +  M^2_{\eta_8} $
           &
           & $\delta m_l, $
           & $\delta m_l^2, $
           & $ \delta ml^3 $                                             \\
        27 & $ - M_\pi^2 + 4 M_K^2 - 3  M^2_{\eta_8} $
           &   
           & 
           & $\delta m_l^2, $
           & $\delta m_l^3 $                                             \\
      \hline
      \hline
        1  & $3 M_\rho + 4 M_{K^*} + M_{\phi_8} $
           & \quad 1,
           &
           & $\delta m_l^2, $
           & $\delta m_l^3 $                                             \\
        8  & $-3 M_\rho + 2 M_{K^*} + M_{\phi_8} $
           &
           & $\delta m_l, $ 
           & $\delta m_l^2, $ 
           & $\delta m_l^3  $                                            \\
        27 & $- M_\rho + 4 M_{K^*} - 3 M_{\phi_8} $
           &
           &
           & $\delta m_l^2, $
           & $\delta m_l^3 $                                             \\
      \hline
   \end{tabular}
\end{center}
\caption{Meson mass combinations classified by $SU(3)$ representation, 
         in the $2+1$ case. Octet-singlet mixing is not taken into account.}
\label{meso2p1}
\end{table} 
remove the unknown $M_{\eta_8}$, $M_{\phi_8}$ masses are
given in Table~\ref{mesomix2p1}.
\begin{table}[p]
\begin{center} 
   \begin{tabular}{crcccc}
      $SU(3)$ & Mass Combination & \multicolumn{4}{c}{Expansion}          \\
      \hline 
      $1,\; 27$ & $ 2 M_K^2 + M^2_\pi $    
                & \quad 1,
                &
                & $\delta m_l^2,$ & $\delta m_l^3$                        \\
      $8,\; 27$ & $ M^2_K -  M^2_{\pi} $
                &
                & $\delta m_l,$ 
                & $\delta m_l^2,$
                & $\delta m_l^3$                                          \\
      \hline
      \hline
      $1,\; 27$ & $ 2 M_{K^*} + M_\rho $
                & \quad 1,
                &
                & $\delta m_l^2, $
                & $\delta m_l^3$                                          \\
      $8,\; 27$ & $ M_{K^*} - M_{\rho} $
                &
                & $\delta m_l, $
                & $\delta m_l^2, $
                & $\delta m_l^3$                                          \\
      \hline
   \end{tabular}
\end{center}
\caption{Meson mass combinations free from mixing problems, 
         classified by $SU(3)$ representation. These combinations
         have been chosen to eliminate the $\eta_8$ and $\phi_8$
         states, so they now contain mixtures of different $SU(3)$
         representations.}
\label{mesomix2p1}
\end{table} 

We can see how well this works in practice by looking,
for example, at the physical masses of the decuplet baryons.
If we consider the physical values of the four decuplet
mass combinations in Table~\ref{bary2p1} and using mass values
given later in Table~\ref{hadron_masses_av}, we get
\begin{eqnarray}
   4 M_\Delta + 3 M_{\Sigma^*} + 2 M_{\Xi^*} + M_{\Omega}
      &=& 13.82 {\rm \ GeV}
          \qquad \quad \ {\rm singlet}
          \quad \propto (\delta m_l)^0 \label{num1}           \nonumber  \\
   - 2 M_\Delta \qquad\ \quad  + M_{\Xi^*} + M_{\Omega}
      &=& \ 0.742 {\rm \ GeV} 
          \qquad \quad \ {\rm octet}
          \quad\,\,\, \propto \delta m_l \label{num8}      \label{num64} \\
   4 M_\Delta - 5 M_{\Sigma^*} - 2 M_{\Xi^*} + 3 M_{\Omega}
      &=& -0.044  {\rm \  GeV} 
          \qquad \quad {\rm 27-plet}
          \quad \propto \delta m_l^2 \label{num27}            \nonumber  \\
   - M_\Delta + 3 M_{\Sigma^*} - 3 M_{\Xi^*} +  M_{\Omega}
      &=& -0.006 {\rm \  GeV}  \qquad \quad {\rm 64-plet}
          \quad \propto \delta m_l^3 \,,
                                                             \nonumber
\end{eqnarray}
with a strong hierarchy in values, corresponding to the leading term
in the Taylor expansion. Each additional factor of $\delta m_l$
reduces the value by about an order of magnitude, the $6$4-plet
combination is more than $2000$ times smaller than the singlet
combination. This suggests that the Taylor expansion converges
well all the way from the symmetric point to the physical point.
(Though of course it is possible that the singlet and octet
curvature terms are larger than those in the $27$ and $64$.)
Unfortunately it may be very difficult to see a signal
in the $64$-plet channel. 

We can now `invert' the results in Tables~\ref{bary2p1},
\ref{meso2p1} to give the expansion for each hadron mass%
\footnote{Alternatively, of course, the methods of
Appendix~\ref{had_mass_matrices} can be directly applied.}
from the symmetry point $\overline{m} = m_0$. This inversion
is made easier by orthogonality relations between the different
$SU(3)$ representations. We can simply read off the answers from the 
Tables~\ref{mat10_2p1}, \ref{mat8_2p1} and \ref{mat8mes_2p1}. 
This leads to the constrained fit formulae
\begin{eqnarray}
   M_\pi^2      &=& M_0^2 + 2\alpha\delta m_l 
                         + (\beta_0 + 2\beta_1)\delta m_l^2
                                                             \nonumber \\
   M_K^2        &=& M_0^2 - \alpha\delta m_l 
                         + (\beta_0 + 5\beta_1 + 9\beta_2)\delta m_l^2
                                                             \nonumber \\
   M_{\eta_8}^2  &=& M_0^2 - 2\alpha\delta m_l
                    + (\beta_0 + 6\beta_1 + 12\beta_2 + \beta_3)\delta m_l^2 \,,
\label{fit_mpsO}
\end{eqnarray}
\begin{eqnarray}
   M_\rho       &=& M_0 + 2\alpha\delta m_l 
                         + (\beta_0 + 2\beta_1)\delta m_l^2
                                                             \nonumber \\
   M_{K^*}      &=& M_0 - \alpha\delta m_l 
                         + (\beta_0 + 5\beta_1 + 9\beta_2)\delta m_l^2
                                                             \nonumber \\
   M_{\phi_8}   &=& M_0  - 2\alpha\delta m_l
                    + (\beta_0 + 6\beta_1 + 12\beta_2 + \beta_3)\delta m_l^2 \,,
\label{fit_mvO}
\end{eqnarray}
\begin{eqnarray}
   M_N      &=& M_0 + 3A_1\delta m_l 
                    + (B_0+3B_1)\delta m_l^2
                                                             \nonumber \\
   M_\Lambda &=& M_0  + 3A_2\delta m_l 
                    + (B_0+6B_1-3B_2+9B_4)\delta m_l^2
                                                             \nonumber \\
   M_\Sigma &=& M_0  - 3A_2\delta m_l 
                    + (B_0+6B_1+3B_2+9B_3)\delta m_l^2
                                                             \nonumber \\
   M_\Xi    &=& M_0 - 3(A_1-A_2)\delta m_l 
                    + (B_0+9B_1-3B_2+9B_3)\delta m_l^2 \,,
\label{fit_mNO}
\end{eqnarray}
\begin{eqnarray}
   M_\Delta     &=& M_0 + 3A\delta m_l 
                       + (B_0 + 3B_1)\delta m_l^2
                                                             \nonumber \\
   M_{\Sigma^*} &=& M_0  + 0
                       + (B_0 + 6B_1 + 9B_2)\delta m_l^2
                                                             \nonumber \\
   M_{\Xi^*}    &=& M_0 - 3A\delta m_l 
                       + (B_0 + 9B_1 + 9B_2)\delta m_l^2
                                                             \nonumber \\
   M_\Omega    &=& M_0 - 6A\delta m_l 
                       + (B_0 + 12B_1)\delta m_l^2 \,.
\label{fit_mDD}
\end{eqnarray}
(The values of the constants are obviously different for each
octet or decuplet.)
We see that the linear terms are highly constrained. The decuplet baryons
have only one slope parameter, $A$, while the octet baryons have two slope
parameters, $A_1$, $A_2$. Mesons have fewer slope parameters than octet
baryons because of constraints due to charge conjugation,
leaving again just one slope parameter.

The quadratic terms are much less constrained; indeed only for the
baryon decuplet is there any constraint. The coefficients of the
$\delta m_l^2$ terms appear complicated; there seem to be
too many for the meson and baryon octets,
eqs.~(\ref{fit_mpsO}) -- (\ref{fit_mNO}). In the next section,
we generalise these formulae to the case of different valence
quark masses to sea quark masses or `partial quenching'
when this choice of coefficients becomes relevant.
Note that not all the coefficients can thus be determined: for 
$\eta_8$ and $\phi_8$ the $\beta_3$-coefficient cannot be found
from partially quenched results.

An $s \bar{s}$ meson state has charge $0$, isospin $0$, the
same quantum numbers as a $(u \bar{u} + d \bar{d})$ meson.
In the real world we should therefore expect that $I=0$ (isoscalar)
mesons will always be mixtures of $s \bar{s}$ and $l \bar{l}$, to
a greater or lesser extent. The mixing has been investigated in
detail in \cite{dudek10a}.

On the lattice we can remove the mixing by just dropping
disconnected contributions, and only keeping the connected
part of the meson propagator. (In fact this calculation is
easier and cheaper than the full calculation.) Theoretically
the resulting pure $s \bar{s}$ meson is best treated in the
context of partial quenching. We can get its mass formula
simply by making the substitution for the valence quark
mass $\delta \mu_s \to - 2\delta \mu_l$ in the mass formula
for the partially quenched $I=1$ mesons as described in
section~\ref{partial_quenching}, eq.~(\ref{PSoct})
to give
\begin{eqnarray}
   M_{\eta_s}^2  &=& M_0^2 - 4\alpha\delta m_l 
                         + (\beta_0 + 8\beta_1)\delta m_l^2
                                                             \nonumber \\
   M_{\phi_s}   &=& M_0 - 4\alpha\delta m_l 
                         + (\beta_0 + 8\beta_1)\delta m_l^2 \,.
\label{fit_strange}
\end{eqnarray}
(These are, as previously mentioned, the masses that can be (easily)
found from lattice simulations rather than $M_{\eta_8}$ and $M_{\phi_8}$.)
The pseudoscalar $s \bar{s}$ state, the $\eta_s$ or `strange pion',
does not correspond to any real-world state, but the vector
$s \bar{s}$ meson, the $\phi_s$, is very close to the real-world
$\phi$. Phenomenologically, the observed fact that the
$\phi$ almost always decays to $ K \bar{K}$ (84 \%) rather than
$\rho \pi$ \cite{nakamura10a} is best explained by saying that
the $\phi$ is almost purely $s \bar{s}$. 

As eqs.~(\ref{fit_mpsO})--(\ref{fit_mDD}) have been derived using
only group theoretic arguments, they will be valid for results
derived on any lattice volume (though the coefficients are still
functions of the volume). 

Finally there is the practical question of whether fits should be
against the (light) quark mass or alternatively against the pseudoscalar
pion mass. In Appendix~\ref{coordinate_choice} we argue that
`internally' at least the fits should be made against the quark mass.
Of course this is only a useful observation when quadratic or
higher terms are involved. To leading order
eqs.~(\ref{fit_mvO})--(\ref{fit_mDD}) can be written as
\begin{eqnarray}
   \delta M \equiv M - M_0 = c_M\delta m_l \,,
\label{LO_qm_M_expansion}
\end{eqnarray}
(together with $\delta M^2 \equiv M^2 - M_0^2 = c_M\delta m_l$
in the case of pseudoscalar mesons, eq.~(\ref{fit_mpsO})).
The coefficients $c_M$ can be found from these equations.
Thus an expansion in $\delta m_l$ is equivalent to an
expansion in $\delta M$ or $\delta M^2$.


\section{Partial quenching}
\label{partial_quenching}


In partial quenching (PQ) measurements are made with the mass of the
valence quarks different from the sea quarks. In this case
the sea quark masses $m_l$, $m_s$ remain constrained by
$\overline{m} = \mbox{constant}$, but the valence quark masses
$\mu_l$, $\mu_s$ are unconstrained. We define
\begin{eqnarray}
   \delta\mu_q = \mu_q - \overline{m} \,, \quad q = l, s \,.
\end{eqnarray}
When $\mu \to m$ (i.e.\ return to the `unitary line') then the following
results collapse to the previous results of
eqs.~(\ref{fit_mpsO}) -- (\ref{fit_mDD}).
Here we sketch some results, see~\cite{pq} for more details.

In Table~\ref{cubic} we have often used the identity 
$\delta m_u + \delta m_d + \delta m_s = 0$ to simplify the
symmetric polynomials. Since we are not going to keep
$\mu_u + \mu_d + \mu_s$ constant, we write out our basis polynomials in full,
without the constraint $\sum \delta \mu_q = 0$, in Table~\ref{PQpoly}.
We can check that the polynomials in Table~\ref{PQpoly} reduce to multiples
of those in Table~\ref{cubic} when the identity  $\sum \delta \mu_q = 0$
is applied. For example, $2 \delta\mu_s - \delta\mu_u - \delta\mu_d 
= 3 \delta\mu_s - (\delta\mu_u + \delta\mu_d + \delta\mu_s)$, 
so when the zero-sum identity is applied, and $\mu_q \to m_q$ the
polynomial $2 \delta\mu_s - \delta\mu_u - \delta\mu_d$ in Table~\ref{PQpoly} 
\begin{table}[htb]
   \begin{center}
   \begin{tabular} {ccccc}
   Polynomial  & $\qquad S_3 \qquad$ & \multicolumn{3}{c}{$SU(3)$}      \\
   \hline
   $1$  & $A_1$ & $1$ &  &                                              \\
   \hline
   \hline
   $\delta\mu_u + \delta\mu_d +\delta\mu_s$ &  $A_1$ & $1$ &  &         \\
   \hline
   $2 \delta\mu_s - \delta\mu_u - \delta\mu_d $ &  $E^+$ &  & $8$ &     \\
   $  \delta\mu_u - \delta\mu_d$ &  $E^-$ &  & $8$ &                    \\
   \hline 
   \hline
   $(\delta\mu_u + \delta\mu_d +\delta\mu_s)^2$ &  $A_1$ & 1 &  &       \\
   $(\delta\mu_u + \delta\mu_d +\delta\mu_s)
      ( 2 \delta\mu_s - \delta\mu_u - \delta\mu_d) $ &   $E^+$ &  & $8$ &    \\
   $(\delta\mu_u + \delta\mu_d +\delta\mu_s)( \delta\mu_u - \delta\mu_d)$  &
                                                       $E^-$ &  & $8$ & \\
   \hline
   $(\delta\mu_s -\delta\mu_u)^2 + (\delta\mu_s-\delta\mu_d)^2
                 + (\delta\mu_u-\delta\mu_d)^2 $ & $A_1$ & 1 &  &  $27$ \\
   $(\delta\mu_s -\delta\mu_u)^2 + (\delta\mu_s-\delta\mu_d)^2-2 
                   (\delta\mu_u-\delta\mu_d)^2$ & $E^+$ & & $8$ &  $27$ \\
   $ (\delta\mu_s -\delta\mu_u)^2 -  (\delta\mu_s-\delta\mu_d)^2 $ & 
                                              $E^-$ &  & $8$ & $27$    \\
   \hline 
   \hline
   $\delta m_u^2 + \delta m_d^2 +\delta m_s^2$ & $A_1$ & 1 &  &  $27$   \\
   \hline \hline
\end{tabular}
\end{center}
\caption{All the quark-mass polynomials needed for 
         partially quenched masses, classified by
         symmetry properties. The table includes entries up to
         $O(\mu_q^2)$.}
\label{PQpoly}
\end{table}
reduces to the corresponding polynomial $\delta m_s$ in Table~\ref{cubic}.


\subsection{PQ decuplet baryons}


In the partially quenched case we know that the hadron
mass formulae should have an $SU(3)$ symmetry for interchanging
the sea quarks, and another $SU(3)$ symmetry for operations
on the valence quarks. The sea quark symmetry will always
be singlet, the valence quark terms for a hadron in the octet
can be in any representation which occurs in $\overline{8}\otimes 8$
and the decuplet can be in any representation which occurs in
$10 \otimes \overline{10}$.

Let us see what sort of mass relations symmetry allows us, taking
the decuplet baryons as our example. Starting with linear terms
in the quark masses, we can form two polynomials of the valence masses,
a singlet combination $(2 \delta\mu_l + \delta\mu_s)$ and an octet
combination with $E^+$ symmetry, $(\delta\mu_s - \delta\mu_l)$.
These are the polynomials on line 2 and line 3 of Table~\ref{PQpoly}.
(A first-order term in the sea quark masses is ruled out
because we are keeping $2 m_l + m_s$ constant.)
We can read off the coefficients each polynomial must have
by looking for the $A_1$ singlet and $E^+$ octet entries in 
Fig.~\ref{decumat}, or in Tables~\ref{mat10} or \ref{mat10_2p1}. 
Thus a singlet polynomial must have the same flavour coefficients
for every baryon and the octet polynomial must have a coefficient
proportional to the hypercharge.

So, at first sight we would expect
\begin{eqnarray}
   M_\Delta    &=& M_0 + \alpha_1 (2 \delta\mu_l + \delta\mu_s)
                  - \alpha_2 (\delta\mu_s - \delta\mu_l)
                                                              \nonumber \\
   M_{\Sigma^*}&=& M_0 + \alpha_1 (2 \delta\mu_l + \delta\mu_s)
                                                              \nonumber \\
   M_{\Xi^*}  &=& M_0 + \alpha_1 (2 \delta\mu_l + \delta\mu_s)
                 + \alpha_2 (\delta\mu_s - \delta\mu_l)
                                                              \nonumber \\
   M_\Omega  &=& M_0 + \alpha_1 (2 \delta\mu_l + \delta\mu_s)
                 + 2 \alpha_2 (\delta\mu_s - \delta\mu_l) \,,
\end{eqnarray}
with no connection between $\alpha_1$ and $\alpha_2$.
However, it is clear that the $\Delta$ mass cannot know anything
about the strange valence mass, and the $\Omega$ mass must
similarly be independent of $\delta\mu_l$. These constraints are
both satisfied if
\begin{equation}
   \alpha_1 = \alpha_2 \equiv A \,,
\end{equation}
giving us a leading-order formula
\begin{eqnarray}
   M_\Delta    &=& M_0 + 3 A \delta\mu_l
                                                              \nonumber \\
   M_{\Sigma^*}&=& M_0 + A (2\delta\mu_l +\delta\mu_s)
                                                              \nonumber \\
   M_{\Xi^*}   &=& M_0 + A (\delta\mu_l + 2\delta\mu_s)
                                                              \nonumber \\
   M_\Omega   &=& M_0 + 3 A \delta\mu_s \,.
\label{PQlead} 
\end{eqnarray}
We can continue this procedure to the quadratic level. Again, the
number of terms is reduced by keeping the sum of the sea quark masses
fixed; and we again find the number of coefficients reduced by the constraint
that the $\Delta$ mass is independent of $\delta\mu_s$,
and the $\Omega$ mass independent of $\delta\mu_l$. The most
general quadratic formula, consistent with the above constraint,
and with $SU(3)$ symmetry, is
\begin{eqnarray}
   M_\Delta  &=& M_0 + 3 A \delta\mu_l + B_0 \delta m_l^2 + 3 B_1 \delta\mu_l^2
                                                               \nonumber  \\
   M_{\Sigma^*}&=& M_0 + A (2\delta\mu_l +\delta\mu_s)
                + B_0 \delta m_l^2 + B_1 (2\delta\mu_l^2 + \delta\mu_s^2)
                + B_2(\delta\mu_s - \delta\mu_l)^2
                                                               \nonumber  \\
   M_{\Xi^*}   &=& M_0 + A (\delta\mu_l + 2\delta\mu_s)
                 + B_0 \delta m_l^2 + B_1 (\delta\mu_l^2 + 2\delta\mu_s^2)
                 + B_2(\delta\mu_s - \delta\mu_l)^2
                                                               \nonumber  \\
   M_\Omega  &=& M_0 + 3 A \delta\mu_s 
                    + B_0 \delta m_l^2 + 3 B_1 \delta\mu_s^2 \,.
\label{PQOmega}
\end{eqnarray}
These formulae apply when the sum of the sea quark masses is
held constant, $\third(2 m_l + m_s) =  m_0$, but the valence quark
masses are completely free, because at this level (terms up to
second order) a restriction on valence masses would not lead to
any reduction in the number of free parameters.

We can check the formulae~(\ref{PQOmega}) by forming the $SU(3)$ 
mass combinations of Table~\ref{bary2p1}, and checking that in 
each case only polynomials of the appropriate symmetry 
appear in the answer. 

There are some combinations of the partially quenched
masses, eqs.~(\ref{PQOmega}), which have simpler
dependences on the valence quark masses and only involve a small
number of fit parameters. Examples include 
\begin{eqnarray}
   -M_\Delta + M_{\Sigma^*} + M_{\Xi^*} - M_\Omega
      &=& 2 B_2 ( \delta\mu_s - \delta\mu_l)^2
                                                                \nonumber \\
   M_{\Xi^*} - M_{\Sigma^*}
      &=& A ( \delta\mu_s - \delta\mu_l ) + B_1(\delta\mu_s^2 - \delta\mu_l^2)
                                                                \nonumber \\
   M_\Omega - M_\Delta
      &=& 3 A ( \delta\mu_s - \delta\mu_l )
          + 3 B_1 (\delta\mu_s^2 - \delta\mu_l^2) \,.
\end{eqnarray}

Let us use these formulae to illustrate how partially quenched
measurements might help us fit masses on the constant sea-quark mass line.
If we only use unitary data, with $\third(2 m_l^R + m_s^R) =  m_0^R$, 
we are limited to points on the line between the endpoints
$(m_l^R, m_s^R) = (0, 3 m_0^R)$ and $(m_l^R, m_s^R) = (\frac{3}{2} m_0^R, 0)$,  
which means that the quark mass splitting is limited to the range
$ -\, {3 \over 2} m_0^R <  (m_s^R - m_l^R) <  3 m_0^R$. 
In the partially quenched case, we can increase $\mu_s$ without
having to decrease $\mu_l$, so we can make the splitting 
$\mu_s^R - \mu_l^R$ much larger than $m_s^R - m_l^R$, 
which gives a much better lever-arm to determine $B_1$ and $B_2$. 

Because the decuplet baryon has a particularly high degree of symmetry
we can give an alternative derivation of eq.~(\ref{PQOmega}), with less
explicit reference to the flavour group. 
Consider a decuplet baryon made from quarks of type $a, b$ and $c$.
Since the decuplet is a fully symmetric representation, the
mass of the baryon must be a symmetric function of
$\delta\mu_a, \delta\mu_b$ and $\delta\mu_c$.

The mass can also depend on the sea quark masses
$\delta m_u, \delta m_d, \delta m_s$, but this dependence must be
in a flavour singlet way. If we keep the sum of the sea quark masses fixed,
the first singlet polynomial allowed (see Table~\ref{cubic}) is
${1 \over 6}(\delta m_u^2 + \delta m_d^2 + \delta m_s^2)$,
where we have chosen the prefactor ${1 \over 6}$ to lead to a
tidy expression in the limit when $m_u = m_d$.  Since this mass
polynomial is flavour singlet, it can only appear in the coefficient
of the identity matrix -- i.e.\ it must make the same contribution to every
decuplet baryon. We therefore know that the mass formula for the
$abc$ decuplet baryon must have the form
\begin{eqnarray}
   M(abc) &=& M_0 + B_0 \sixth (\delta m_u^2 + \delta m_d^2 + \delta m_s^2)
              + F_{sym}( \delta\mu_a, \delta\mu_b, \delta\mu_c)
                                                           \nonumber       \\
          &\to&  M_0 + B_0 \delta m_l^2 
                 + F_{sym}( \delta\mu_a, \delta\mu_b, \delta\mu_c) \,,
\label{decsym}
\end{eqnarray}
in the $2+1$ limit.

Now let us consider what terms are possible in the completely symmetric
function $F_{sym}$. There is only one symmetric linear term,
$\delta\mu_a + \delta\mu_b + \delta\mu_c$. There are two symmetric
quadratic terms we can write down,
$\delta\mu_a^2 + \delta\mu_b^2 + \delta\mu_c^2$ and
$\delta\mu_a \delta\mu_b + \delta\mu_a \delta\mu_c + \delta\mu_b \delta\mu_c $,
or linear combinations of these two terms. We get simpler final expressions
if we choose the basis $\delta\mu_a^2 + \delta\mu_b^2 + \delta\mu_c^2$ and
$\delta\mu_a^2 + \delta\mu_b^2 + \delta\mu_c^2
-\delta\mu_a \delta\mu_b - \delta\mu_a \delta\mu_c - \delta\mu_b \delta\mu_c $,
giving us the general mass formula
\begin{eqnarray}
   M(abc) &=& M_0 + A(\delta\mu_a + \delta\mu_b + \delta\mu_c) + B_0\delta m_l^2
              + B_1 (\delta\mu_a^2 + \delta\mu_b^2 + \delta\mu_c^2)
                                                           \nonumber       \\
          & & + B_2 ( \delta\mu_a^2 + \delta\mu_b^2 + \delta\mu_c^2
              - \delta\mu_a \delta\mu_b - \delta\mu_a \delta\mu_c 
              - \delta\mu_b \delta\mu_c) \,.
\end{eqnarray}
Calculating $M(lll), M(sll), M(ssl)$ and $M(sss)$ gives the
partially quenched mass formulae in eq.~(\ref{PQOmega}).
Remember that these formulae are only complete for the case
$m_u + m_d + m_s$ held constant. If the average sea quark mass is
allowed to vary, more terms become possible, including `mixed'
polynomials which contain both $\delta m_q$ and $\delta \mu_q$.
With fixed average sea quark mass, such mixed polynomials do not
arise until the cubic order.

This argument gives us the same result as the full group-theory 
derivation, though in some sense it explains less. For instance, 
it does not immediately explain why the mass combinations of 
Table~\ref{bary2p1} give particularly tidy polynomials, or
where the hierarchy in eq.~(\ref{num64}) comes from.


\subsection{PQ octet mesons}


As in the previous section, we find the meson mass formula by
constructing the most general matrix consistent with $SU(3)$
symmetry, and the constraint that the partially quenched 
pion mass must know nothing about $\mu_s$. 
The resulting mass formulae for partially quenched mesons take the form:
\begin{eqnarray}
   M^2_\pi &=& M_0^2 + 2 \alpha \delta\mu_l + \beta_0 \delta m_l^2
                     + 2 \beta_1 \delta\mu_l^2
                                                                \nonumber \\
   M^2_K   &=& M_0^2 + \alpha (\delta\mu_l + \delta\mu_s )
                     + \beta_0 \delta m_l^2
                     + \beta_1 ( \delta\mu_l^2 +  \delta\mu_s^2 )
                     + \beta_2 (\delta\mu_s - \delta\mu_l)^2
                                                                \nonumber \\
   M^2_{\eta_s}
           &=& M_0^2 + 2 \alpha \delta\mu_s
                     + \beta_0 \delta m_l^2
                     + 2 \beta_1 \delta\mu_s^2 \,.
\label{PSoct}
\end{eqnarray}
Again the $\eta_s$ is the meson made of a partially quenched
$\overline{s}_{val}s_{val}$ quarks (i.e.\ the `strange pion')
which in the partially quenched framework can be observed and can
yield useful information about the extrapolation constants.
The PQ $M^2_{\eta_s}$ can thus be obtained from the PQ $M^2_\pi$
by simply changing $\mu_l \to \mu_s$ which changes the top row of
eq.~(\ref{PSoct}) into the bottom row.

Some useful combinations, which avoid the delicate $\eta$ sector, are
\begin{eqnarray}
   M^2_K - M^2_\pi &=& \alpha ( \delta\mu_s - \delta\mu_l )
                       + \beta_1 ( \delta\mu_s^2 - \delta\mu_l^2)
                       + \beta_2 ( \delta\mu_s - \delta\mu_l )^2
                                                                \nonumber \\
   2  M^2_K + M^2_\pi
                  &=& 3 M_0^2  + \alpha ( 4 \delta\mu_l + 2 \delta\mu_s )
                      + 3 \beta_0 \delta m_l^2
                                                                \nonumber \\
                  & & + \beta_1 (  4 \delta\mu_l^2  + 2 \delta\mu_s^2 )
                      + 2 \beta_2 ( \delta\mu_s - \delta\mu_l )^2  \,.
\end{eqnarray}
$ M^2_K - M^2_\pi$ is useful as a measure of the quark mass splitting,
$ 2  M^2_K + M^2_\pi$ as a quantity which is nearly constant along
our trajectory.

The same form, mutatis mutandis, applies to the other meson
octets, e.g.\ the $\rho$, $K^*$, $\phi$ system. We thus have
\begin{eqnarray}
   M_{\rho} &=& M_0 + 2 \alpha \delta\mu_l + \beta_0 \delta m_l^2
                   + 2 \beta_1 \delta\mu_l^2
                                                                \nonumber \\
   M_{K^*}  &=& M_0 + \alpha (\delta\mu_l + \delta\mu_s )
                   + \beta_0 \delta m_l^2
                   + \beta_1 ( \delta\mu_l^2 +  \delta\mu_s^2 )
                   + \beta_2 (\delta\mu_s - \delta\mu_l)^2
                                                                \nonumber \\
   M_{\phi_s}
           &=& M_0 + 2 \alpha \delta\mu_s
                   + \beta_0 \delta m_l^2
                   + 2 \beta_1 \delta\mu_s^2 \,,
\label{PQoctvec}
\end{eqnarray}
following the pattern of eq.~(\ref{PSoct}).

We can give a similar elementary argument to derive the
partially quenched mass formula for mesons with different quarks
(e.g. the $K, K^*$ and the charged $\pi$ and $\rho$). Consider
an $a \bar{b}$ meson. It must have the same mass as its antiparticle,
the $b \bar{a}$ meson. So, by the same argument as in eq.~(\ref{decsym})
we will have a mass formula
\begin{equation}
   M(a \bar{b}) 
      = M( b \bar{a})
      =  M_0 + \beta_0 \sixth (\delta m_u^2 + \delta m_d^2 + \delta m_s^2)
             + F_{sym}(\delta\mu_a , \delta\mu_b) \,.
 \end{equation}
The only linear symmetric polynomial is $\delta\mu_a +  \delta\mu_b$,
while there are two independent quadratic possibilities, which can
be chosen to be
$\delta\mu_a^2 +  \delta\mu_b^2$ and $(\delta\mu_a - \delta\mu_b)^2$.
With this choice we get the meson mass formula
\begin{equation}
   M(a \bar{b})
      = M_0 + \alpha (\delta\mu_a +  \delta\mu_b)
        + \beta_0 \sixth (\delta m_u^2 + \delta m_d^2 + \delta m_s^2)
        + \beta_1 (\delta\mu_a^2 +  \delta\mu_b^2) 
        + \beta_2 (\delta\mu_a -  \delta\mu_b)^2 \,,
\label{mesform}
\end{equation}
which reduces to eqs.~(\ref{PSoct}) and (\ref{PQoctvec}).
(For pseudoscalar mesons we expect $M^2$ to have a smoother Taylor
expansion than $M$ itself, so we keep the form of eq.~(\ref{mesform}),
but apply it to the squares of the pseudoscalar masses.)
This formula will also apply to the $s \bar{s}$ meson with annihilation
`switched off', i.e.\ with disconnected diagrams dropped.

We cannot see a way to extend this argument to include 
the $\eta_8$ or $\phi_8$ mesons. Using our full group argument we find
the formulae eq.~(\ref{fit_mpsO}) and eq.~(\ref{fit_mvO}), which involve an 
additional quadratic parameter, $\beta_3$. Physically it is very 
reasonable that the $\eta_8$ and $\phi_8$ should have a term 
that cannot be linked by symmetry to the mass of a meson with two
different valence quarks --- the $\eta_8$ and $\phi_8$ have
contributions from $q \bar{q}$ annihilation, which is absent in 
the other mesons of the multiplet, so it would be surprising if
symmetry could completely determine the masses of these `central' mesons.


\subsection{PQ octet baryons}


The number of free coefficients in the meson case was reduced
by the requirement that $K$ and $\overline{K}$ have the same masses.
However, there is no similar constraint linking $N$ and $\Xi$,
so more coefficients are allowed, both at the linear
and quadratic levels. Arguing as before, we find
\begin{eqnarray}
   M_N &=& M_0 + 3A_1\delta\mu_l + B_0\delta m_l^2 + 3B_1\delta\mu_l^2
                                                             \nonumber \\
   M_\Lambda
       &=& M_0 + A_1(2\delta\mu_l + \delta\mu_s)
               - A_2(\delta\mu_s - \delta\mu_l)
               + B_0\delta m_l^2
                                                             \nonumber \\
       & &     + B_1(2\delta\mu_l^2 + \delta\mu_s^2)
               - B_2(\delta\mu_s^2 - \delta\mu_l^2)
               + B_4(\delta\mu_s - \delta\mu_l)^2
                                                             \nonumber \\
   M_\Sigma
       &=& M_0 + A_1(2\delta\mu_l + \delta\mu_s)
               + A_2(\delta\mu_s - \delta\mu_l)
               + B_0\delta m_l^2
                                                             \nonumber \\
       & &     + B_1(2\delta\mu_l^2 + \delta\mu_s^2)
               + B_2(\delta\mu_s^2 - \delta\mu_l^2)
               + B_3(\delta\mu_s - \delta\mu_l)^2
                                                             \nonumber \\
   M_\Xi
       &=& M_0 + A_1(\delta\mu_l + 2\delta\mu_s)
               - A_2(\delta\mu_s - \delta\mu_l)
               + B_0\delta m_l^2
                                                             \nonumber \\
       & &     + B_1(\delta\mu_l^2 + 2\delta\mu_s^2)
               - B_2(\delta\mu_s^2 - \delta\mu_l^2)
               + B_3(\delta\mu_s - \delta\mu_l)^2 \,.
\label{Noct}
\end{eqnarray}
As usual, the nucleon mass has been made independent of $\delta\mu_s$.
Some useful combinations, which only depend on a few parameters, are
\begin{eqnarray}
   2 M_N - M_\Sigma - 3 M_\Lambda + 2 M_\Xi
     &=& (B_3 - 3 B_4) ( \delta\mu_s - \delta\mu_l )^2
                                                                       \\
   M_\Xi- M_\Sigma
     &=& ( A_1 - 2 A_2) ( \delta\mu_s - \delta\mu_l )
         + (B_1 - 2 B_2) ( \delta\mu_s^2 - \delta\mu_l^2 ) \,.
                                                             \nonumber
\end{eqnarray}
As mentioned previously, we can check that when $\mu \to m$
(i.e.\ return to the `unitary line') then these results return
to the previous results of
eqs.~(\ref{fit_mpsO}) -- (\ref{fit_mDD}).

As with the decuplet baryons, and the partially quenched mesons, 
we can give an alternative derivation, with less use of explicit 
group theory. However, the argument for partially quenched
octet baryons is slightly more complicated, because there are
fewer symmetry constraints. We will first consider the baryons
of the type $aab$, two valence quarks of flavour $a$, and one of
flavour $b$. These form the outer hexagon of the octet diagram.
As before, the sea quarks must contribute equally to all masses
in the octet, the mass formula must take the form
\begin{equation}
   M(aab)
      = M_0 + B_0 \sixth (\delta m_u^2 + \delta m_d^2 + \delta m_s^2)
            + F(\delta\mu_a , \delta\mu_b) \,,
\end{equation}
but now the dependence on the valence quark masses is not
symmetric under $a \leftrightarrow b$, so the function $F$ need
not be symmetric. This means there are two independent linear
terms (we choose $(2 \delta\mu_a + \delta\mu_b)$ and
$(\delta\mu_b - \delta\mu_a)$). There are three independent
quadratic terms, we choose $(2 \delta\mu_a^2 + \delta\mu_b^2)$ and
$(\delta\mu_b^2 - \delta\mu_a^2)$ (to mirror the pattern of
the linear terms) and $(\delta\mu_b - \delta\mu_a)^2 $.
Thus, the general formula for the $aab$ baryons can be written
\begin{eqnarray}
   M(aab) &=& M_0 + A_1 (2 \delta\mu_a + \delta\mu_b) 
                  + A_2 (\delta\mu_b - \delta\mu_a)
                  + B_0 \sixth (\delta m_u^2 + \delta m_d^2 + \delta m_s^2 )
                                                           \nonumber    \\
          & & + B_1 (2 \delta\mu_a^2 + \delta\mu_b^2) 
              + B_2 (\delta\mu_b^2 - \delta\mu_a^2)
              + B_3 (\delta\mu_b - \delta\mu_a)^2 \,.
\end{eqnarray}
Taking the cases $M(lll), M(lls)$ and $M(ssl)$ gives the
$N$, $\Sigma$ and $\Xi$ masses of eq.~(\ref{Noct}).

We have not found an equally simple argument for the mass of the $\Lambda$.
The result of the group theoretical calculation, as set out in
Appendix~\ref{group_theory}, is
\begin{eqnarray}
   M_\Lambda &=& M_0 + A_1 (2 \delta\mu_l + \delta\mu_s) 
                    - A_2 (\delta\mu_s - \delta\mu_l)
                    + B_0 \sixth (\delta m_u^2 + \delta m_d^2 + \delta m_s^2 )
                                                           \nonumber    \\
            & & + B_1 (2 \delta\mu_l^2 + \delta\mu_s^2) 
                       - B_2 (\delta\mu_s^2 - \delta\mu_l^2)
                       + B_4 (\delta\mu_s - \delta\mu_l)^2 \,.
\end{eqnarray}
Most terms are related to terms in the $\Sigma$ mass, the
$M_0, A_1, B_0$ and $B_1$ terms are the same for $\Lambda$ and $\Sigma$,
while the $A_2$ and $B_2$ terms have opposite signs for the
$\Lambda$ and $\Sigma$. However, for the term $(\delta\mu_s - \delta\mu_l)^2$
there is no connection between the coefficient in the $\Lambda$ mass
and the coefficient of this term in the other masses, and we need to
introduce a new parameter, that can only be determined by simulating
the $\Lambda$. We can understand this from Table~\ref{mat8_2p1}.
There is a particular combination of singlet, $8_b$ and $27$-plet
matrices which gives $0$ for all the baryons in the outer ring,
and only acts on the central baryons. Clearly, the coefficient
of this matrix only appears in the $\Lambda$ mass,
so at the quadratic level, we can no longer predict the $\Lambda$
mass from the other masses in the octet. We have a similar situation
with the mesons -- there is a quadratic coefficient that we only see
in the $\eta_8$ mass formula.


\subsection{Generalising a constant $\mathbf{\overline{m}}$ formula}
\label{mbar_varies}


We have stressed the advantages of keeping the average sea-quark mass,
$\overline{m}$, constant when approaching the physical point.
This leads to simpler extrapolation formulae, and results closer
to the physical values for flavour singlets and for partially quenched
calculations. If we want to move away from the surface $\overline{m} = m_0$, 
for example to consider a completely different trajectory,
such as $m_s = \mbox{constant}$ or $m_s^{\R} = \mbox{constant}$,
then it would be useful to know how to generalise our constant
$\overline{m}$  formulae to cover the full parameter space.

The procedure is simple; every constant parameter in our 
formulae becomes a function of $\overline{m}$, which we can 
then Taylor expand around $\overline{m} = m_0$. Taking as a first
example our cubic formula eq.~(\ref{r0_expansion})
for a flavour singlet quantity (such as $r_0$)
\begin{eqnarray}
   {r_0 \over a}
   &=& \alpha + \beta (\delta m_u^2 + \delta m_d^2 + \delta m_s^2)  
       + \gamma \, \delta m_u \delta m_d \delta m_s \,,
                                                    \nonumber    \\
   &\to& \alpha + \alpha^\prime (\overline{m} - m_0)
      + {1 \over 2!} \alpha^{\prime \prime} (\overline{m} - m_0)^2 
      + {1 \over 3!} \alpha^{\prime \prime \prime}
                                          (\overline{m} - m_0)^3 
                                                    \nonumber    \\
   & & + \beta (\delta m_u^2 + \delta m_d^2 + \delta m_s^2)
       + \beta^\prime (\overline{m} - m_0) 
               (\delta m_u^2 + \delta m_d^2 + \delta m_s^2) 
                                                    \nonumber    \\
   & & + \gamma \, \delta m_u \delta m_d \delta m_s \,,
\end{eqnarray} 
yielding a cubic formula with $7$ parameters. The extra polynomials
appearing in this formula are the `unticked' $A_1$ polynomials
in Table~\ref{cubic}.

We can take a slightly more complicated example, the 
partially quenched formula for the $\Sigma$ baryon. 
In eq.~(\ref{Noct}) we give the quadratic formula, valid with 
$\overline{m}$ held constant. Thus to construct the quadratic 
formula without this constraint, we must Taylor expand the 
parameters $M_0, A_1, A_2$, giving 
\begin{eqnarray} 
   M_\Sigma 
      &=& M_0 + M_0^\prime (\overline{m} - m_0)  
          + {1 \over 2!} M_0^{\prime \prime} (\overline{m} - m_0)^2 
          + A_1 (2 \delta\mu_l + \delta\mu_s )
                                                    \nonumber    \\
      & & + A_2 (\delta\mu_s - \delta\mu_l) 
          + A_1^\prime (\overline{m} - m_0)
                                 (2 \delta\mu_l + \delta\mu_s ) 
          + A_2^\prime  (\overline{m} - m_0)
                                 (\delta\mu_s - \delta\mu_l )
                                                    \nonumber    \\
      & & + B_0 \delta m_l^2 
          + B_1 (2 \delta\mu_l^2 + \delta\mu_s^2)
          + B_2 (\delta\mu_s^2 - \delta\mu_l^2) 
          + B_3 (\delta\mu_s - \delta\mu_l)^2 \,.
 \end{eqnarray} 
Note that this formula contains `mixed' polynomials such as
$(\overline{m} - m_0) (\delta\mu_s - \delta\mu_l)$,
involving both valence and sea  quarks. If we restrict
ourselves to the constant $\overline{m}$ surface such mixed
polynomials only show up at the cubic level.


\subsection{The usefulness of PQ}


There are several possible advantages to considering PQ results.
\begin{enumerate}

   \item The coefficients that appear in the expansions about
         the flavour symmetric line in the PQ case are the same
         as those that appear on the `unitary' case. Hence
         this may be a computationally cheaper way of obtaining them.

   \item PQ results can be helpful in choosing the
         next point to simulate, because the meson masses measured
         in the partially quenched approximation are very close to
         those found in a full calculation, giving us a preview of
         results on the next simulation point. We understand theoretically
         why this works well on our trajectory, with $\overline{m}$ held fixed.
         The reason is that the effect on the sea of making the $u$ and $d$
         quarks lighter is largely cancelled by the effect of making the
         $s$ quark heavier (the cancellation is perfect at the
         flavour symmetric point). Therefore partial quenching works best
         when only the non-singlet part of the quark mass matrix is varied
         (as is the case here). If the singlet part (the average sea-quark mass)
         is changed, there is no compensation, and the partially quenched
         results are less reliable.

   \item We can use partial quenching to get a good estimate of results
         at the physical point, by taking configurations generated with
         quark masses some distance short of the physical point, and then
         at the measurement stage using valence quarks chosen to give the
         physical $\pi$ and $K$ masses. Important physical effects, such as
         the light pion cloud, would be incorporated in the results.
         The effects of partially quenching can be further reduced by
         repeating the calculation with several choices of
         sea-quark masses, and making an extrapolation towards
         the physical sea-quark mass values.
         
   \item It is necessary in the determination of strange (or
         $s\overline{s}$) mesons without disconnected pieces.

\end{enumerate}

We can also show that on this trajectory the errors of the partially
quenched approximation are much smaller than on other trajectories.
In leading order $\chi$PT (terms linear in the quark mass), the
suggested procedure (valence quarks at the physical value, 
sea quarks anywhere on the physical constant $\overline{m}_{sea}$)
is exact. See Table VIII of \cite{walker-loud04a} for the
leading order formulae for both octet and decuplet baryon masses.
At this order partial quenching moves all the octet baryons
by the same amount, and all decuplets also move together. 
The leading order partial quenching errors are
\begin{eqnarray} 
   M_{oct}^{\PQ} - M_{oct}^{*} 
      &=& 6 \sigma_M ( \overline{m} - \overline{m}^{*} )
                                                             \nonumber \\
   M_{dec}^{\PQ} - M_{dec}^{*}
      &=& - 6 \overline{\sigma}_M ( \overline{m} - \overline{m}^{*} ) \,,
\label{PQoct_error}
\end{eqnarray} 
(using the notation of \cite{walker-loud04a} for quark masses
and the $\sigma$ coefficients). The superscript $^*$ denotes
quantities at the physical point. Since we have tuned $\overline{m}$
to be equal to the physical value, the partial quenching error
vanishes on our trajectory, but not on other trajectories,
which vary $\overline{m}$. 
   
We can give a partial derivative argument, like that of
section~\ref{introduction} or \cite{bietenholz10a},
which explains why this is so. Take the proton mass as an example, 
but any quantity will work the same way. The proton  mass will
depend on the valence quark masses and the sea quark masses, 
so we can write
\begin{eqnarray}
   M_p^{\PQ}(\mu_u, \mu_d ; m_u, m_d, m_s ) \,.
\end{eqnarray}
$M^{\PQ}$ is the mass of a partially quenched hadron calculated
on a sea background. The dependence on the three sea masses must be
completely symmetrical, unlike the dependence on valence masses.
At the symmetric point
\begin{eqnarray}
   {\partial M_p^{\PQ} \over \partial  m_u } =
   {\partial M_p^{\PQ} \over \partial  m_d } =
   {\partial M_p^{\PQ} \over \partial  m_s } \,, 
\end{eqnarray}
so if the sea quark masses are changed in a way which preserves 
$\overline{m}$, while the valence masses are held constant, 
$M_p^{\PQ}$ will not change (to leading order).
We can see these benefits of the constant $\overline{m}$
procedure in the mass formulae of this section. If we make
our valence quark masses equal to the quark masses at the
physical point, the only difference between the partially
quenched hadron mass and the physical hadron mass comes from
the $B_0$ or $\beta_0$ term in eqs.~(\ref{PQOmega}), (\ref{PSoct}),
(\ref{Noct}) which gives a quadratic mass shift
\begin{equation}
   M^* - M^{\PQ} = B_0 [ (\delta m_l^*)^2 - \delta m_l^2 ] \,,
\end{equation}
(where $\delta m_l^*$ refers to $\delta m_l$ at the physical
sea point), which is one power higher in the quark mass than the
usual result on other trajectories, eq.~(\ref{PQoct_error}).
This partial quenching shift is the same for all particles
in a multiplet, so splittings are unaffected by partial
quenching at this order - we would have to expand to
cubic terms to see partial quenching errors in the splittings.

  
\section{Applications to chiral perturbation theory} 
\label{applications_chipt}


Almost all LO (i.e.\ leading order or zero loop) chiral
perturbation theory ($\chi$PT) results follow simply from
flavour blindness, without any input from chiral symmetry.  
The linear terms in $m_q$, which are usually called LO $\chi$PT 
results, were originally discovered by Gell-Mann and Okubo
\cite{gell-mann62a,okubo62a}, using the (non-chiral) $SU(3)$
argument we are using in this article.

The only case where we need to invoke chiral symmetry is for the
pseudoscalar meson mass formula, where it is chiral symmetry 
which tells us that we have massless Goldstone bosons if $2$ or more
quark masses vanish. 

Beyond leading order we cannot derive the $\chi$PT result in full
solely from flavour blindness, but we can still make useful 
statements about the form that higher order contributions must take. 


\subsection{Decuplet baryon masses at $\mathbf{O(m_q^{3/2})}$} 


$O(m_q^{3/2})$ $\chi$PT is based on one-loop graphs, all 
with a pseudo-Goldstone boson. So we should expect 
that the individual terms in the $\chi$PT answer will
be functions of $M_\pi$ or of $M_K$ or of $M_{\eta_8}$, 
with no mixed terms (such as $M_\pi^2 M_K^2$), which can
only arise at the two-loop level. 

As an example, let us examine the $2+1$ next to leading order
(NLO) results for the decuplet baryon masses, \cite{tiburzi04a}.
Taking the formulae for individual masses, and grouping them into the 
multiplets of Table~\ref{bary2p1}, we know that in each case we 
are only allowed chiral perturbation theory expressions in
the corresponding multiplet:
\begin{eqnarray} 
   4 M_\Delta + 3 M_{\Sigma^*} + 2 M_{\Xi^*} + M_{\Omega}
   &=& 10 M_0 + 20 (\gamma_M - 3 \overline{\sigma}_M ) \overline{m}
                                                             \label{d1}\\
   & & - { 5 {\cal{H}}^2 \over  72 \pi f^2 } \, {5 \over 3}
       \left[ 3 M^3_\pi + 4 M^3_K + M^3_{\eta_8} \right]
                                                             \nonumber \\
   & & - {{\cal C}^2 \over (4 \pi f)^2} \, {5 \over 3}
       \left[ 3 {\cal F}_-(M_\pi)+ 4 {\cal F}_-(M_K)
              + {\cal F}_-(M_{\eta_8}) \right]
                                                             \nonumber \\
   - 2 M_\Delta + M_{\Xi^*} + M_{\Omega}
   &=& - 10 \gamma_M \delta m_l
                                                             \label{d8}\\
   & & - { 5 {\cal{H}}^2 \over 72 \pi f^2} \,  {1 \over 2}
       \left[ - 3 M^3_\pi + 2  M^3_K +  M^3_{\eta_8} \right]
                                                             \nonumber \\
   & & - { {\cal C}^2 \over (4 \pi f)^2} \,  {1 \over 3}
       \left[ - 3 {\cal F}_-(M_\pi)+ 2 {\cal F}_-(M_K)
              + {\cal F}_-(M_{\eta_8}) \right]
                                                             \nonumber \\
   4 M_\Delta - 5 M_{\Sigma^*} - 2 M_{\Xi^*} + 3 M_{\Omega}
   &=& { 5 {\cal{H}}^2 \over 72 \pi f^2} \,  {7 \over 9}
       \left[ - M^3_\pi + 4  M^3_K -3  M^3_{\eta_8} \right]
                                                            \label{d27}\\
   & &    - { {\cal C}^2 \over (4 \pi f)^2} \,  {7 \over 9}
       \left[ - {\cal F}_-(M_\pi)+ 4 {\cal F}_-(M_K)
              - 3 {\cal F}_-(M_{\eta_8}) \right]
                                                             \nonumber \\
   - M_\Delta + 3 M_{\Sigma^*} - 3 M_{\Xi^*} + M_{\Omega}
   &=& 0 \,,
\label{d64}
\end{eqnarray} 
using the notation of \cite{tiburzi04a} (In particular 
${\cal F}_-(M_i)$ is short-hand for the function
${\cal F}(M_i, -\Delta, \mu)$ defined there).
The coefficients on the right-hand side of eq.~(\ref{d64})
must follow the pattern of Table~\ref{meso2p1},
but any function of the meson masses is allowed.
We proved a weaker version of this result in \cite{bietenholz10a},
using the permutation group instead of full $SU(3)$. 

The meson mass matrix, $8 \otimes 8$, contains no $64$, 
there are no possible $1$-loop terms to place on the right-hand side
of eq.~(\ref{d64}), so this mass combination must be zero in 
NLO $\chi$PT. We have already noted in Table~{\ref{bary2p1}}
that this combination has a Taylor expansion beginning at
$O(\delta m_l^3)$ and is very small experimentally, eq.~(\ref{num27}).


\subsection{Relationships between expansion coefficients}


We now investigate the relation between the parameters of
$\chi$PT and the Taylor coefficients in our approach,
eqs.~(\ref{fit_mpsO}) -- (\ref{fit_mDD}).

For example for the pseudoscalar octet, using the $2+1$ results in
\cite{allton08a} and assuming their validity up to the point on the
flavour symmetric line (the kaon mass is always smaller here than
the physical kaon mass), we find
\begin{eqnarray}
   M_0^2 &=& \overline{\chi}\,\left[
                1 - {16\overline{\chi}\over f_0^2}
                       \left( 3L_4 + L_5 -6L_6 -2L_8 \right)
                        + {\overline{\chi}\over 24\pi^2 f_0^2}
                          \ln {\overline{\chi}\over \Lambda_\chi^2}
                             \right]
                                                              \nonumber \\
  \alpha &=& Q_0 \,\left[
                1 - {16\overline{\chi}\over f_0^2}
                       \left( 3L_4 + 2L_5 -6L_6 -4L_8 \right)
                        + {\overline{\chi}\over 8\pi^2 f_0^2}
                          \ln {\overline{\chi}\over \Lambda_\chi^2}
                             \right]
                                                              \nonumber \\
  \beta_0&=& - {Q_0^2 \over 6\pi^2 f_0^2}
                                                              \nonumber \\
  \beta_1&=& {Q_0^2 \over f_0^2}\, \left[
                - 32\left(L_5 - 2L_8 \right)
                + {1\over 24\pi^2}
                    \left(7 + 4\ln {\overline{\chi}\over \Lambda_\chi^2} \right)
                                  \right]
                                                              \nonumber \\
  \beta_2&=& {Q_0^2 \over f_0^2}\, \left[
                 16\left(L_5 - 2L_8 \right)
                 - {1\over 24\pi^2}
                   \left(3 + 2\ln {\overline{\chi}\over \Lambda_\chi^2} \right)
                                  \right] \,,
\label{LECs}
\end{eqnarray}
where $\chi_q = 2Q_0m_q$ with $Q_0 = B_0^{\R}Z_m^{\NS}$ so that here
we have $\overline{\chi} = 2Q_0(1+\alpha_Z)\overline{m}$ which is
kept constant. The $L_i$s are appropriate low energy constants or LECs.

We first note that when expanding the $\chi$PT about a point on
the $SU(3)$ flavour symmetry line gives to leading order
only one parameter, $\alpha$ as expected. (This means, in particular,
that the flavour singlet combination, $X_\pi^2$, vanishes to leading order,
as discussed previously.) Secondly, while we can fit to $\alpha$
and $\beta_0$, $\beta_1$ and $\beta_2$, it will be difficult
to determine the individual LECs. The best we can probably
hope for are these combinations.


\subsection{Chiral non-analytic behaviour}


We briefly discuss the question of how chiral logs,
or other chiral singularities, fit with the Taylor expansion.
The answer is that the chiral singularity should show
up in the large-$n$ behaviour of the coefficients of $\delta m_q^n$.

For example if we make a Taylor expansion about $\overline{\chi}$
of the singular term $\chi_l^2 \ln( \chi_l /\Lambda^2_\chi)$ which occurs
in the chiral expansion of $M_\pi^2$, we would find
\begin{equation} 
   \chi_l^2 \ln \chi_l 
      = \left[ \overline{\chi} + \delta\chi_l \right]^2
        \ln \overline{\chi} + \overline{\chi}\delta\chi_l
        + {3 \over 2}\delta\chi_l^2
        + \overline{\chi}^2 
            \sum_{n=3}^\infty {2  (-1)^{n-1} \over n(n-1)(n-2)} 
               \left({\delta\chi_l \over \overline{\chi}}\right)^n \,,
\end{equation}
where $\delta\chi_l = \chi_l - \overline{\chi}$. 
So at large $n$ the coefficients of $\delta\chi_l^n$ are
proportional to $\overline{\chi}^{2-n} / n^3$.
If we look at the first singular term in the baryon mass formula,
$\chi_l^{3/2}$, and expand, we would get a series
$\sim \overline{\chi}^{3/2 -n} \delta\chi_l^n / n^{5/2}$
at large $n$.

This is general, the power of $n$ with which the terms
drop off tells us the chiral singularity. If the singularity
is $\chi_l^p \ln (\chi_l/\Lambda_\chi^2)$, with integer $p$,
then the series drops like $1/n^{p+1}$. If the singularity
is $\chi_l^q$, with non-integer $q$, then the series drops
like $1/n^{q+1}$. (If we have a singularity in
$\ln (\chi_\eta/\Lambda_\chi^2)$ or $\ln (\chi_K/\Lambda_\chi^2)$,
where $\chi_\eta = (\chi_l + 2\chi_s)/3$, $\chi_K = (\chi_l + \chi_s)/2$
we just change $\chi_l$ to $\chi_\eta$ or $\chi_K$ in the Taylor series.)
Needless to say, this large-$n$ behaviour would prove difficult
to see in practice, because the coefficients are small these terms
only become important close to the chiral limit.


\section{The path to the physical point}
\label{path}


In section~\ref{introduction} the proposed path to the physical
point was introduced. We shall now discuss this a little further.

For the simulation it is easiest to keep the (bare) singlet quark
mass fixed,
\begin{eqnarray}
   \overline{m} = \third(2m_l + m_s)  =  m_0 =  \mbox{constant} \,,
\label{Lo_path}
\end{eqnarray}
starting from some reference point $(m_l,m_s) = (m_0,m_0)$
on the flavour symmetric line. We can use the singlet
combinations from Table~\ref{perminv_2p1} to locate the
starting point of our path to the physical point by fixing
a dimensionless ratio such as
\begin{eqnarray}
  {X_\pi^2 \over X_N^2} = \mbox{physical value} 
                        = \left. {X_\pi^2 \over X_N^2} \right|^* \,.
\label{XpioXN}
\end{eqnarray}
Note also that at the flavour symmetric point $X_\pi = M_\pi|_0$
so this determines our starting pion mass (from the experimental
values given later in Table~\ref{hadron_masses_singlet})
to be $\sim 410\,\mbox{MeV}$.

However the equivalence of eqs.~(\ref{Lo_path}), (\ref{XpioXN})
is only strictly true at lowest order. While at this order it
does not matter whether we kept the quark mass singlet constant,
eq.~(\ref{Lo_path}), or a particle mass singlet constant,
eq.~(\ref{XpioXN}), higher order terms mean that it now does.
If we make different choices of the quantity we keep constant
at the experimentally measured physical value, for example
\begin{eqnarray}
   { X_\pi^2 \over X_N^2 }\,, \quad
   { X_\pi^2 \over X_\Delta^2 }\,, \quad
   { X_\pi^2 \over X_\rho^2 }\,, \quad
   \ldots \,,
\label{some_other_XpioXs}
\end{eqnarray}
we get slightly different trajectories. The different trajectories
begin at slightly different points along the flavour $SU(3)$ symmetric
line. Initially they are all parallel with slope $-2$, but away from the
symmetry line they can curve, but will all meet at the physical 
point. (Numerically we shall later see that this seems to be a small
effect.)

An additional effect comes from the choice of Wilson lattice fermions.
The physical domain is defined by
\begin{eqnarray}
   m_l^{\R} &\ge& 0
                                                             \nonumber \\
   m_s^{\R} &\ge& 0 \,,
\end{eqnarray}
which using eq.~(\ref{mr2mbare}) translates to
\begin{eqnarray}
   m_l \ge - { \third\alpha_Z \over (1 + \twothird\alpha_Z) } m_s
                                 \,, \qquad
   m_s \ge - { \twothird\alpha_Z \over (1 + \third\alpha_Z) } m_l \,,
\label{phy_dom_b}
\end{eqnarray}
leading to a non-rectangular region and possibly negative bare quark mass.
(These features disappear of course for chiral fermions when
$\alpha_Z = 0$.)

\begin{figure}[htb]
   \vspace*{0.15in}
   \begin{center}
      \includegraphics[width=7.0cm]{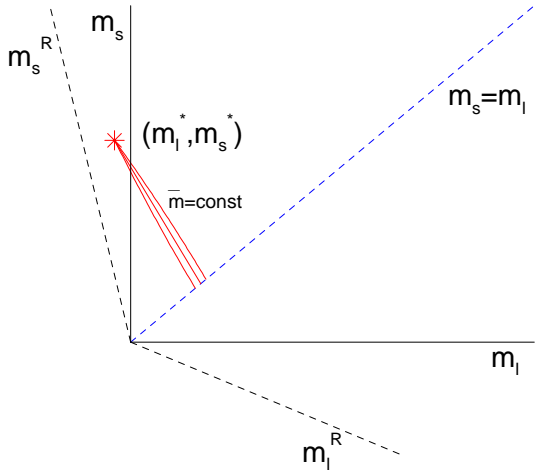}
   \end{center} 
   \caption{Sketch of some possible paths (red lines) in the $m_l$--$m_s$
            plane to the physical point $(m_l^*, m_s^*)$.
            Due to eq.~(\protect\ref{phy_dom_b}) the $m_l^{\R}$
            and $m_s^{\R}$ axes are not orthogonal when plotted
            in the bare quark mass plane.}
   \label{sketch_ml_ms_bLO+paths}
\end{figure}

These two features are sketched in Fig.~\ref{sketch_ml_ms_bLO+paths},
which shows possible paths in the $m_l$--$m_s$ plane starting from the
flavour symmetric line. In this figure the horizontal and vertical axes
are the bare quark masses, $m_l$ and $m_s$ respectively. Because of
renormalisation effects, eq.~(\ref{phy_dom_b}), lines of constant
renormalised mass, $m_l^R = \mbox{constant}$ or $m_s^R = \mbox{constant}$,
will be at an angle. This is in contrast with Fig.~\ref{sketch_mlR_msR+path},
where renormalised masses were used as the axes.


\section{The lattice -- generalities}
\label{lat_generalities}


After the general discussion of $SU(3)$ flavour expansions
described in sections~\ref{theory_1p1p1}--\ref{applications_chipt}
(which are lattice independent), we now turn to more specific lattice
considerations.


\subsection{Lattice simulations}
\label{simulations}


We use a clover action for $2+1$ flavours with a single step of
mild stout smearing as described in Appendix~\ref{slinc_action}.
Further details are given in \cite{cundy09a} together with a
non-perturbative determination of the improvement coefficient
for the clover term, using the Schr\"odinger functional method.

The bare quark masses are defined as
\begin{eqnarray}
   am_q = {1 \over 2} 
            \left ({1\over \kappa_q} - {1\over \kappa_{0;c}} \right) \,,
\label{kappa_bare}
\end{eqnarray}
where vanishing of the quark mass along the $SU(3)$ flavour 
symmetric line determines $\kappa_{0;c}$.
We then keep $\overline{m} = \mbox{constant} \equiv  m_0$ which gives
\begin{eqnarray}
   \kappa_s = { 1 \over { {3 \over \kappa_0} - {2 \over \kappa_l} } } \,.
\label{kappas_mbar_const}
\end{eqnarray}
So once we decide on a $\kappa_l$ this then determines $\kappa_s$.

How accurately must we satisfy eq.~(\ref{kappas_mbar_const})?
In choosing suitable $(\kappa_l, \kappa_s)$, the natural scale
is to say that changes in $\overline{m}$ should be small when
compared to $\overline{m}$ itself, i.e.\ $|\overline{m} - m_0| \ll m_0$
which gives
\begin{equation}
   \left| {1\over 3}\left({2\over\kappa_l} + {1\over\kappa_s}\right)
    - {1 \over \kappa_0}\right| \ll {1\over\kappa_0} - {1\over\kappa_{0;c}} \,,
\end{equation}
which is satisfied if we give all our $\kappa$s to $6$
significant figures.

Furthermore note that we are not expanding about the chiral limit,
but have expansions around a flavour symmetric point which
does not require knowledge of $\kappa_{0;c}$. This follows as
\begin{eqnarray}
   \delta m_q 
      &=& m_q - \overline{m}
                                                             \nonumber \\
      &=& {1 \over 2a} 
          \left( {1 \over \kappa_q} - {1 \over \kappa_0} \right) \,.
\end{eqnarray}

HMC (hybrid Monte Carlo) and RHMC (rational HMC) were used for the
$2$ and $1$ fermion flavours respectively, \cite{nakamura10b},
to generate the gauge configurations. We note the following in
connection with the simulations and our path choice:

\begin{itemize}

   \item The simulations should equilibrate quickly from one point
         to another along this path, because the effects of making
         the strange quark mass heavier tend to be cancelled by making
         the $u$ and $d$ quarks lighter. 

   \item The simulation cost change should be moderate for this path.
         This may be motivated by the following crude cost argument.
         Modelling the cost, $C$, as
         \begin{eqnarray}
            C \propto { 1 \over am^{\R}_l} + { k \over am^{\R}_s} \,,
         \end{eqnarray}
         where $k$ is the relative cost of the two algorithms,
         gives on the line $a\overline{m} = \mbox{constant}$
         \begin{eqnarray}
            C(\xi) \propto {1 \over (1+\alpha_Z) + \xi} 
                           + {k \over (1+\alpha_Z)-2\xi } \,,
         \end{eqnarray}
         with
         \begin{eqnarray}
            \xi = \delta m_l / \overline{m}  \,,
         \end{eqnarray}
         (alternatively we could consider $M_\pi^2/X_\pi^2$).
         The cost $C(\xi)/C(0)$ from the symmetric point
         $\xi = 0$, is plotted in Fig.~\ref{cost_sketch}.
         \begin{figure}[tb]
            \hspace*{1.50in}
            \includegraphics[width=8.00cm]{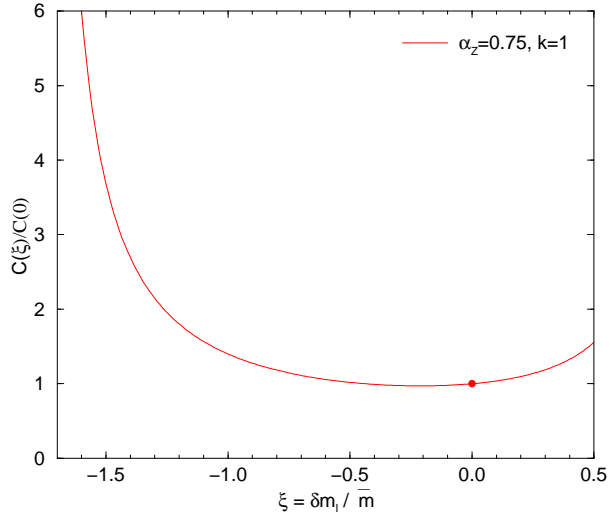}
            \caption{Simulation cost $C(\xi)/C(0)$ against $\xi$ where 
                     $\xi =  \delta m_l / \overline{m}$ (with, for example
                     $\alpha_Z = 0.75$, cf. eq.~(\protect\ref{alphaZ_num}),
                     and $k = 1$). The symmetric point is denoted by a
                     (red) filled circle. Very roughly we need to
                     reach a region where $\xi \lsim -1.5$.}
            \label{cost_sketch}
         \end{figure}
         There is little change in a reasonably large range of $\xi$
         starting from $\xi = 0$.

\end{itemize}
Both these points are indeed found in practice (at least approximately).


\subsection{$\mathbf{O(a)}$ improvement of the coupling constant}


$O(a)$ improvement leads to a change in the coupling constant via
\cite{luscher96a},
\begin{equation}
   g_0^2 \to \tilde{g}_0^2 = g_0^2 \, ( 1 + b_g a\overline{m} ) \,,
\label{g0imp}
\end{equation}
where $b_g$ is some function of $g_0^2$. Not much is known about
the value of $b_g$. For Wilson glue and $\alpha = 0$
(i.e.\ no-stout smeared links) Wilson-Dirac fermions the lowest
order perturbative result is $b_g = 0.01200n_fg_0^2 + O(g_0^4)$,
\cite{luscher96a}, which is small but increasing with $n_f$
(here $n_f = 3$). A crude estimate was made in \cite{bakeyev03a}
and indicated a possible $1$--$2\%$ effect (but with considerable
uncertainty).

In general from eq.~(\ref{g0imp}) when we vary a quark mass then
$g_0$ must be changed to keep $\tilde{g}_0$ constant. However
for our choice of path ($\overline{m} = \mbox{constant}$)
the relation between $g_0$ and $\tilde{g}_0$ is fixed,
so only a small overall shift of results might be
necessary -- nothing else changes as we traverse our path.
Therefore in the following we shall not consider the effect of $b_g$
any further.


\subsection{Hadron `sources' and `sinks'}


The operators (or interpolators) used to determine the hadron masses
are uniformly taken to be Jacobi smeared (\cite{allton93a} and
\cite{best97a} (Appendix C)) and to be non-relativistic, NR,
(\cite{billoire85a}, \cite{gockeler94a} and \cite{gockeler04b}
(Appendix C)). Specifically, we consider the following hadron
sources and sinks:
\begin{itemize}

   \item Pseudoscalar meson octet
         \begin{eqnarray}
            {\cal M}_{\pi}(x)  &=& \overline{d}(x) \gamma_5 u(x)
                                                             \nonumber \\
            {\cal M}_K(x)      &=& \overline{s}(x) \gamma_5 u(x)
                                                             \nonumber \\
            {\cal M}_{\eta_s}(x)&=& \overline{s}(x) \gamma_5 s(x)
         \end{eqnarray}

   \item Vector meson octet
         \begin{eqnarray}
            {\cal M}_{\rho \, i}(x)&=& \overline{d}(x) \gamma_i u(x)
                                                             \nonumber \\
            {\cal M}_{K^* \, i}(x) &=& \overline{s}(x) \gamma_i u(x)
                                                             \nonumber \\
            {\cal M}_{\phi_s \, i}(x)&=& \overline{s}(x) \gamma_i s(x)
         \end{eqnarray}

   \item Baryon octet
         \begin{eqnarray}
            {\cal B}_{N \, \alpha}(x)
                        &=& \epsilon^{abc} u_\alpha^a(x)
                            \left[ u^b(x)^{T_D}C\gamma_5 d^c(x) \right]
                                                             \nonumber \\
            {\cal B}_{\Lambda\, \alpha}(x)
                        &=& \epsilon^{abc} \left(
                     2s_\alpha^a(x)\left[u^b(x)^{T_D}C\gamma_5 d^c(x)\right]
                                           \right.
                                                             \nonumber \\
                        & & \hspace*{0.25in}\left.
                     +d_\alpha^a(x)\left[u^b(x)^{T_D}C\gamma_5 s^c(x)\right]
                     -u_\alpha^a(x)\left[d^b(x)^{T_D}C\gamma_5 s^c(x)\right]
                                           \right)
                                                             \nonumber \\
            {\cal B}_{\Sigma \, \alpha}(x)
                        &=& \epsilon^{abc} u_\alpha^a(x)
                            \left[ u^b(x)^{T_D}C\gamma_5 s^c(x) \right]
                                                             \nonumber \\
            {\cal B}_{\Xi \, \alpha}(x)
                        &=& \epsilon^{abc} s_\alpha^a(x)
                            \left[ s^b(x)^{T_D}C\gamma_5 u^c(x) \right]
         \end{eqnarray}

   \item Baryon decuplet
         \begin{eqnarray}
            {\cal B}_{\Delta \, \alpha}(x)
                        &=& \epsilon^{abc}
                            \left( 2u_\alpha^a(x)
                                   \left[u^b(x)^{T_D}C\gamma_- d^c(x)\right]
                                  + d_\alpha^a(x) 
                                   \left[u^b(x)^{T_D}C\gamma_- u^c(x)\right]
                            \right)
                                                             \nonumber \\
            {\cal B}_{\Sigma^* \,\alpha}(x)
                        &=& \epsilon^{abc}
                            \left( 2u_\alpha^a(x)
                                   \left[u^b(x)^{T_D}C\gamma_- s^c(x)\right]
                                  + s_\alpha^a(x) 
                                   \left[u^b(x)^{T_D}C\gamma_- u^c(x)\right]
                            \right)
                                                             \nonumber \\
            {\cal B}_{\Xi^* \,\alpha}(x)
                        &=& \epsilon^{abc}
                            \left( 2s_\alpha^a(x)
                                   \left[s^b(x)^{T_D}C\gamma_- u^c(x)\right]
                                  + u_\alpha^a(x) 
                                   \left[s^b(x)^{T_D}C\gamma_- s^c(x)\right]
                            \right)
                                                             \nonumber \\
            {\cal B}_{\Omega \,\alpha}(x)
                        &=& \epsilon^{abc} s_\alpha^a(x)
                            \left[ s^b(x)^{T_D}C\gamma_- s^c(x) \right]
         \end{eqnarray}

\end{itemize}
where $C = \gamma_2\gamma_4$ and
$\gamma_- = {1 \over 2}(\gamma_2 + i\gamma_1)$ and the superscript $^{T_D}$
is a transpose in Dirac space. The $u$ and $d$ quarks are considered as distinct,
but of degenerate mass.

The correlation functions (on a lattice of temporal extension $T$) we
use are given by
\begin{eqnarray}
   C_{\pi_O}(t) &=& {1 \over V_s} \, \left\langle
                 \sum_{\vec{y}} {\cal M}_{\pi_O}(\vec{y},t)
                 \sum_{\vec{x}} {\cal M}_{\pi_O}^\dagger(\vec{x},0)
              \right\rangle
                                                             \nonumber \\
          &\propto& A\left(e^{-M_{\pi_O}t} + e^{-M_{\pi_O}(T-t)}\right) \,,
                    \qquad \pi_O = \pi, K, \eta_s
                                                             \nonumber \\
   C_{\rho_O}(t)
          &=& {1 \over 3V_s} \, \sum_i \, \left\langle
                 \sum_{\vec{y}} {\cal M}_{\rho_O\,i}(\vec{y},t)
                 \sum_{\vec{x}} {\cal M}_{\rho_O\,i}^\dagger(\vec{x},0)
              \right\rangle
                                                             \nonumber \\
          &\propto& A\left(e^{-M_{\rho_O}t} + e^{-M_{\rho_O}(T-t)}\right) \,,
                    \qquad \rho_O = \rho, K^*, \phi_s
                                                             \nonumber \\
   C_{N_O}(t)
          &=& {1 \over V_s} \, \mbox{Tr}_D \Gamma_{unpol} \, \left\langle
                 \sum_{\vec{y}} {\cal B}_{N_O}(\vec{y},t)
                 \sum_{\vec{x}} \overline{\cal B}_{N_O}(\vec{x},0)
              \right\rangle
                                                             \nonumber \\
          &\propto& A e^{-M_{N_O}t} + B e^{-M^\prime_{N_O}(T-t)} \,,
                    \qquad N_O = N, \Sigma, \Xi
                                                             \nonumber \\
   C_{N_\Lambda}(t)
          &=& {1 \over V_s} \, \mbox{Tr}_D \Gamma_{pol} \, \left\langle
                 \sum_{\vec{y}} {\cal B}_{N_\Lambda}(\vec{y},t)
                 \sum_{\vec{x}} \overline{\cal B}_{N_\Lambda}(\vec{x},0)
              \right\rangle
                                                             \nonumber \\
          &\propto& A e^{-M_{N_\Lambda}t} + B e^{-M^\prime_{N_{\Lambda}}(T-t)} \,,
                                                             \nonumber \\
   C_{\Delta_O}(t)
          &=& {1 \over V_s} \, \mbox{Tr}_D \Gamma_{pol} \, \left\langle
                 \sum_{\vec{y}} {\cal B}_{\Delta_O}(\vec{y},t)
                 \sum_{\vec{x}} \overline{\cal B}_{\Delta_O}(\vec{x},0)
              \right\rangle
                                                             \nonumber \\
          &\propto& A e^{-M_{\Delta_O}t} + B e^{-M^\prime_{\Delta_O}(T-t)} \,,
                    \qquad \Delta_O = \Delta, \Sigma^*, \Xi^*, \Omega
\end{eqnarray}
with $\Gamma_{unpol} = {1 \over 2}(1 + \gamma_4)$ and 
$\Gamma_{pol} = \Gamma_{unpol}(1 + i\gamma_3\gamma_5)$. $M^\prime$ 
is the lowest excited state with opposite parity to $M$.


\section{The lattice -- results}
\label{lat_results}


All the results given in this article will be at 
$\beta \equiv 10/g_0^2 = 5.50$, $\alpha = 0.1$, together with
$c_{sw} = 2.65$, see Appendix~\ref{slinc_action}. (This $\beta$ value
was located by an initial series of short degenerate quark mass runs,
to give a rough idea of the associated scale.)
The hadron masses will be given below in a series of
Tables~\ref{table_run_M_sym}--\ref{table_run_M_BDoX}.
The runs on $24^3\times 48$ lattices have $O(2000)$ trajectories,
while the runs on $32^3\times 64$ lattices have $O(1500)$--$O(2000)$
trajectories. Errors are determined using the bootstrap method.
Experimental values of the hadron masses are given in
section~\ref{hadron_masses}.


\subsection{Locating $\mathbf{\kappa_0}$ and the
            $\mathbf{m_s^{\R}}$--$\mathbf{m_l^{\R}}$ plane}
\label{location}


From the discussion in section~\ref{path} for our path choice,
we must first determine the starting value on the flavour symmetric line.
A series of runs along the $SU(3)$ flavour line determines this
point, $\kappa_0$, by checking when $X_\pi^2/X_S^2$,
$S = N$, $\Delta$, $\rho$ are equal to their physical values,
see eqs.~(\ref{XpioXN}), (\ref{some_other_XpioXs}). (This would
also include $S = r$ if we have previously determined the physical value
of $r_0$.) On the flavour symmetric line obviously all the particles
in the multiplet are mass degenerate, so for example taking $S = N$
means that
\begin{eqnarray}
   { (aM_\pi)^2 \over (aM_N)^2 } = \left. { X_\pi^2 \over X_N^2 } \right|^* \,,
\end{eqnarray}
where, to emphasise that the left-hand side are the lattice measurements,
we temporarily include the lattice spacing. (Again, the star denotes the
experimental value.)

\begin{figure}[htb]
   \begin{center}
      \includegraphics[width=11.0cm,angle=270]
          {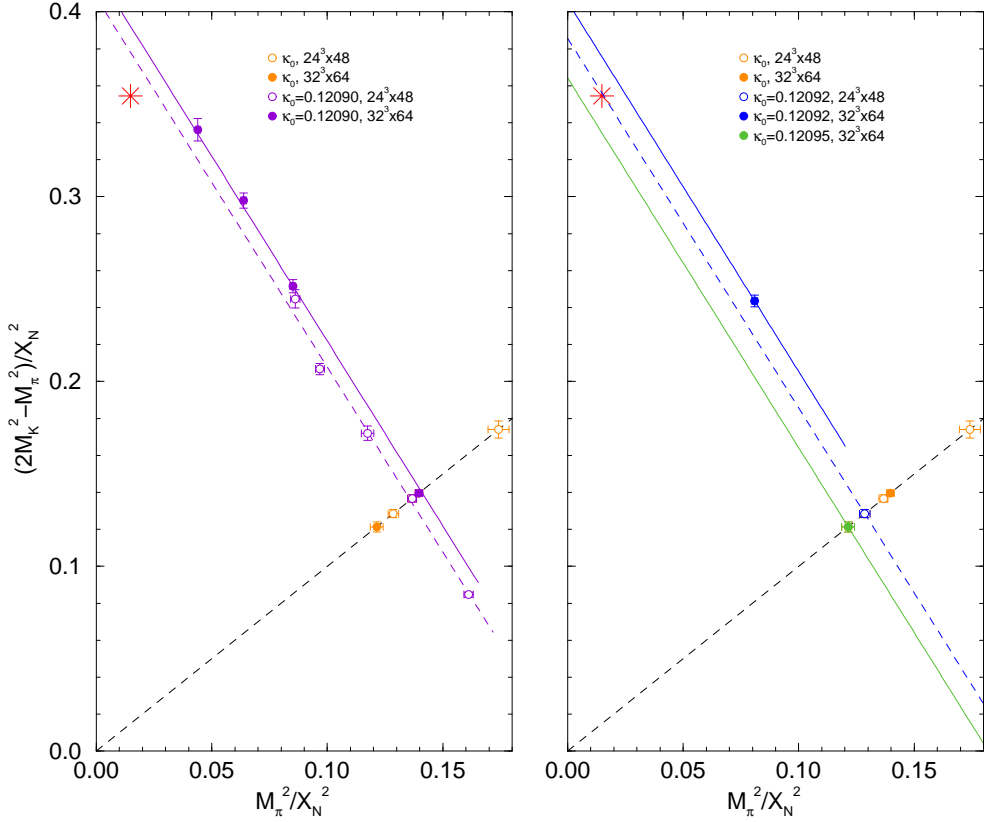}
   \end{center} 
   \caption{$(2M_K^2 - M_\pi^2)/X_N^2$ versus $M_\pi^2/X_N^2$ for
            $\kappa_0 = 0.12090$ (left panel) and $0.12092$,
            $0.12095$ (right panel). The dashed black line,
            $y = x$ represents the $SU(3)$ flavour symmetric line.
            Filled points are on $32^3\times 64$ lattices
            while open points are on a $24^3\times 48$ sized lattice.
            Shown are points on the flavour symmetric line
            (coloured orange) followed by results with
            $\overline{m} = \mbox{constant}$
            (coloured violet, left panel; blue and green,
            right panel). The fits are from
            eq.~(\protect\ref{const_mbar_fit})
            with $c_N$ a free parameter.
            The physical value is denoted by a (red) star.}
\label{b5p50_mps2oXn2_2mpsK2-mps2oXn2}
\end{figure}

Once we have located a promising $\kappa_0$ (or better a small range
of $\kappa_0$s) then we keep $\overline{m} = \mbox{constant}$ and pick
appropriate $(\kappa_l, \kappa_s)$ values, using eq.~(\ref{kappas_mbar_const}).
Again, setting $X_\pi^2/X_N^2 = \mbox{physical value}$,
eq.~(\ref{XpioXN}) can be re-written as
\begin{eqnarray}
   {2 M_K^2 - M_\pi^2 \over X_N^2}
      = c_N - 2 {M_\pi^2 \over X_N^2} \,, \qquad 
         c_N = 3 \left. { X_\pi^2 \over X_N^2 } \right|^* \,,
\label{const_mbar_fit}
\end{eqnarray}
considering for the present only lowest order in the flavour 
expansion. In Fig.~\ref{b5p50_mps2oXn2_2mpsK2-mps2oXn2} we plot 
$(2M_K^2 - M_\pi^2)/X_N^2$ versus $M_\pi^2/X_N^2$. This is equivalent
to plotting $m_s^{\R}$ against $m_l^{\R}$ because from LO $\chi$PT,
$M_\pi^2 \propto m_l^{\R}$ and $2M_K^2-M_\pi^2 \propto m_s^{\R}$.

Note that simulations with a `light' strange quark mass and heavy
`light' quark mass are possible -- here the right most points
in Fig.~\ref{b5p50_mps2oXn2_2mpsK2-mps2oXn2}. In this inverted
strange world we would expect the weak interaction decays 
$p \to \Sigma$ or $\Lambda$.

Also shown in Fig.~\ref{b5p50_mps2oXn2_2mpsK2-mps2oXn2} are fits
using eq.~(\ref{const_mbar_fit}) leaving $c_N$ as a free parameter
starting from the flavour symmetric points
\begin{eqnarray}
   \kappa_0 = 0.12090\,, \quad
   \kappa_0 = 0.12092\,, \quad
   \kappa_0 = 0.12095 \,,
\end{eqnarray}
(the latter two points are reference points). It is seen from
the figure that this range covers the possible paths to the physical
point. There are two observations to be made. Firstly we note
that there does not seem to be much non-linearity in the data,
i.e.\ the leading order in the expansion about the flavour
symmetric line already seems sufficient. So if
$c_N = 3 ( X_\pi^2 / X_N^2 )|^*$ then the lines would go exactly
through the physical point. Also this means from the discussion
in section~\ref{path} that using other singlet scale choices
should lead to a similar result. Secondly, as noted before
at the end of section~\ref{theory_2p1}, as the expansions have been
derived using only group theoretic arguments, they will be
valid for results derived on any lattice volume (though the
coefficients of the expansion are still functions of the volume).
So here, to test this, we have made separate fits for the two
volumes --- $24^3\times 48$ and $32^3\times 64$. Indeed this shows
that finite size effects are present but small.

Thus our present conclusion is that $\kappa$ in the range
$0.12090$ -- $0.12095$ is within a few percent
of the reference $\kappa_0$. Most of the results reported
here will be at $\kappa_0 = 0.12090$.


\subsection{Determination of $\mathbf{\kappa_{0;c}}$, $\mathbf{\alpha_Z}$}
\label{kappasymc}

\begin{figure}[htb]
   \begin{center}
      \includegraphics[width=11.0cm]
             {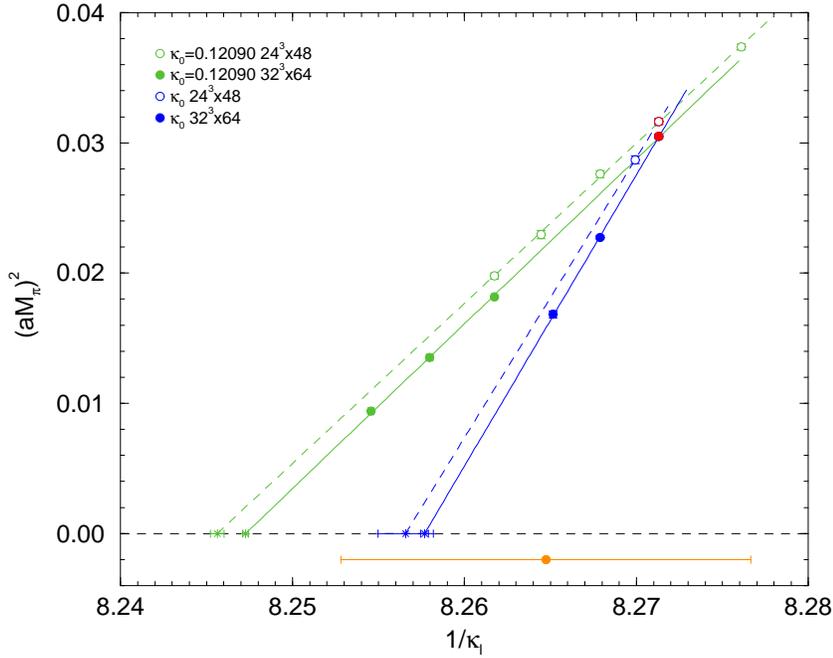}
   \end{center} 
   \caption{$(aM_\pi)^2$ versus $1/\kappa_l$ for both the flavour
            symmetric case (blue points) and keeping $\overline{m}$
            constant (green points). $24^3\times 48$ volumes are
            opaque circles and $32^3\times 64$ volumes are filled
            circles. The $\kappa_0 = 0.12090$ points are highlighted 
            in red. The chirally extrapolated values from the
            linear fits are denoted by stars. The horizontal
            (orange) filled circle is the Schr\"odinger
            functional estimate.}
\label{b5p50_amps2_ookl_ksymp12090+ksym}
\end{figure}

Although not strictly necessary, we briefly indicate here the
determination of $\kappa_{0;c}$ and $\alpha_Z$ to illustrate some
of the discussion in section~\ref{path}. Using lowest order $\chi$PT 
(i.e.\ the fact that the pion mass vanishes if the masses of the
light quarks vanish) and
\begin{eqnarray}
   (aM_\pi)^2 \propto am_l^{\R} \propto am_l + \alpha_Z a\overline{m} \,,
\end{eqnarray}
where the constant of proportionality from eq.~(\ref{LECs})
is $2a\alpha = 2aQ_0 = 2aB_0^{\R}Z_m^{\NS}$,
we first determine $\kappa_{0;c}$ (the critical hopping parameter
on the flavour symmetric line) as defined in eq.~(\ref{kappa_bare}).
In Fig.~\ref{b5p50_amps2_ookl_ksymp12090+ksym} we show the
plot of $(aM_\pi)^2$ versus $1/\kappa_l$ together with linear fits.
For the flavour symmetric case, from the blue points we find
\begin{eqnarray}
   {1 \over \kappa_{0;c}} = 8.25768(23) \,, \qquad \mbox{or} \quad
             \kappa_{0;c} = 0.121099(4) \,,
\end{eqnarray}
which is in good agreement with the Schr\"odinger functional
determination, see Appendix~\ref{slinc_action}.
Note that for $\kappa_l < \kappa_{0;c}$ the bare $am_q$ is negative,
but the renormalised $m_q^{\R}$ is always positive, eq.~(\ref{phy_dom_b})
and Fig.~\ref{sketch_ml_ms_bLO+paths}. This occurs for the last point
on the $32^3\times 64$ line in Fig.~\ref{b5p50_amps2_ookl_ksymp12090+ksym}.

$\alpha_Z$ can then be estimated using the $\overline{m} = \mbox{constant}$
line as here $(aM_\pi)^2$ vanishes at $\kappa_c$ giving
\begin{equation}
   \alpha_Z = - { am_q|_{\kappa = \kappa_c}\over a\overline{m} }
            =
              { \left( {1 \over \kappa_{0;c}} - {1 \over \kappa_c} \right)
                \over
                \left( {1 \over \kappa_0} - {1 \over \kappa_{0;c}}
                \right)  } \,.
\end{equation}
Using the $32^3\times 64$ data only (green points) gives
\begin{eqnarray}
   {1 \over \kappa_c} = 8.24727(17) \,, \qquad \mbox{or} \quad
            \kappa_c  = 0.121252(3) \,.
\end{eqnarray}
Hence this gives here 
\begin{eqnarray}
   \alpha_Z \sim 0.76 \,.
\label{alphaZ_num}
\end{eqnarray}
Note that the determination is quite sensitive to small changes in
$\kappa_{0;c}$ and $\kappa_c$. We conclude that for clover
fermions at our lattice spacing $\alpha_Z$ is indeed non-zero.


\subsection{Singlet quantities and the scale}

\begin{figure}[htb]
   \vspace*{0.15in}
   \begin{center}
      \includegraphics[width=10.0cm,angle=270]
          {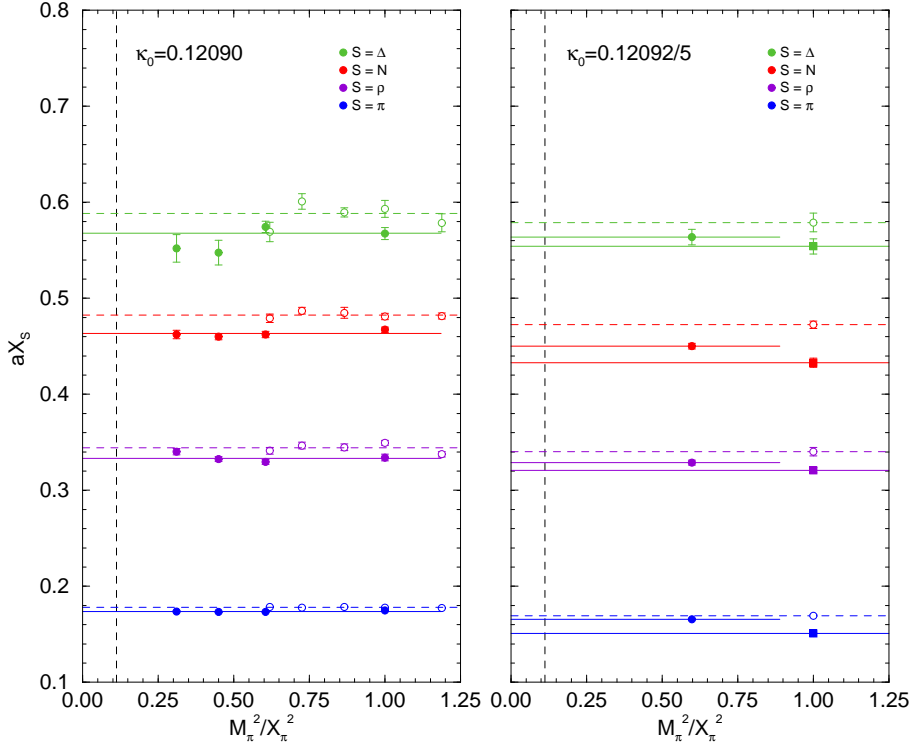}
   \end{center} 
   \caption{$aX_S$ for $S = \Delta$, $N$, $\rho$ and $\pi$ versus
            $M_\pi^2/X_\pi^2$ for $\kappa_0 = 0.12090$ (left panel,
            circles) and $0.12092$, $0.12095$ (right panel, circles
            and squares respectively) together with constant fits.
            Filled points and lines are for $32^3\times 64$ lattices,
            while opaque points and dashed lines are for
            $24^3\times 48$ lattices. (In the right panel the lower
            filled points and lines are for $\kappa_0 = 0.12095$.)
            The physical point corresponds to the dashed line
            at $M_\pi^2 / X_\pi^2 = M_\pi^2 / X_\pi^2|^*$, while
            the symmetric point corresponds to
            $M_\pi^2 / X_\pi^2 = 1$ in the figure.}
\label{b5p50_mpsO2o2mpsK2+mps2o3_aX}
\end{figure}

We take Fig.~\ref{b5p50_mps2oXn2_2mpsK2-mps2oXn2} as a
sign that singlet quantities are very flat and the fluctuations
are due to low statistics. We now investigate this further.
In Fig.~\ref{b5p50_mpsO2o2mpsK2+mps2o3_aX}
we show $aX_S$ for $S = \Delta$, $N$, $\rho$ and $\pi$
against $M_\pi^2/X_\pi^2$ for $\kappa_0 = 0.12090$ (left panel)
and comparison results for $0.12092$, $0.12095$ (right panel)
together with constant fits. This indicates that other singlet
quantities are also rather flat (we interpret variations
in $X_\Delta$ to be due to statistical fluctuations). Again
fits are made for each lattice volume separately.

We showed in section~\ref{gen_strat} that singlet quantities
must have zero derivative at the symmetric point. A second
derivative would be allowed, but we see from the left-hand
panel of Fig.~\ref{b5p50_mpsO2o2mpsK2+mps2o3_aX} that it must
be very small.


\subsubsection{Finite size effects}
\label{finite_size_effects}


In Fig.~\ref{b5p50_mpsO2o2mpsK2+mps2o3_aX} there are again indications
of relatively small finite size effects. We now briefly investigate
this a little more. While we do not attempt to formally
derive a formula here, we do have the obvious constraint that
the finite size $X_S$ must also be flat at the symmetry point
(symmetry arguments apply in any volume). The various possibilities
are given in eq.~(\ref{stationary_fun}). The first $f$-term counts
the contributions of the kaons and charged pions; the second $g$-term
is irrelevant because the strange pion is fictitious. The third
$h$-term accounts for the $\eta$ and $\pi^0$. So it is likely that
the first term is dominant because there are more particles
exchanged (the functional forms are all likely to be similar).
So if we only want a rough estimate (for the $x$-axis of the
plot) then we shall just choose the first term.

\begin{figure}[htb]
   \vspace*{0.15in}
   \begin{center}
      \includegraphics[width=11.0cm]
      {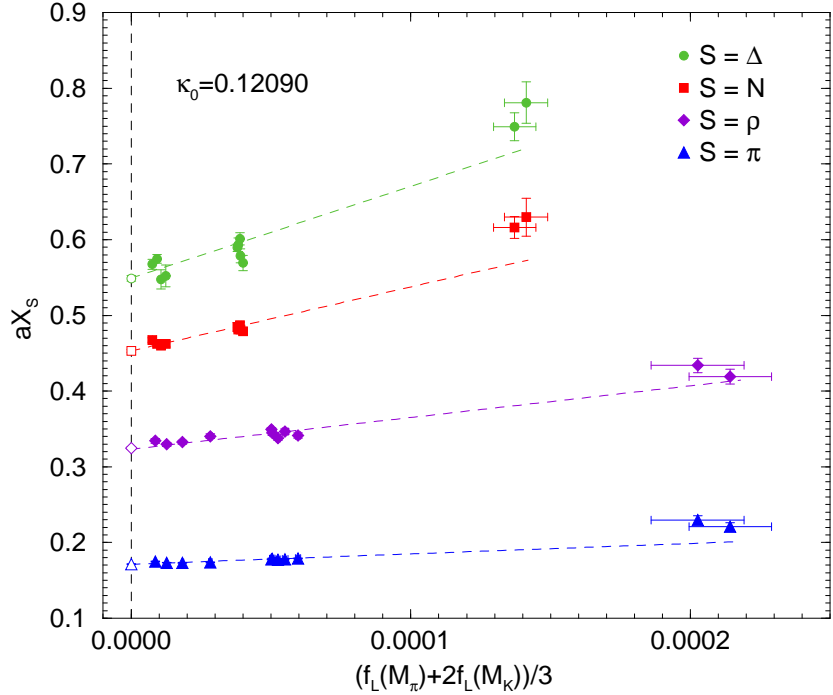}
   \end{center} 
   \caption{$aX_S$ versus $(f_L(M_\pi) + 2f_L(M_K))/3$ for
            $\kappa_0 = 0.12090$, with
            $S = \Delta$ (circles), $N$ (squares), $\rho$ (diamonds)
            and $\pi$ (upper triangles). The left-most clusters of
            points are from the $32^3 \times 64$ lattices
            ($L = 32a$), then follow $24^3 \times 48$ and finally
            $16^3 \times 32$ lattices. The dashed lines are linear fits.}
\label{b5p50_ampi2ompiL-3o2exp-mpiL+2amK2omKL-3o2exp-mKLo3_aX}
\end{figure}

Thus from eq.~(\ref{stationary_fun}) and as we shall consider only
the lowest order term from eq.~(\ref{LO_qm_M_expansion}),
we expect the finite size functional form to be
\begin{eqnarray}
   X_S(L) = X_S \left( 1 + c_S \third [ f_L(M_\pi) + 2 f_L(M_K) ] \right) \,.
\end{eqnarray}
Lowest order $\chi$PT, \cite{colangelo05a,alikhan03a}
indicates that a suitable form for $f_L(M)$ is
\begin{eqnarray}
   f_L(M) &=& (aM)^2 {e^{-ML} \over (ML)^{\threehalf}} \,, \qquad \mbox{meson}\,,
                                                             \nonumber \\
   f_L(M) &=& (aM)^2 {e^{-ML} \over (X_NL)} \,, \qquad \mbox{baryon} \,.
\end{eqnarray}
In Fig.~\ref{b5p50_ampi2ompiL-3o2exp-mpiL+2amK2omKL-3o2exp-mKLo3_aX}
we plot $(f_L(M_\pi) + 2 f_L(M_K))/3$ against $aX_S$ for 
$S = \Delta$, $N$, $\rho$ and $\pi$ on $32^3 \times 64$, $24^3 \times 48$
and $16^3 \times 32$ lattices for $\kappa_0 = 0.12090$.
The fits are linear. A reasonable agreement is seen. (The noisiest
signal is for $S = \Delta$.) We see that the extrapolated
(i.e.\ $ L \to \infty$) results are very close to the largest
lattice results (i.e.\ $32^3\times 64$), so we conclude that using the
largest lattice size available should only introduce small errors.
We shall also go a little further and assume that finite size effects
for masses are similar to those of $X_S$ for each mass of the
appropriate multiplet. Thus we shall later consider ratios $M/X_S$
for all the available lattice data; finite size results then tend
to cancel in the ratio.


\subsubsection{Scale estimation}
\label{scale_est}


One advantage of our method is that $X_S$ remains constant and can be
used to determine the scale. We do not have to first extrapolate to
the physical limit in distinction to other methods.

\begin{figure}[htb]
   \vspace*{0.15in}
   \begin{center}
      \includegraphics[width=11.0cm]
             {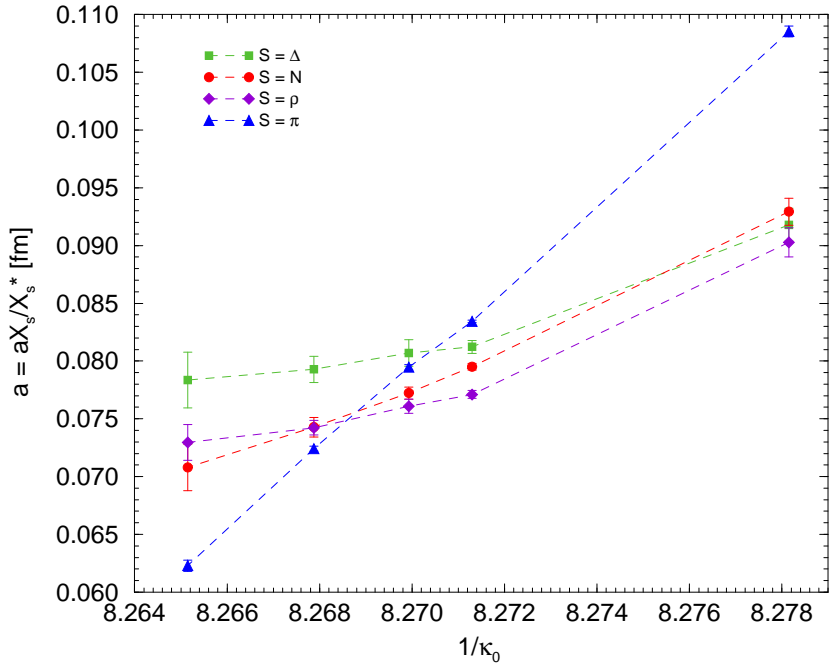}
   \end{center} 
   \caption{$aX_S / X_S^*$ against $1/\kappa_0$ for
            $S = \Delta$ (green squares), $N$ (red circles),
            $\rho$ (violet diamonds), $\pi$ (blue upper triangles)
            with $\kappa_0 = 0.12080$, $0.12090$, $0.12092$, $0.12095$
            and $0.12099$.}
\label{b5p50_ookl_aXoXexpt}
\end{figure}

The result of section~\ref{finite_size_effects} is that the largest volumes
available seem to have small finite size effects, so we now
simply take the largest volume available. In Fig.~\ref{b5p50_ookl_aXoXexpt}
we plot $aX_S/X_S^*$, for $S = N$, $\Delta$, $\rho$, $\pi$
using the largest volume fitted results from
Fig.~\ref{b5p50_mpsO2o2mpsK2+mps2o3_aX} (together with smaller data
sets for $\kappa_0 = 0.12080$ and $\kappa_0 = 0.12099$).
The experimental values of $X_S^*$ are given in
section~\ref{expt_hadron_masses}. This ratio gives estimates
for the lattice spacing $a$ for the various scales%
\footnote{For example for $\kappa_0 = 0.12090$ we find that
$a = 0.0834(1)\,\mbox{fm}$, $0.0812(6)\,\mbox{fm}$,
$0.0795(3)\,\mbox{fm}$ and $0.0771(3)\,\mbox{fm}$ using
$X_\pi$, $X_\Delta$, $X_N$, $X_\rho$ to set the scale respectively.}.
We would expect most variation of the ratio with $X_\pi$
and convergence to a common scale where the lines cross, assuming 
\begin{itemize}
   \item the simulation statistics are sufficient 
   \item all $O(a^2)$ corrections are negligible
   \item there is little (or no) curvature present in $X_S$.
\end{itemize}
This appears to be the case, with the possible exception of the
decuplet scale. However this is the channel with the worst signal,
and may be showing some curvature (we cannot at present say whether
there might be large $O(a^2)$ effects), so presently we just consider
the approximate crossing of the other lines giving
$a \sim 0.075$ -- $0.078\,\mbox{fm}$.

\begin{figure}[htb]
   \vspace*{0.15in}
   \begin{center}
      \includegraphics[width=11.0cm]
             {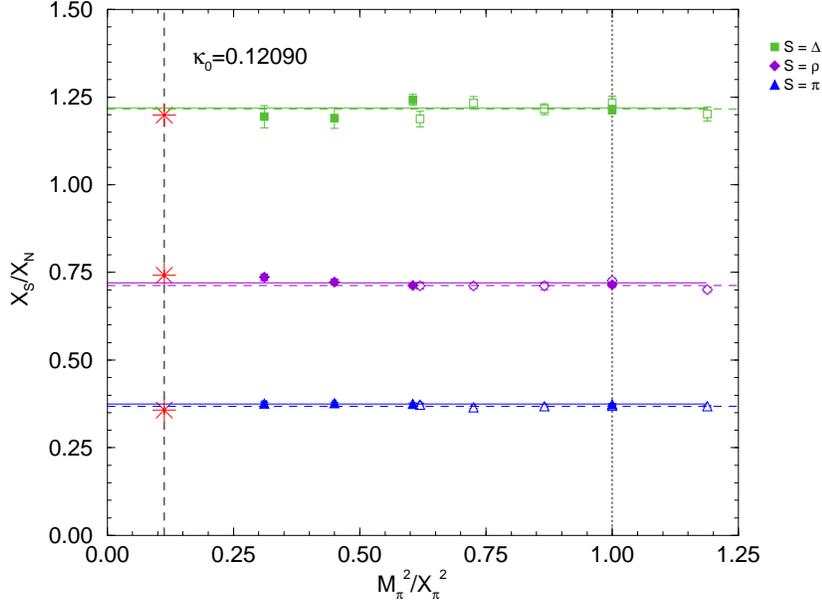}
   \end{center} 
   \caption{$aX_S/aX_N$ versus $M_\pi^2/X_\pi^2$ for
            $S = \Delta$ (squares), $\rho$ (diamonds) and
            $\pi$ (upper triangles), top to bottom,
            for $\kappa_0=0.12090$. The dashed vertical line
            represents the physical value, while the dotted
            line gives the $SU(3)$ flavour symmetric point.
            Filled points are on $32^3\times 64$ lattices
            while open points are on $24^3\times 48$ sized lattices.
            Horizontal lines and dashed horizontal lines represent
            constant fits to either the $32^3\times 64$ or
            $24^3\times 48$ results respectively. For illustration,
            we also show the physical values -- denoted by stars.}
\label{b5p50_mps2oX2_XoXNO_ksymp12090}
\end{figure}

As discussed in section~\ref{finite_size_effects} we expect a
(partial) cancellation of finite size effects (and also statistical
fluctuations) within the same multiplet so we shall adopt the
philosophy when considering the hadron spectrum of first finding
the ratio of the mass to the singlet quantity from the same multiplet.
For example, we can take as our base singlet quantity as the baryon
octet $X_N$ (not only are these stable particles under QCD interactions
and so might physically be considered a good choice, but
$X_N$ also has smaller numerical errors than $X_\Delta$ on the lattice).
To translate from one scale to another we then need the ratio $aX_S/aX_N$.
In Fig.~\ref{b5p50_mps2oX2_XoXNO_ksymp12090} we plot $X_S/X_N$ 
for various $X_S$ (with $S = \Delta$, $\rho$, $\pi$).
Also shown are constant fits to the two volumes --- $24^3\times 48$ and
$32^3\times 64$. The change in the ratios between the two volumes
is seen to be small. Note also that all ratios are close to their
physical values. We use the results of the largest volume,
which are given in Table~\ref{scale_conversion}.
\begin{table}[htb]
   \begin{center}
      \begin{tabular}{ccc}
         Ratio            & $\kappa_0=0.12090$ &  $\times (X_N/X_S)|^*$  \\
         \hline
         $aX_\pi    / aX_N$   & 0.3751(13)    &  1.049(4)  \\  
         $aX_\rho   / aX_N$   & 0.7200(38)    &  0.971(5)  \\  
         $aX_\Delta / aX_N$   & 1.219(9)      &  1.017(8)  \\  
         \hline
      \end{tabular}
   \end{center}
\caption{Lattice ratios of singlet quantities $aX_S/aX_N$, $S = \pi$,
         $\rho$, $\Delta$ from $32^3\times 64$ lattices. In the last
         column we have multiplied by the experimental inverse
         ratio, taken from Table~\protect\ref{hadron_masses_singlet}.
         If we had perfect scaling then this ratio should be $1.0$.}
\label{scale_conversion}
\end{table}
In the last column of this table we have used the experimental
values of $X_S$ (as given in Table~\ref{hadron_masses_singlet})
to form the ratio $aX_S/aX_N \times (X_N/X_S)|^*$. This should be one.
As can be seen from Fig.~\ref{b5p50_mps2oX2_XoXNO_ksymp12090}, this
is the case and Table~\ref{scale_conversion} confirms that
the ratios are $1$ within a few percent. All this shows that
$\kappa_0 = 0.12090$ has $\overline{m}$ very close to the
correct physical value.


\section{Spectrum results for $\mathbf{2+1}$ flavours}
\label{hadron_masses}


We shall now discuss our lattice results.


\subsection{Experimental values}
\label{expt_hadron_masses}


As we will compare our lattice results with the experimental
results, we first give the experimental masses from
the Particle Data Group tables \cite{nakamura10a}.

To minimise $u$--$d$ quark mass differences (and also electromagnetic
effects) for the experimental data, we average the particle masses
over isospin $I_3$ (i.e.\ horizontally in Figs.~\ref{meson_mults},
\ref{baryon_mults}). This gives the experimental values in
Table~\ref{hadron_masses_av} (we postpone giving them here
in order to display them with the lattice values in
Table~\ref{hadron_masses_av}). Using these experimental numbers,
the experimental values for the hadron singlet quantities
used here are then given in Table~\ref{hadron_masses_singlet}.
\begin{table}[h]
   \begin{center}
      \begin{tabular}{ll}
         Singlet                               & GeV         \\
         \hline
         $X_\pi^* = \sqrt{(M_\pi^2 + 2M_K^2)/3} |^*$  & 0.4109      \\  
         $X_\rho^* = (M_{\rho} + 2M_{K^*})/3 |^*$       & 0.8530     \\
         $X_N^* = (M_N + M_\Sigma + M_\Xi)/3 |^*$     & 1.1501      \\
         $X_\Delta^* = (2M_\Delta + M_\Omega)/3|^*$   & 1.3788      \\
         \hline
      \end{tabular}
   \end{center}
\caption{Experimental values for the $X_S$ singlet quantities, $X_S^*$,
         $S = \pi$, $\rho$, $N$ and $\Delta$.}
\label{hadron_masses_singlet}
\end{table}


\subsection{Mass hierarchy}

\begin{figure}[htb]
   \begin{center}
      \includegraphics[width=11.0cm]
             {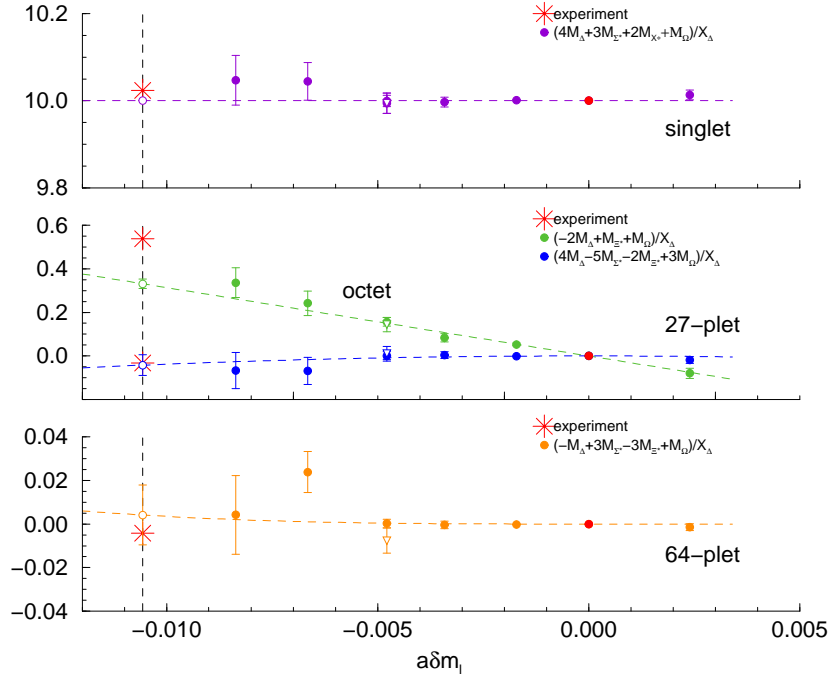}
   \end{center} 
   \caption{$(4M_\Delta + 3M_{\Sigma^*}+2M_{\Xi^*}+M_\Omega)/X_\Delta$,
            $(-2M_\Delta + M_{\Xi^*}+M_\Omega)/X_\Delta$,
            $(4M_\Delta - 5M_{\Sigma^*}-2M_{\Xi^*}+3M_\Omega)/X_\Delta$
            and
            $(-M_\Delta + 3M_{\Sigma^*} - 3M_{\Xi^*} + M_\Omega)/X_\Delta$
            (filled circles) against $\delta m_l$ together with a
            constant, linear, quadratic and cubic term in $\delta m_l$
            respectively, as given in eq.~(\protect\ref{num64_reduced}).
            Extrapolated values are shown as opaque circles.
            Experimental values are denoted by stars.
            The opaque triangle corresponds to a run at the same
            $(\kappa_l,\kappa_s)$, but on a $24^3\times 48$ lattice
            rather than a $32^3\times 64$ lattice. The vertical line
            is at the value of $\delta m_l$ -- $\delta m_l^*$ obtained
            from a quadratic fit to the pseudoscalar octet, 
            eqs.~(\protect\ref{fit_mpsO}), (\protect\ref{fit_strange})
            as described in section~\protect\ref{theory_2p1}
            and Fig.~\protect\ref{b5p50_mpsO2oXpi2-jnt_mpsO2oXpi2-jnt}.}
\label{mps2o2mpsK2+mps2o3_su3NDsymo2mDelta+mOmDo3}
\end{figure}

We now consider the lattice results for the mass spectrum.
First we check whether there is a strong hierarchy due to the
$SU(3)$ flavour symmetry as found in eq.~(\ref{num64}), namely
\begin{eqnarray}
   4 M_\Delta + 3 M_{\Sigma^*} + 2 M_{\Xi^*} + M_{\Omega}
      &\propto& (\delta m_l)^0 \quad \quad \ {\rm singlet}
                                                              \nonumber  \\
   - 2 M_\Delta  \phantom{+ 3 M_{\Sigma^*}} + M_{\Xi^*} + M_{\Omega}
      &\propto& \delta m_l  \qquad \quad \ {\rm octet}
                                                              \nonumber  \\
   4 M_\Delta - 5 M_{\Sigma^*} - 2 M_{\Xi^*} + 3 M_{\Omega}
      &\propto& \delta m_l^2 \qquad \quad {\rm 27-plet}
                                                              \nonumber  \\
   - M_\Delta + 3 M_{\Sigma^*} - 3 M_{\Xi^*} +  M_{\Omega}
      &\propto& \delta m_l^3 \qquad \quad {\rm 64-plet} \,.
\label{num64_reduced}
\end{eqnarray}
In Fig.~\ref{mps2o2mpsK2+mps2o3_su3NDsymo2mDelta+mOmDo3}
we plot these mass combinations (over $X_\Delta$) against $a\delta m_l$
for $\kappa_0 = 0.12090$. Also shown are the experimental values
using the values from Table~\ref{hadron_masses_av}.
Note the change of scale between the axes. There is reasonable
agreement with these numbers. Well reproduced, as expected,
is the order of magnitude drop in the hadron mass
contributions with each additional power of $\delta m_l$.
(See \cite{beane06a} for a similar investigation of octet baryons.)
It is also seen that while $(-2M_\Delta + M_{\Xi^*}+M_\Omega)/X_\Delta$
has a linear gradient in $\delta m_l$, in the other fits
any gradient is negligible as expected. To check for possible
finite size effects we also plot a run at the same $(\kappa_l,\kappa_s)$
but using a $24^3\times 48$ lattice rather than $32^3\times 64$.
There is little difference and so it appears that considering
ratios of quantities within the same multiplet leads to (effective)
cancellation of finite size effects.


\subsection{`Fan' plots}
\label{fan_hadron_masses}


We now show a series of plots of the hadron masses from a small
quark mass just above the flavour symmetric line down to the
physical point. As the masses (of a particular octet or
decuplet) are all degenerate at a point on the flavour
symmetric line, then we would expect a `fanning' out of masses
from this point. We consider second order fits in the quark mass,
but show plots using the pseudoscalar mass on the $x$-axis,
i.e.\ from eq.~(\ref{fit_mpsO}). Thus we are using the quark mass
as an `internal parameter'. As discussed previously at the
end of section~\ref{theory_1p1p1} and in more detail in
Appendix~\ref{coordinate_choice} this is the natural choice.

\begin{figure}[htb]
   \begin{center}
      \includegraphics[width=10.0cm]
         {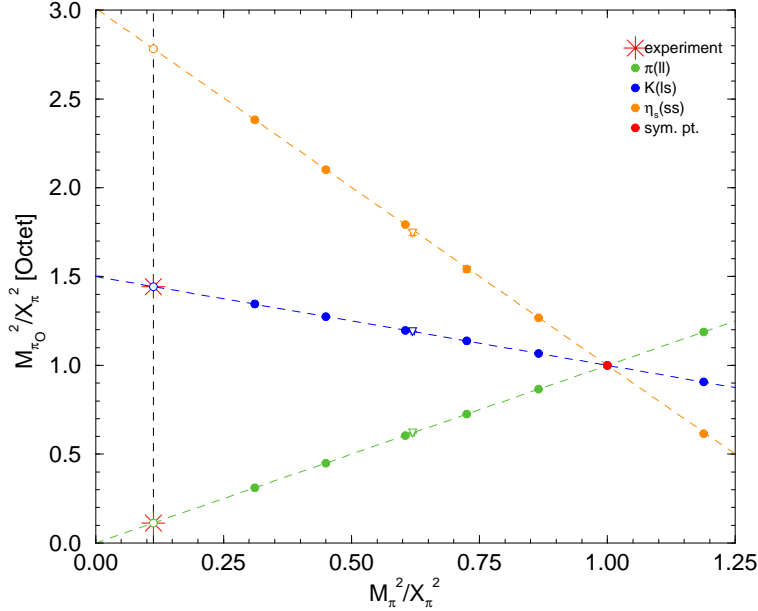}
   \end{center} 
   \caption{$M^2_{\pi_O}/ X^2_\pi$ ($\pi_O = \pi$, $K$, $\eta_s$)
            against $M_\pi^2/X_\pi^2$ together with the combined fit
            of eq.~(\protect\ref{fit_mpsO}) for both the
            $32^3\times 64$ and $24^3\times 48$ lattices.
            The flavour symmetric point (`sym. pt.') when
            $\kappa_0 = 0.12090$ is denoted as a red point. 
            Experimental values are denoted by stars.
            The opaque triangle corresponds to a run at the same
            mass but on a $24^3\times 48$ lattice rather
            than $32^3\times 64$.}
\label{b5p50_mpsO2oXpi2-jnt_mpsO2oXpi2-jnt}
\end{figure}

In Fig.~\ref{b5p50_mpsO2oXpi2-jnt_mpsO2oXpi2-jnt}
we begin with the pseudoscalar octet and show
$M^2_{\pi_O}/ X^2_\pi$ ($\pi_O = \pi$, $K$, $\eta_s$)
against $M_\pi^2/X_\pi^2$ together with the combined fit
of eqs.~(\ref{fit_mpsO}), (\ref{fit_strange}).
A typical `fan' structure is seen with masses radiating from the
common point on the symmetric line. Note that the right-most
point has a small strange quark mass and a large `light'
quark mass, so that the order of the meson masses is inverted.

There is however little real content in this plot -- the
$\pi_O = \pi$ line is obviously trivial, for the $\pi_O = K$
line the chiral limit and gradient are known as we have
\begin{equation}
   {M_K^2 \over X_\pi^2} = {3\over 2} - {1 \over 2} {M_\pi^2 \over X_\pi^2} \,.
\end{equation}
(This can also be seen to $O(\delta m_l^3)$ by using eq.~(\ref{fit_mpsO})
to form $M_\pi^2/X_\pi^2$ and $M_K^2/X_\pi^2$ to $O(\delta m_l^2)$.)
An inspection of Fig.~\ref{b5p50_mpsO2oXpi2-jnt_mpsO2oXpi2-jnt}
shows that the numerical results indeed follow very well this line,
with a gradient of $-1/2$ and having in the chiral limit a value of $3/2$. 

However the graph does tell us that for the fictitious $\eta_s$
particle, there is very little curvature which, as this is a
constrained fit, must hold for all the pseudoscalar octet particles,
including the fictitious one. We also note that ratios within
the same multiplet do indeed tend to give cancellation of
finite size effects.

\begin{figure}[htb]
   \begin{center}
      \includegraphics[width=10.0cm]
             {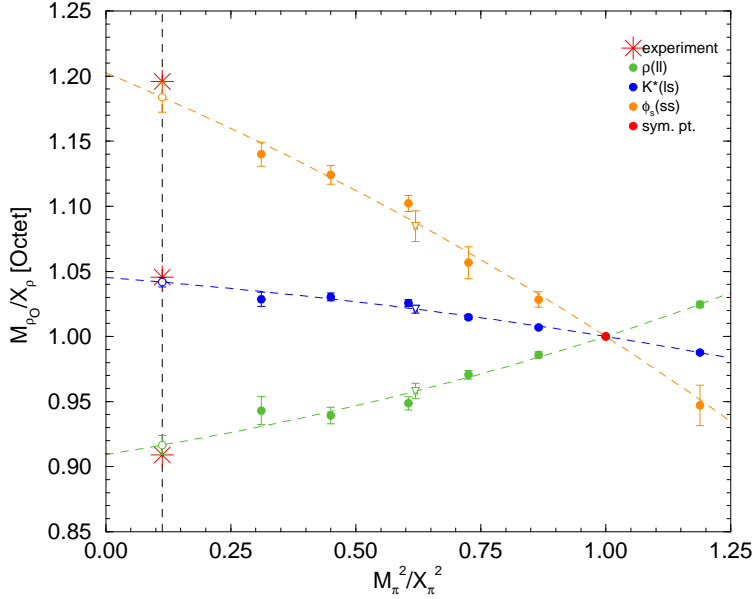}
   \end{center} 
   \caption{$M_{\rho_O}/ X_{\rho}$
            ($\rho_O = \rho$, $K^*$, $\phi_s$)
            against $M_\pi^2/X_\pi^2$ together with the combined fit of
            eqs.~(\protect\ref{fit_mvO}), (\protect\ref{fit_mpsO})
            (the dashed lines). Same notation as in
            Fig.~\protect\ref{b5p50_mpsO2oXpi2-jnt_mpsO2oXpi2-jnt}.}
\label{b5p50_psOX2-jnt_mvOoX-jnt}
\end{figure}

In Fig.~\ref{b5p50_psOX2-jnt_mvOoX-jnt}
we plot the vector octet multiplet $M_{\rho_O}/ X_\rho$ against
$M_\pi^2/X_\pi^2$ for $\rho_O = \rho$, $K^*$, $\phi_s$.
Again finite volume effects tend to cancel in the ratio
(normalising with the singlet quantity from the same octet)
and so both volumes have again been used in the fit. The combined
fit uses eqs.~(\ref{fit_mvO}), (\ref{fit_mpsO}) again with the
bare quark mass being an `internal' parameter. Some moderate curvature
is now seen in the extrapolations. Note that as
$M_{\phi_s} \approx M_{\phi}$, the physical $\phi$ must indeed
be almost a perfect $s\overline{s}$ state, i.e.\ we almost have
`ideal' mixing.

\begin{figure}[htbp]
   \begin{center}
      \includegraphics[width=10.0cm]
    {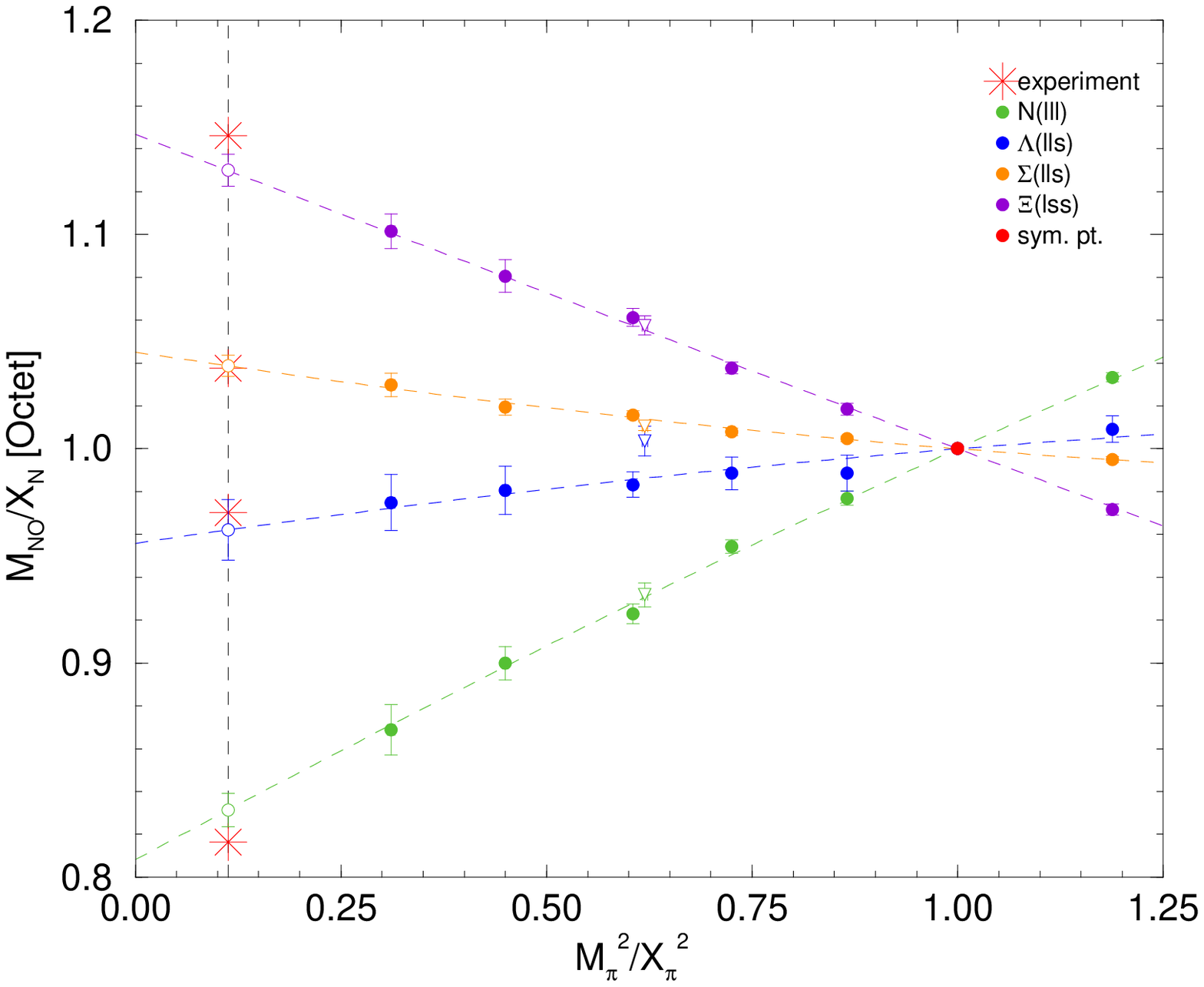}
   \end{center} 
   \caption{$M_{N_O}/ X_N$ ($N_O = N$, $\Lambda$, $\Sigma$, $\Xi$)
            against $M_\pi^2/X_\pi^2$ together with the combined fit
            of eqs.~(\protect\ref{fit_mNO}), (\protect\ref{fit_mpsO})
            (the dashed lines).
            Same notation as in Fig.~\protect
            \ref{b5p50_mpsO2oXpi2-jnt_mpsO2oXpi2-jnt}.}
\label{b5p50_mpsO2o2mpsK2+mps2o3-jnt_mNOomNOpmSigOpmXiOo3-boot-jnt}
\end{figure}

\begin{figure}[htbp]
   \begin{center}
      \includegraphics[width=10.0cm]
     {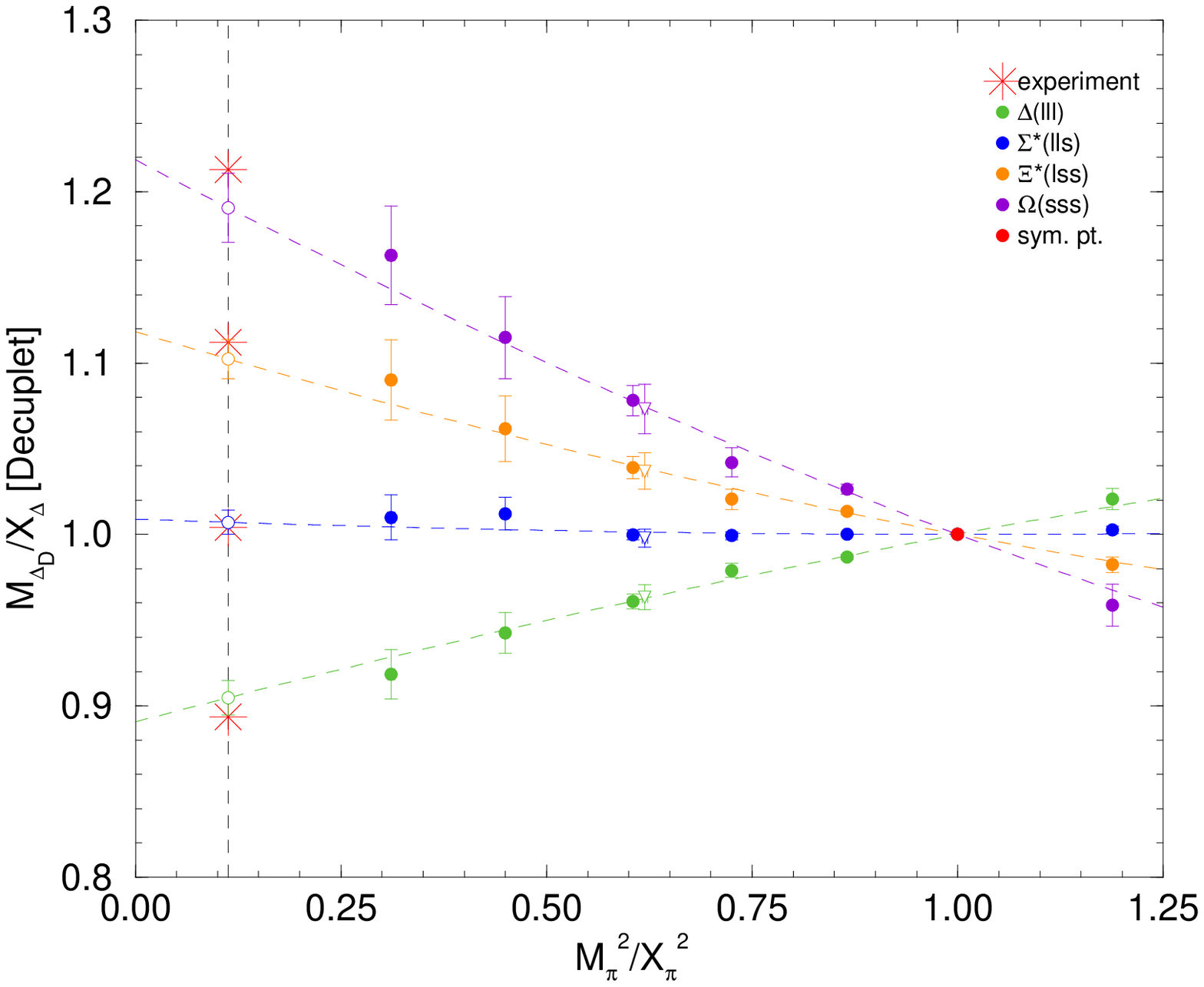}
   \end{center}
   \caption{$M_{\Delta_O}/ X_{\Delta}$
            ($\Delta_O = \Delta$, $\Sigma^*$, $\Xi^*$, $\Omega$)
            against $M_\pi^2/X_\pi^2$ together with the combined fit of
            eqs.~(\protect\ref{fit_mDD}), (\protect\ref{fit_mpsO}).
            Same notation as in Fig.~\protect
            \ref{b5p50_mpsO2oXpi2-jnt_mpsO2oXpi2-jnt}.}
\label{b5p50_mpsO2o2mpsK2+mps2o3-jnt_mDDo2mDeltaD+mOmDo3-boot-jnt}
\end{figure}

Continuing in
Fig.~\ref{b5p50_mpsO2o2mpsK2+mps2o3-jnt_mNOomNOpmSigOpmXiOo3-boot-jnt}
we plot the baryon octet $M_{N_O}/ X_N$ for $N_O = N$, $\Lambda$,
$\Sigma$, $\Xi$ against $M_\pi^2/X_\pi^2$ and similarly in
Fig.~\ref{b5p50_mpsO2o2mpsK2+mps2o3-jnt_mDDo2mDeltaD+mOmDo3-boot-jnt}
we plot the corresponding baryon decuplet $M_{\Delta_O}/ X_\Delta$
for $\Delta_O = \Delta$, $\Sigma^*$, $\Xi^*$, $\Omega$ against
$M_\pi^2/X_\pi^2$. Although we have included quadratic terms in the
fit, there is really very little curvature in the results.
In both these pictures the correct ordering of masses is
achieved (also the reverse order behind the symmetric point
when we have heavy $l$ quark masses and light $s$ quark masses).
In particular in
Fig.~\ref{b5p50_mpsO2o2mpsK2+mps2o3-jnt_mNOomNOpmSigOpmXiOo3-boot-jnt}
we see that the $\Lambda$--$\Sigma$ splitting is correct.
This is a dynamical effect because $\Lambda$ and $\Sigma$ both
have the same quark content. Also in
Fig.~\ref{b5p50_mpsO2o2mpsK2+mps2o3-jnt_mDDo2mDeltaD+mOmDo3-boot-jnt}
$M_{\Sigma^*}$ is indeed constant as expected. 

These results show that the Gell-Mann--Okubo relations
work all the way from the symmetry point to the physical
point.

The masses (using the scale determined by the appropriate $X_S$)
are given in Table~\ref{hadron_masses_av}.
\begin{table}[h]
   \begin{center}
      \begin{small}
      \begin{tabular}{lllll}
         particle      &      & expt[GeV] & $aM/aX_S$ & result[GeV] \\
         \hline
         $M_\pi = (M_{\pi^+} + M_{\pi^0} +  M_{\pi^-})/3$
                       & $ll$ & 0.1380 & $0.3359^*$& 0.145(1)     \\
         $M_K = M_{K^+} = M_{K^-}$
                       & $ls$ & 0.4937 & $1.2015^*$& 0.518(1)     \\
         $M_{\eta_s}$  & $ss$ & $\sim 0.685$
                                       & 1.668(3)  & 0.720(3)     \\
         \hline
         $M_\rho = M_{\rho^+} = M_{\rho^-}$
                       & $ll$ & 0.7755 & 0.9166(73)& 0.759(7)     \\
         $M_{K^*} = M_{K^{*+}} = M_{K^{*-}}$
                       & $ls$
                       & 0.8917        & 1.042(4)  &  0.863(6)     \\
         $M_{\phi_s} \sim M_\phi$
                       & $ss$ 
                       & 1.0195        & 1.184(12) & 0.980(11)     \\
         \hline
         $M_N = (M_p + M_n)/2$
                       & $lll$  & 0.9389 & 0.8313(77)  & 0.956(9)  \\
         $M_\Lambda$
                       & $lls$  & 1.1157 & 0.9621(142) & 1.107(16) \\
         $M_{\Sigma} = (M_{\Sigma^+}+M_{\Sigma^0}+M_{\Sigma^-})/3$
                       & $lls$  & 1.1932 & 1.039(5)    & 1.195(6)  \\
         $M_{\Xi} = (M_{\Xi^0}+M_{\Xi^-})/2$
                       & $lss$  & 1.3183 & 1.130(7)    & 1.300(9)  \\
         \hline
         $M_\Delta$    & $lll$  & 1.232  & 0.9047(100)  & 1.269(17) \\
         $M_{\Sigma^*} = (M_{\Sigma^{*+}}+M_{\Sigma^{*0}}+M_{\Sigma^{*-}})/3$
                       & $lls$  & 1.3846 & 1.007(7)    & 1.413(14)  \\
         $M_{\Xi^*} = (M_{\Xi^{*0}}+M_{\Xi^{*-}})/2$
                       & $lss$  & 1.5334 & 1.102(11)   & 1.546(20)  \\
         $M_\Omega = M_{\Omega^-}$
                       &  $sss$ & 1.6725 & 1.191(20)   & 1.670(31)  \\
         \hline
      \end{tabular}
      \end{small}
   \end{center}
\caption{The hadron masses. The third column, `expt', gives the
         isospin averaged masses. (The $\eta_s$ mass is taken from
         \protect\cite{burakovsky97a}.) The fourth column,  $aM/aX_S$,
         gives the numerical results from
         Figs.~\protect\ref{b5p50_mpsO2oXpi2-jnt_mpsO2oXpi2-jnt}
         -- \protect
             \ref{b5p50_mpsO2o2mpsK2+mps2o3-jnt_mDDo2mDeltaD+mOmDo3-boot-jnt}.
         (The $aM/aX_\pi$ values for $M_\pi$, $M_K$ are exact.)
         The last column, `result', has used
         eq.~(\protect\ref{convert_base_XN}) to convert the scale
         to the base scale $X_N$.}
\label{hadron_masses_av}
\end{table}
The results are rather close to their experimental values.
($M_{\eta_s}$ is taken from \cite{burakovsky97a} which uses a quadratic
mass formula and ideal mixing which is in agreement with the prediction
of LO $\chi$PT).) However at present we are effectively
using a different scale, $X_S$, for each multiplet. If we wish to convert
these numbers to a base scale, say $X_N$, then they can be converted using
\begin{eqnarray}
   M_{S_O} = {a M_{S_O} \over aX_N} \times X_N|^*
          = \left( {a M_{S_O} \over aX_S} \times X_S|^* \right) \times
            \left( {a X_S \over aX_N} 
                      \times \left. { X_N \over X_S} \right|^* \right)\,,
\label{convert_base_XN}
\end{eqnarray}
where the second factor is given in the last column of
Table~\ref{scale_conversion}%
\footnote{If using $X_N$ there is an additional factor from
the setting of $ M_\pi^2 / X_\pi^2|^*$ on the $x$-axis,
\begin{eqnarray}
   \left. {M_\pi^2 \over X_\pi^2} \right|^* \sim 0.1128
      \to  \left. {M_\pi^2 \over X_N} \right|^* 
              \times \left( aX_N \over aX_\pi \right)^2 
           \sim 0.1023 \,.
                                                           \nonumber
\end{eqnarray}
The change in the hadron mass due to this is very small
(very much smaller than the error bar), so we will ignore this here.}.
These numbers are all $\sim 1$ (within a few percent).
However it is to be noted that this causes the largest discrepancy
to the experimental value. So the largest source of error appears
to come from the uncertainty in the consistency of different
flavour singlet quantities used to determine the common scale.


\subsection{Partially quenched results}
\label{pq_results}


We illustrate partial quenching using baryon splittings as an
example. The splittings depend mainly on $\mu_s - \mu_l$
and only weakly (at second order) on other quark combinations.
In the PQ data shown here, we have points with a large splitting
between $\mu_s$ and $\mu_l$ reaching up to points where $\mu_s - \mu_l$
is equal to its physical value. We can therefore make partially
quenched splitting plots reaching down to the physical point.
   
\begin{figure}[htbp]
   \begin{center}
      \includegraphics[width=8.0cm,angle=270]{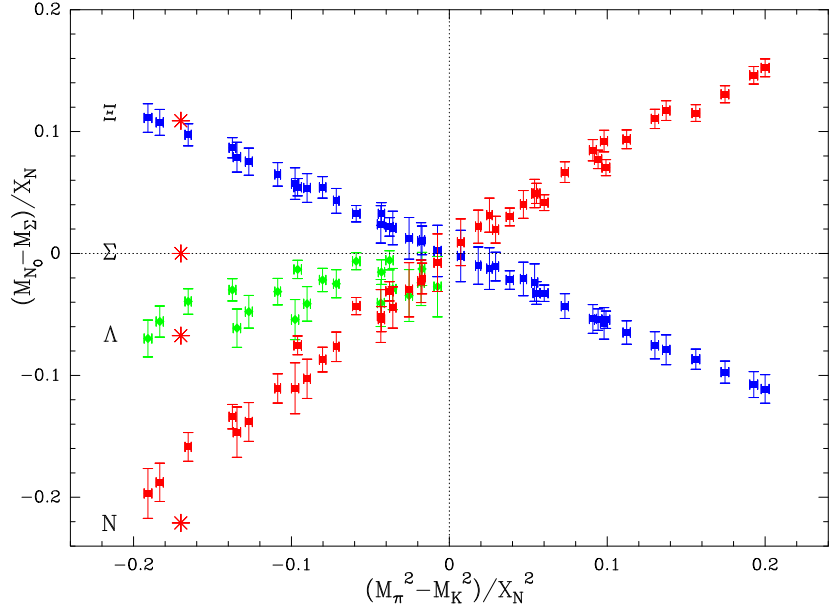}
   \end{center} 
   \caption{Partially quenched data, $(M_{N_O}-M_\Sigma)/X_N$
            versus $(M_\pi^2-M_K^2)/X_N^2$. The experimental points
            are denoted by red stars.}
\label{octsplit}
\end{figure}

\begin{figure}[htbp]
   \begin{center}
      \includegraphics[width=8.0cm,angle=270]{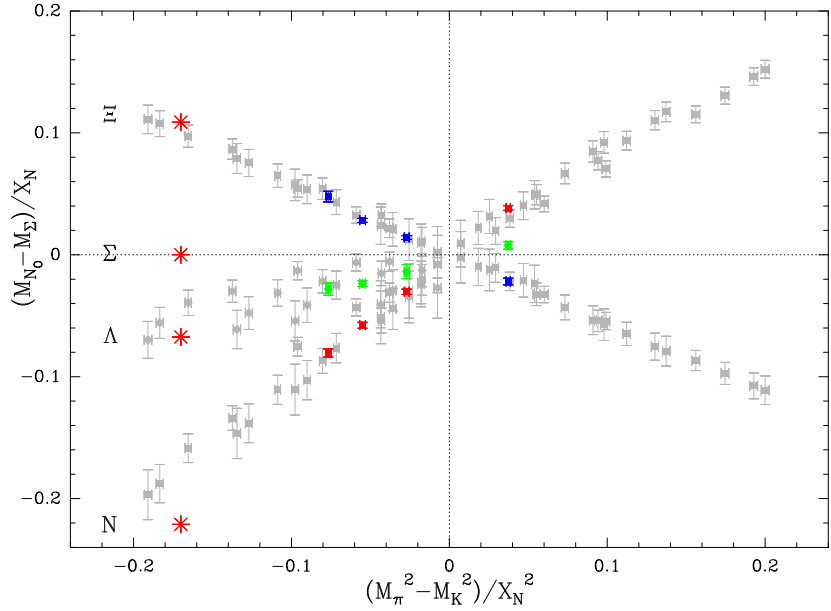}
   \end{center}
   \caption{A comparison between partially quenched and full data
            from $24^3\times 48$ lattices,
            $(M_{N_O}-M_\Sigma)/X_N$ versus $(M_\pi^2-M_K^2)/X_N^2$.
            Same notation as in Fig.~\protect\ref{octsplit}.}
\label{octsplitfull}
\end{figure}

We have generated partially quenched results on an ensemble with
$\kappa_0 = 0.12090$ and lattice volume $24^3\times 48$.
The first baryon octet splitting `flag' diagram, Fig.~\ref{octsplit},
shows just the PQ data. The second, Fig.~\ref{octsplitfull},
shows the PQ data in grey, compared with the unitary $24^3\times 48$
data in colour. While only to be taken as an illustration, it shows
that the PQ data has the potential to be a good predictor of real data. 

Of course partially quenched data is not a complete substitute
for simulations at the physical point, even for splittings. 
If we take the $\Sigma$-$N$ splitting as an example, we find
from eq.~(\ref{Noct}),
\begin{eqnarray}
   M_N - M_\Sigma 
      &=& (A_1 + A_2) (\delta\mu_l - \delta\mu_s) 
                                                              \nonumber  \\
      & & \vspace*{0.25in}
          + (B_1 + B_2) (\delta\mu_l^2 - \delta\mu_s^2) 
          - B_3 (\delta\mu_l - \delta\mu_s)^2
                                                              \nonumber  \\
      &=& (A_1 + A_2) (\delta\mu_l - \delta\mu_s)
                                                              \nonumber  \\
      & & \vspace*{0.25in}
          + (B_1 + B_2) (\delta\mu_l - \delta\mu_s)(\delta\mu_l + \delta\mu_s) 
          - B_3 (\delta\mu_l - \delta\mu_s)^2 \,.
 \end{eqnarray} 
If we plot this baryon splitting against the quark mass splitting 
$(\delta\mu_l - \delta\mu_s)$ the $A_1$, $A_2$ and $B_3$ terms give
a simple parabola (as does the $B_4$ term if we consider a splitting
involving the $\Lambda$). However the $B_1$ and $B_2$ terms do not
depend purely on the splitting $(\delta\mu_l - \delta\mu_s)$,
they also depend on the quark-mass sum  $(\delta\mu_l + \delta\mu_s)$.
Thus the $B_1$ and $B_2$ terms lead to a broadening of data bands
in Fig.~\ref{octsplit} (two data points with the same value of
$(\delta\mu_l - \delta\mu_s)$ may have differing values of 
$(\delta\mu_l + \delta\mu_s)$), and can lead to the partially
quenched data missing the physical point slightly. Although
we reach splittings  $(\delta\mu_l - \delta\mu_s)$ equal to
and even a little larger than the physical quark mass 
splitting, we do this with our light valence quark still 
noticeably heavier than the real $u$ and $d$ quarks, so 
$(\delta\mu_l + \delta\mu_s) > (\delta m_l^* + \delta m_s^*)$
at our end-point. The above argument still applies (with minor
modifications) if we use the partially quenched meson mass
difference $M_\pi^2 - M_K^2$ as a substitute for
$(\delta\mu_l - \delta\mu_s)$ on the figure's $x$-axis.


\section{Conclusions}
\label{conclusions}


We have outlined a programme to systematically approach the
physical point in simulations of QCD with three flavours starting
from a point on the $SU(3)$ flavour symmetric line by keeping the singlet
quark mass constant. As we move from the symmetric point
$(m_u, m_d, m_s) = (m_0, m_0, m_0)$ towards the physical point
along our $\overline{m} = \mbox{constant}$ path, the $s$ quark becomes
heavier while the $u$ and $d$ quarks become lighter.
These two effects tend to cancel in any flavour singlet quantity.
The cancellation is perfect at the symmetric point, and we have found that
it remains good down to the lightest points we have simulated.

Since gluonic properties are also flavour singlet, this means that
all properties of our configurations, from simple ones such as the
plaquette, to more complicated ones such as the potential and $r_0$,
vary slowly along the trajectory. Compared with other paths,
the properties of our configurations are already very close to those
at the physical point. This has many advantages, from technical ones
such as the rapid equilibration when we move to a new mass point,
to physically useful results, such as the closeness between
partially quenched and full physical results. In addition it also
enables the lattice spacing to be determined without an extrapolation
to the physical point, and indeed allows the consistency of
various definitions to be discussed.

The flavour symmetry expansion is developed here, by classifying
up to $O(\delta m_q^3)$ how quark-mass polynomials behave under the
$S_3$ permutation group and the $SU(3)$ flavour group,
leading to Table~\ref{cubic}, given for $1+1+1$ quark flavours.
We also show that for non-chiral (e.g.\ clover) fermions,
where we have different renormalisation for the singlet
and non-singlet pieces and also have $O(a)$ improvement,
that all the additional terms that appear are just these mass polynomials. 
In section~\ref{class_mass_matrix} we classify the hadron mass
matrices, and show that certain combinations, for example
the Coleman-Glashow relation, have only small violations
(in terms of the quark mass).

Turning now to $2+1$ quark flavours, we have found that the
flavour symmetry expansion (again when holding the average quark mass,
$\overline{m}$, constant) leads to highly constrained extrapolations
(i.e.\ fits) for non-singlet quantities, such as hadronic masses here,
and reduce the number of free parameters drastically.
(There is a short discussion of this point at the end of 
section~\ref{taylor_expansion} and in section~\ref{mbar_varies}.)
It is also to be noted that a $2+1$ simulation is sufficient
to determine most of the expansion coefficients for the $1+1+1$
case (one exception being the particle at the centre of
the octet multiplet).

In section~\ref{applications_chipt} we discussed 
the relationship of the flavour symmetry expansion to the
chiral perturbation expansion. Lattice simulations are at
somewhat large pion masses which juxtaposes well with the
flavour symmetry expansion presented here, while chiral
perturbation theory is an expansion about a zero pion mass
which lattice simulations strive to reach.
In section~\ref{applications_chipt} we give, as an example,
the relationship between these expansions for the
pseudoscalar octet. We also briefly discuss how a chiral
singularity would show up in the flavour symmetric expansion
and show that at large $n$ the coefficient would drop like
a higher power in $1/n$. (In practice this would be difficult
to determine.)

We have also extended these results in section~\ref{partial_quenching}
to the partially quenched case (when the masses of the valence 
quarks do not have to be the same as the sea quark masses, but
we still have the constraint for the sea quarks that $\overline{m}$
remains constant). In general we have shown (at least to quadratic
quark mass order) that the number of expansion coefficients does not increase.
Thus a (cheaper) simulation with partially quenched hadron masses
may potentially be of help in determining these coefficients.
We also show that on the trajectory $\overline{m} = \mbox{constant}$
the partially quenched error vanishes.

\begin{figure}[htb]
   \begin{center}
      \includegraphics[width=10.0cm]
         {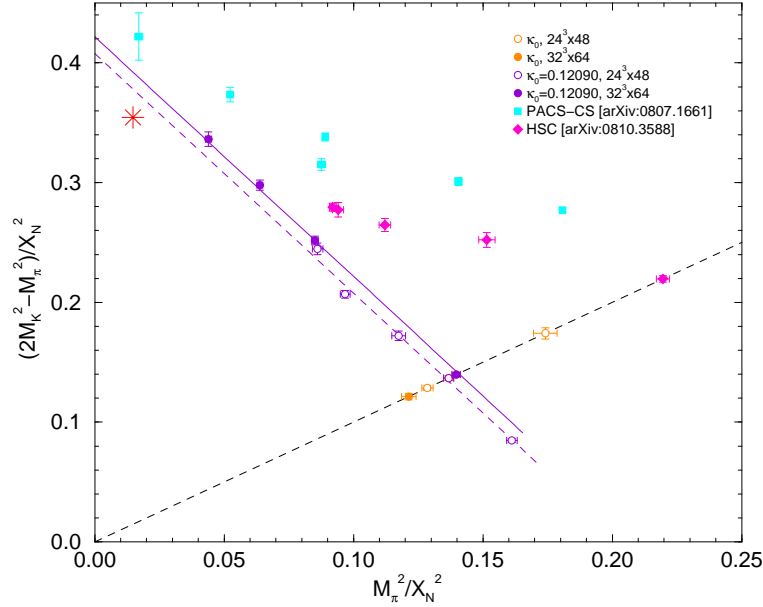}
   \end{center} 
   \caption{$(2M_K^2 - M_\pi^2)/X_N^2$ versus $M_\pi^2/X_\pi^2$ for
            $\kappa_0 = 0.12090$. The dashed black line,
            $y = x$ represents the $SU(3)$ flavour symmetric line.
            Filled violet circles are on $32^3\times 64$ lattices
            while open violet circles are on a $24^3\times 48$ sized lattice.
            Shown are also points on the flavour symmetric line
            (open and filled orange circles).
            The fits are from eq.~(\protect\ref{const_mbar_fit}).
            Results from the PACS-CS Collaboration,
            \cite{aoki08a} and the HS Collaboration, \cite{lin08a}
            are given by cyan coloured squares and 
            magenta coloured diamonds respectively.
            The physical value is denoted by a (red) star.}
\label{b5p50_mps2oXn2_2mpsK2-mps2oXn2_bj+pacs-cs+hsc_cut}
\end{figure}

In sections~\ref{path}--\ref{hadron_masses} numerical results are presented.
We first locate in section~\ref{expt_hadron_masses} a suitable point
on the flavour symmetric line, which we take as our initial
point for the trajectory $\overline{m} = \mbox{constant}$.
This path can be compared with the trajectory from other collaborations.
In Fig.~\ref{b5p50_mps2oXn2_2mpsK2-mps2oXn2_bj+pacs-cs+hsc_cut}
we show the left panel of Fig.~\ref{b5p50_mps2oXn2_2mpsK2-mps2oXn2}
again together with results from the PACS-CS Collaboration,
\cite{aoki08a} and the HS Collaboration, \cite{lin08a}.
The strategy of the HS Collaboration was to keep the strange
quark mass constant (by considering $(2M_K^2 - M_\pi^2)/M_\Omega^2$
versus $M_\pi^2/M_\Omega^2$, the `JLAB' plot). We see that their points
are indeed approximately constant on our plot.

We then show numerically that flavour singlet quantities, $aX_S$ 
($S = \Delta$, $N$, $\rho$ and $\pi$) remain constant on the
path $\overline{m} = \mbox{constant}$. As the linear term
is not present, this is a sensitive test of the presence of
higher order terms in the flavour symmetry expansion and indicates
that they appear to be small. This also allows an
estimation of the scale, $a$, and a discussion of its consistency.
As can be seen from 
Fig.~\ref{b5p50_mps2oXn2_2mpsK2-mps2oXn2_bj+pacs-cs+hsc_cut}
our trajectory does not reach the physical point exactly. This is 
reflected in the fact that different definitions of the scale
in Table~\ref{scale_conversion} (last column) give results
varying by a few percent.

Results for the hadron mass spectrum are then shown. Numerically
we first see a mass hierarchy, which confirms our theoretical
expectation from the flavour symmetric expansion.
A series of `fan' plots for the various multiplets
are then given, with fits which use the flavour symmetric expansion
and show that indeed all fits for the pseudoscalar, vector and
baryon octets and baryon decuplet are highly linear.
The higher order terms are very small -- one early
hint of this is the fact that the Gell-Mann--Okubo relations
work so well for hadron masses. We also note that simulations
with a `light' strange quark mass and heavy `light' quark mass
are possible -- here the right most points in
Figs.~\ref{b5p50_mpsO2oXpi2-jnt_mpsO2oXpi2-jnt} --
\ref{b5p50_mpsO2o2mpsK2+mps2o3-jnt_mDDo2mDeltaD+mOmDo3-boot-jnt}.
In this inverted strange world we would expect the weak interaction decays 
$p \to \Sigma$ or $\Lambda$.

\begin{figure}[htb]
   \begin{center}
      \includegraphics[width=10.0cm]
         {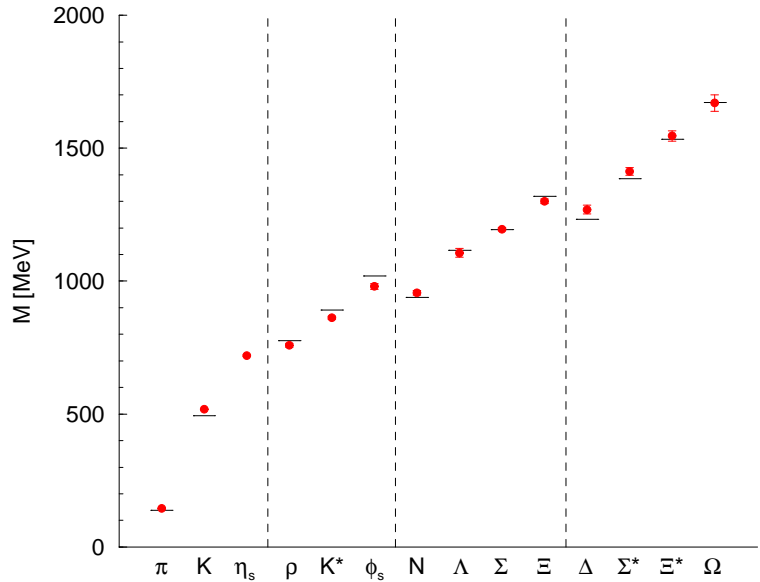}
   \end{center} 
   \caption{The masses of the octet and decuplet multiplets
            as given in Table~\protect\ref{hadron_masses_av}
            using $X_N$ to determine the scale, together
            with the experimental values (short horizontal lines).}
\label{spectrum_XNscale}
\end{figure}

In Fig.~\ref{spectrum_XNscale}
we plot the results from Table~\ref{hadron_masses_av} and compare
them with the experimental results (also given in this table)
for the octet and decuplet hadron multiplets. This means that our
physical input necessary to determine the hadron mass spectrum is
$\kappa_0$ (i.e.\ ideally the value corresponding to 
$\overline{m}^{\R} = \overline{m}^{\R\,*}$),
together with $m_\pi^2/X_\pi^2 |^*$ and $X_N|^*$.

Exploratory partially quenched results for the baryon octet
spectrum are shown,  using the heavier $24^3\times 48$ data.
It is illustrated that they contain useful information and allow
for the possibility that partially quenched results can help in
the determination of coefficients in the flavour symmetry expansion.
(We plan to discuss this further in \cite{pq}.)

We are also applying this method to the computation of matrix elements,
\cite{me}, some initial numerical results are given in
\cite{zanotti10a,winter10a,horsley10a}.


\section*{Acknowledgements}


The numerical calculations have been performed on the IBM
BlueGeneL at EPCC (Edinburgh, UK), the BlueGeneL and P at
NIC (J\"ulich, Germany), the SGI ICE 8200 at HLRN (Berlin-Hannover, Germany)
and the JSCC (Moscow, Russia). We thank all institutions.
The BlueGene codes were optimised using Bagel, \cite{boyle09a}.
This work has been supported in part by the EU grants
227431 (Hadron Physics2), 238353 (ITN STRONGnet)
and by the DFG under contract SFB/TR 55 (Hadron Physics from Lattice QCD).
JMZ is supported by STFC grant ST/F009658/1.


\clearpage

\appendix

\section*{Appendix}


\section{The permutation group $\mathbf{S_3}$} 
\label{perm_group}


If we have three quarks $u$, $d$ and $s$ with different masses, 
physics should be unchanged if we simply permute the names
we give to the quarks. The permutation group is not the 
complete symmetry group -- for example we could also 
perform $U(1)$ phase rotations on any particular quark flavour
-- but it is already enough to tell us something useful. 
The permutation group of $3$ objects, $S_3$, is the same as the 
symmetry group of an equilateral triangle, $C_{3v}$.
There are 6 group operations
\begin{enumerate}

   \item The identity
         \begin{eqnarray}
            u \to u \,, \,\, d \to d \,, \,\, s \to s \,,
         \end{eqnarray}

   \item Two cyclic permutations
         \begin{eqnarray}
            u \to d \,, \,\, d \to s \,, \,\, s \to u 
            \qquad {\rm and } \qquad 
            u \to s \,, \,\, s \to d \,, \,\, d \to u \,,
         \end{eqnarray}
         which correspond to rotations of the triangle 
         through $\pm 120^{\rm o}$, and for a diagram in the
         $I_3$--$Y$ plane rotations through $\pm 120^{\rm o}$.

   \item Three pair interchanges
         \begin{eqnarray}
            u \leftrightarrow d \,, \,\, s \to s ; 
            \qquad u \leftrightarrow s \,, \,\, d \to d ;
            \qquad d \leftrightarrow s \,, \,\, u \to u \,,
         \end{eqnarray}
         which correspond to the $3$ reflection symmetries
         of the triangle, and reflections of a diagram in the
         $I_3$--$Y$ plane.

\end{enumerate}
If an equation is to respect flavour blindness, both sides of 
the equation should transform the same way under all $6$
operations. The representations of the group allow us to
arrange for this to hold. 

The permutation group $S_3$ is a subgroup of $SU(3)$
and has $3$ irreducible representations \cite{atkins70a}:
two different singlets, $A_1$ and $A_2$; and a doublet $E$.
The group properties of these are briefly summarised in
Table~\ref{S3_simplified} and discussed at greater length below.


\subsection{ Singlet representation $\mathbf{A_1}$}


The representation $A_1$, the trivial representation, includes objects
which are invariant under all $6$ group operations. Examples include gluonic
quantities, such as glueball masses, $r_0$, $V(r)$,
as well as certain averages over hadron multiplets. (We shall
collectively denote these objects by $X$.) Examples of quark mass polynomials
with $A_1$ symmetry (complete up to $O(m_q^3)$) are
\begin{eqnarray}
   &  & 1                                                         \\
   &  & \overline{m}
                                                       \nonumber  \\
   &  & \overline{m}^2 \,, \qquad \delta m_u^2 + \delta m_d^2 + \delta m_s^2
                                                       \nonumber  \\
   &  & \overline{m}^3 \,, \qquad 
        \overline{m} (\delta m_u^2 + \delta m_d^2 + \delta m_s^2 )
                     \,, \qquad
        \delta m_u \delta m_d \delta m_s \,,
                                                       \nonumber
\end{eqnarray}
and linear combinations of these. These are the $7$ polynomials
of symmetry $A_1$ listed in Table~\ref{cubic}, the only change
is that in the table we have made the replacement 
$\overline{m} \to (\overline{m} - m_0)$, appropriate
for a Taylor expansion about the point $(m_0, m_0, m_0)$.
Any other completely symmetric polynomial is a linear combination
of these, for example
\begin{equation}
   m_u^3 + m_d^3 + m_s^3 =
      3 \overline{m}^3 
       + 3 \overline{m} (\delta m_u^2 + \delta m_d^2 + \delta m_s^2 )
       + 3 \delta m_u \delta m_d \delta m_s \,.
\end{equation}


\subsection{Singlet representation $\mathbf{A_2}$}


This consists of objects which are invariant under
cyclic quark permutations (triangle rotations), but which
change sign under pair exchanges (reflections). $A_2$
quantities automatically vanish if any two quark masses are the same.
The lowest $A_2$ quantity for quark masses is $O(m_q^3)$,
\begin{eqnarray}
   \lefteqn{m_u m_s^2 - m_d m_s^2 
             + m_d^2 m_s - m_u^2 m_s
             + m_u^2 m_d - m_d^2 m_u}
                                                             \nonumber \\
      &=& (\delta m_s - \delta m_u)(\delta m_s-\delta m_d)
             (\delta m_u-\delta m_d)  \,.
\label{A2mq}
\end{eqnarray}
Baryon mass combinations with $A_2$ symmetry are
\begin{eqnarray}
   M_n - M_p - M_{\Sigma^-} + M_{\Sigma^+} + M_{\Xi^-} - M_{\Xi^0} \,,
\label{A2baryon}
\end{eqnarray}
and the corresponding decuplet quantity, with the $p$ and $n$
replaced by $\Delta^+$ and $\Delta^0$, namely
$M_{\Delta^+} - M_{\Delta^0} + M_{\Sigma^{*-}} 
- M_{\Sigma^{*+}} + M_{\Xi^{*0}} - M_{\Xi^{*-}}$.
Because particle and anti-particle have the same mass,
the mesonic analogue of eq.~(\ref{A2baryon}) vanishes.

Group theory tells us that in a $1+1+1$ flavour world,
the splitting, eq.~(\ref{A2baryon}) would be proportional
to eq.~(\ref{A2mq}) and terms of even higher order in $m_q$
(neglecting electromagnetic effects).


\subsection{Doublet representation $\mathbf{E}$}


By considering $A_2$ we have found a mass splitting
formula for the $1+1+1$ case, but by looking at the doublet
$E$ we are able to find some more formulae valid for the
$2+1$ case which is of more interest for this work.

The $E$ representation has two states, which mix under the
cyclic permutations. We can choose to make one state of the
doublet even under the reflection $u \leftrightarrow d$ ,
and the other state odd. (We could just as well choose any
interchange to classify our states, but it makes best sense to
choose $u \leftrightarrow d$, because the hadronic universe
is almost invariant under that operation.) We have called the
even member of the doublet $E^+$, the odd  member $E^-$.
(There does not appear to be a standard notation.)

An example of an $E$ doublet would be the states
\begin{eqnarray}
   {1 \over \sqrt{6}} \left( 2 | s \rangle - | u \rangle 
                                           - |d \rangle \right)
   \qquad {\rm and } \qquad
   {1 \over \sqrt{2}} \left( | u \rangle - | d \rangle \right) \,.
\label{Eexemplar}
\end{eqnarray}
It is easily checked that under any group operation they just
mix with each other, for example under the cyclic operation
$ u \to d$, $d \to s$, $s \to u $:
\begin{eqnarray}
   {1 \over \sqrt{6}} \left( 2 | s \rangle - | u \rangle 
                                           - | d \rangle \right)
      &\to& {1 \over \sqrt{6}}
                      \left( 2 | u \rangle - | d \rangle 
                                           - | s \rangle \right)
                                                                   \\
      & = & {\sqrt{3} \over 2} \, {1 \over \sqrt{2}}
                      \left( | u \rangle - | d \rangle \right)
           - {1 \over 2} \, {1 \over \sqrt{6}}
                      \left( 2 | s \rangle - | u \rangle 
                                           - | d \rangle \right)
                                                       \nonumber
\end{eqnarray}
and so on. In other words, the matrix for a cyclic permutation
has the form
\begin{eqnarray}
   R = \left( \begin{array}{rr}
                 \cos\theta    & \mp\sin\theta \\
                 \pm\sin\theta & \cos\theta    \\
              \end{array}
       \right)
\end{eqnarray}
with $\theta = 120^{\rm o}$. 

Quark mass terms with $E$ doublet symmetry are
\begin{eqnarray}
   \begin{array}{lcl}
      \left\{ {1 \over \sqrt{6}}(2m_s - m_u - m_d) \right.       & , & 
      \left.  {1 \over \sqrt{2}}(m_u - m_d) \right\}                     \\
      \left\{ {1 \over \sqrt{6}}(2m_s^2 - m_u^2 - m_d^2) \right. & , & 
      \left.  {1 \over \sqrt{2}}(m_u^2 - m_d^2) \right\}                 \\
      \left\{ {1 \over \sqrt{6}}(m_um_s + m_dm_s - 2m_um_d) \right. & , & 
      \left.  {1 \over \sqrt{2}}(m_um_s - m_dm_s) \right\}               \\
      \left\{ {1 \over \sqrt{6}}(2m_s^3 - m_u^3 - m_d^3) \right. & , & 
      \left.  {1 \over \sqrt{2}}(m_u^3 - m_d^3) \right\}                 \\
      \left\{ {1 \over 2}(m_um_s^2 + m_dm_s^2 - m_u^2m_d - m_um_d^2) \right.
                                                                & , &    \\
      \multicolumn{3}{c}
      {\hfill
      \left. {1 \over \sqrt{12}}(m_um_s^2 + m_dm_s^2 + 2m_u^2m_s - 2m_d^2m_s 
                                 + m_u^2m_d + m_um_d^2) \right\}}         \\
      \multicolumn{3}{c}
      {\left\{ {1 \over \sqrt{12}}(m_um_s^2 + m_dm_s^2 - 2m_u^2m_s - 2m_d^2m_s 
                                 + m_u^2m_d + m_um_d^2) \right.
                                                 \hfill\mbox{}}          \\
                                                                & , &
      \left. {1 \over 2}(-m_um_s^2 + m_dm_s^2 + m_u^2m_d - m_um_d^2) \right\}
   \end{array}
\end{eqnarray}
The normalisations and phases have been chosen so that each pair transforms
in the same way as eq.~(\ref{Eexemplar}) under all group operations, i.e.\ the
matrices which represent the group operations are the same for every pair.


\section{Some group theory}
\label{group_theory}


If the three quarks have equal masses, the QCD Lagrangian is 
invariant under a global $U(1)$  transformation of the quark fields 
\begin{equation}
   \psi \to e^{i \theta} \psi\,, \qquad 
   \overline{\psi} \to \overline{\psi} e^{-i \theta} \,,
\end{equation}
(corresponding to baryon number conservation) and a global $SU(3)$
flavour transformation 
\begin{equation}
   \psi \to U  \psi\,, \qquad 
   \overline{\psi} \to \overline{\psi} U^\dagger \,,
\end{equation}
with $U$ a unitary matrix with determinant $1$. 

If the quarks are all given different masses
we still have the freedom to change the phase of each 
flavour separately, without changing the action, 
so we have conserved currents for each of the three flavours,
and three independent $U(1)$ symmetries. 
 
When the quarks have different masses, 
flavour $SU(3)$ is no longer a symmetry of the action, 
a global $SU(3)$ rotation no longer leaves the action unchanged, 
but we can still use $SU(3)$ to understand the action. 

An analogy of our argument comes from ordinary mechanics
or quantum mechanics. If we have a quantum mechanical problem 
which is not rotationally symmetric we lose the conservation
of angular momentum. But we do still  have the property
that if we rotate the Hamiltonian, $H$, to give a new problem,
with the Hamiltonian $H^\prime \neq H$, then the
eigenfunctions of the new Hamiltonian $H^\prime$
can be obtained by rotating the eigenfunctions of the original
problem. In the case of broken flavour symmetry, imposing this
(nearly trivial) condition will constrain the way in which
hadron masses can depend on quark masses. 

Consider the transformation of the quark mass matrix
\begin{equation}
   {\cal M } \to U {\cal M} U^\dagger \equiv {\cal M}^\prime \,,
\label{mtrans} 
\end{equation}
(the flavour analogue of a global gauge rotation in colour). 
The quarks may have different masses, ${\cal M}^\prime \ne {\cal M}$, 
although they are physically equivalent in the sense that the
two matrices have the same eigenvalues, but the eigenvectors
are rotated
\begin{equation} 
   \psi^\prime = U  \psi, \qquad \overline{\psi}^\prime 
               = \overline{\psi} U^\dagger \,.
\label{psi_transf}
\end{equation} 
Let us now use these definitions to investigate the group properties
of mass polynomials.


\subsection{Flavour permutations, $\mathbf{S_3}$, as a subgroup
            of $\mathbf{SU(3)}$} 


We want to concentrate initially on a set of $SU(3)$ matrices 
which map a diagonal mass matrix to another diagonal matrix
when used in eq.~(\ref{mtrans}). These are
\begin{itemize}

   \item the identity matrix,  
         \begin{equation} 
            I = \left( \begin{array}{ccc}
                          1 & 0 & 0 \\
                          0 & 1 & 0 \\
                          0 & 0 & 1 \\
                       \end{array}
                \right)
         \end{equation} 

   \item the cyclic permutations of the quark flavours, 
         \begin{eqnarray}
            \left( \begin{array}{ccc}
                      0 & 0 & 1 \\
                      1 & 0 & 0 \\
                      0 & 1 & 0 \\
                   \end{array}
            \right)
            &=& \exp \left\{ i \; {2 \pi \over 3 \sqrt{3}}
                        \left( \begin{array}{ccc}
                                  0  & i  & -i \\
                                  -i & 0  &  i \\
                                  i  & -i &  0 \\
                               \end{array}
                        \right)
                     \right\}
                                                             \nonumber \\
            \left( \begin{array}{ccc}
                      0 & 1 & 0 \\
                      0 & 0 & 1 \\
                      1 & 0 & 0 \\
                   \end{array}
            \right)
            &=& \exp \left\{ - i\; {2 \pi \over 3 \sqrt{3}} 
                        \left( \begin{array}{ccc}
                                  0  & i  & -i \\
                                  -i & 0  &  i \\
                                  i  & -i &  0 \\
                               \end{array}
                        \right)
                     \right\}
         \end{eqnarray}

   \item pair interchanges,
         \begin{eqnarray}
            \left( \begin{array}{ccc}
                      0  & -1 &  0 \\
                      -1 &  0 &  0 \\
                      0  &  0 & -1 \\
                   \end{array}
            \right)
            &=& \exp \left\{ i \; {\pi \over 2}
                        \left( \begin{array}{ccc}
                                  1 &  1 &  0 \\
                                  1 &  1 &  0 \\
                                  0 &  0 & -2 \\
                               \end{array}
                        \right)
                     \right\}
                                                             \nonumber \\
            \left( \begin{array}{ccc}
                      0  &  0 & -1 \\
                      0  & -1 &  0 \\
                      -1 &  0 &  0 \\
                   \end{array}
            \right)
            &=& \exp \left\{ i \; {\pi \over 2}
                        \left( \begin{array}{ccc}
                                  1 &  0 &  1 \\
                                  0 & -2 &  0 \\
                                  1 &  0 &  1 \\
                               \end{array}
                        \right)
                     \right\}
                                                             \nonumber \\
            \left( \begin{array}{ccc}
                      -1 &  0 &  0 \\
                       0 &  0 & -1 \\
                       0 & -1 &  0 \\
                   \end{array}
            \right)
            &=& \exp \left\{ i \; {\pi \over 2}
                        \left( \begin{array}{ccc}
                                  -2 &  0 &  0 \\
                                   0 &  1 &  1 \\
                                   0 &  1 &  1 \\
                               \end{array}
                        \right)
                     \right\}
         \end{eqnarray}

\end{itemize}
Note that when we interchange a quark pair, we also have to
change the sign of the quarks, to keep the determinant equal to $1$,
as required for a matrix in $SU(3)$. These six matrices are all
unitary with determinant $1$, so they are all members of $SU(3)$.
We have also shown that all the matrices can be written in the
canonical $SU(3)$ form $\exp\{ i \sum \alpha_j \lambda_j \}.$
These matrices form a closed set under multiplication,
with a multiplication table matching that of the group $S_3$,
showing that the symmetries of the equilateral triangle are
a subgroup of $SU(3)$.


\subsection{Group classification of quark mass polynomials}
\label{group_class}


This subsection explains how the final column of Table~\ref{cubic}
was calculated. 

We can establish many useful results from the $S_3$ subgroup,
but it has its limitations, it does not connect particles in
different permutation sets, see Fig.~\ref{permset}.
By considering $S_3$ alone we cannot write down a formula
for the mass difference between the $\Sigma^0$ and $\Sigma^-$,
we cannot even show that the two particles have the same mass
in the $2+1$ case. To go further  we need to consider the full
$SU(3)$ group, even though this will involve operations which
make the mass matrix non-diagonal.

We can write any $SU(3)$ rotation as a matrix of the form
\begin{equation}
   U = \exp \left\{ i \sum_{j=1}^8 \alpha_j \lambda_j \right\} \,,
\end{equation}
where the $\lambda_j$ are the 8 Gell-Mann matrices (and $\alpha_j$
are real parameters). Here we only need to consider infinitesimal
transformations
\begin{equation}
   {\cal M} \to U {\cal M} U^\dagger 
      = {\cal M} + i \sum_{j=1}^8 \alpha_j \left( \lambda_j {\cal M}
        - {\cal M} \lambda_j \right) 
      = {\cal M} 
        + i \sum_{j=1}^8 \alpha_j \left [ \lambda_j, {\cal M} \right] \,.
\end{equation}  
We write
\begin{eqnarray} 
   {\cal O}_j \psi     &=& \lambda_j \psi
                                                            \nonumber  \\
   {\cal O}_j {\overline {\psi} } 
                       &=& - {\overline {\psi} } \lambda_j
                                                            \nonumber  \\
   {\cal O}_j {\cal M} &=& \left[ \lambda_j, {\cal M} \right] \,,
\label{group_ops}
\end{eqnarray}
to represent the action of the eight generators of $SU(3)$ on
spinors and on matrices. The eight operators ${\cal O}_j$
are analogous to the three operators $J_j$ in angular momentum. 

In $SU(2)$ we use the eigenvalues of the operator 
\begin{equation} 
   J^2 = \sum_{j=1}^3 J_j^2 \,,
\end{equation} 
to identify the irreducible representations of angular momentum. 
Similarly in $SU(3)$ we can use the eigenvalues of  the quadratic
Casimir operator \cite{pfeifer03a,greiner89}
\begin{equation} 
   {\cal C} = {1 \over 4} \sum_{j=1}^8  {\cal O}_j^2 \,,
\label{Casimir} 
\end{equation} 
to identify irreducible representations of $SU(3)$. (The factor
${1\over 4}$ is a conventional normalisation.) Acting on a matrix
\begin{equation} 
   {\cal C} {\cal M} = {1 \over 4} \sum_{j=1}^8 
      \left[ \lambda_j, \left[ \lambda_j ,  {\cal M} \right] \right] 
     = {1 \over 4} \sum_{j=1}^8
         \left( \lambda_j^2  {\cal M} - 2 \lambda_j  {\cal M}  \lambda_j
                +  {\cal M} \lambda_j^2 \right) \,.
\label{CasiMat} 
\end{equation} 

We can now begin classifying polynomial functions of ${\cal M}$.

At first order, where we have linear functions of mass and ${\cal M}$
can be decomposed as
\begin{equation}
   {\cal M} 
      = I\,\third {\rm Tr}[{\cal M}] 
         + \sum_{j=1}^8 \lambda_j\,\half {\rm Tr}[\lambda_j{\cal M}] \,,
\end{equation}
it is simple. We have
\begin{equation}
   {\rm Tr}[ {\cal M}] ={\cal M}_{11}+{\cal M}_{22}+{\cal M}_{33} \,,
\label{mtrace} 
\end{equation} 
which does not change under $SU(3)$ transformations, so it is singlet.

The other elements of  ${\cal M}$ can be assigned quantum
numbers. For example ${\cal M}_{21}$ takes a $u$ quark and 
changes it to a $d$, so it has $I_3 = -1$ and   hypercharge $Y=0$.   
The $6$ off-diagonal elements of  ${\cal M}$ form the outer ring
of the octet, see for example Fig.~\ref{meson_mults}.
The two central elements of the octet are the combinations 
\begin{eqnarray} 
   2 {\cal M}_{33} - {\cal M}_{11} - {\cal M}_{22} 
      &\propto& {\rm Tr}[ \lambda_8  {\cal M} ] 
                                                             \nonumber  \\
   {\cal M}_{11} - {\cal M}_{22}
     &\propto& {\rm Tr}[ \lambda_3  {\cal M} ] \,.
\label{linEm}
\end{eqnarray}
These both have $I_3 = Y = 0$.
We can check that both are eigenstates of the Casimir operator,
with eigenvalue $3$, showing that both are pure octet quantities.  
If we make the substitutions
\begin{equation} 
   {\cal M}_{11} \to m_u \,, \quad  {\cal M}_{22} \to m_d \,, 
                             \quad {\cal M}_{33} \to m_s  \,,
\end{equation} 
we see that the quantities eqs.~(\ref{mtrace})--(\ref{linEm}) are
proportional  to the three linear polynomials in Table~\ref{cubic},
with the $SU(3)$ assignments given from their behaviour when operated 
on by the Casimir operator. 

It gets more interesting at second order. 
$\left( {\rm Tr} [ {\cal M} ] \right)^2 $ and ${\rm Tr} [ {\cal M}^2 ] $ 
are both flavour-singlet functions of the mass matrix. 
It is more convenient to work with the linear combinations 
\begin{equation}
   \left( {\rm Tr} [ {\cal M} ] \right)^2 \,, \qquad
   3  {\rm Tr} [ {\cal M}^2 ] - \left( {\rm Tr} [ {\cal M} ] \right)^2 \,,
\label{delsing} 
\end{equation}
where we have chosen the coefficients so that the second combination
will be zero at the $SU(3)$ symmetric point. At second order we should
be able to construct functions of the mass matrix that are in the $1$,
$8$ and $27$ representations. One way of constructing a quantity that
is purely $27$-plet is by using the Casimir operator.
If we take an arbitrary quadratic function of ${\cal M}$
it will usually be a mixture of all three representations.
If we multiply by 
\begin{equation} 
   ( {\cal C} - 3)  {\cal C} \,,
\label{proj27} 
\end{equation} 
${\cal C}$ will cancel the singlet part, $( {\cal C} - 3)$ will eliminate
the octet part (see Table~\ref{Caseigen}),
\begin{table}[htb]  
   \begin{center} 
   \begin{tabular}{lcccccc}
      Representation $\phantom{I^{I^{I^X}}}$
      & 1 & 8 & 10 & ${\overline{10}} $ & 27 & 64 \\
      Casimir eigenvalue & 0 & 3 & 6 & 6 & 8 & 15 \\
   \end{tabular} 
   \end{center} 
\caption{The eigenvalues of the quadratic Casimir operator,
         ${\cal C}$, eq.~(\ref{Casimir}), for the $SU(3)$
         representations needed in this article.}
\label{Caseigen}
\end{table} 
so the operator eq.~(\ref{proj27}) leaves a pure $27$-plet function of 
${\cal M}$. Using the eigenvalues in Table~\ref{Caseigen} we can
construct similar operators to project out objects in the other 
representations of $SU(3)$. Of course it would be tedious
to do this by hand: we have programmed the group operations
in Mathematica so that the group theory can be done more easily and rapidly. 

Another useful technique is to use the raising and lowering operators
$I_{\pm}$, $U_{\pm}$, $V_{\pm}$ \cite{gasiorowicz66} to move around within
a multiplet. Once we have one state in a  multiplet,
these operators allow us to construct all the other states.  
Because infinitesimal $SU(3)$ operations do not preserve diagonality, 
a typical eigenstate of the Casimir operator will involve all 
nine elements of the quark mass matrix ${\cal M}$, not just the 
three diagonal elements. For example, if we explicitly 
write out the $SU(3)$ singlet quantity
$3  {\rm Tr} [ {\cal M}^2 ] - ( {\rm Tr} [ {\cal M} ] )^2$ in
eq.~(\ref{delsing}) it is
\begin{eqnarray}
   P_{1} &=&  2{\cal M}_{11} {\cal M}_{11} + 2{\cal M}_{22} {\cal M}_{22} 
                                           + 2{\cal M}_{33} {\cal M}_{33}
                                                           \nonumber   \\
         & & + 6{\cal M}_{12} {\cal M}_{21} + 6{\cal M}_{13} {\cal M}_{31}
                                            + 6{\cal M}_{23} {\cal M}_{32}  
                                                     \label{mtraces}   \\
         & & - 2{\cal M}_{11} {\cal M}_{22} - 2{\cal M}_{11} {\cal M}_{33}
                                            - 2{\cal M}_{22} {\cal M}_{33} \,.
                                                           \nonumber 
\end{eqnarray} 
We can use the techniques discussed in this section to write down 
a pure $SU(3)$ $27$-plet quantity, with the same $S_3$ properties
as eq.~(\ref{mtraces}); the result is
\begin{eqnarray} 
   P_{27}^{A_1} 
        &=&  {\cal M}_{11} {\cal M}_{11} + {\cal M}_{22} {\cal M}_{22} 
            + {\cal M}_{33} {\cal M}_{33}
                                                          \nonumber   \\
        & & - {\cal M}_{12} {\cal M}_{21} - {\cal M}_{13} {\cal M}_{31} 
            - {\cal M}_{23} {\cal M}_{32}
                                                        \label{M27}   \\
        & & - {\cal M}_{11} {\cal M}_{22} - {\cal M}_{11} {\cal M}_{33} 
            - {\cal M}_{22} {\cal M}_{33} \,.
                                                           \nonumber 
\end{eqnarray} 
Expressed this way, the $27$-plet and singlet are clearly 
different functions of the full $9$-element mass matrix. 
However, if we just consider a diagonal mass matrix,
${\cal M}_{ij} = 0$ if $i \ne j$, ${\cal M}_{11} = m_u$,
${\cal M}_{22} = m_d$, ${\cal M}_{33} = m_s$ then the quantities
become indistinguishable: 
\begin{eqnarray}
   P_1 &\to& 2 (m_u^2 +m_d^2 + m_s^2 - m_u m_d - m_u m_s - m_d m_s) 
                                                          \nonumber   \\
       &=&   3 (\delta m_u^2 + \delta m_d^2 + \delta m_s^2)
                                                          \nonumber   \\
  P_{27}^{A_1}
       &\to& m_u^2 +m_d^2 + m_s^2 - m_u m_d - m_u m_s - m_d m_s
                                                          \nonumber   \\
       &=&   {3\over 2} (\delta m_u^2 + \delta m_d^2 + \delta m_s^2) \,.
\end{eqnarray} 
Both collapse to the same quark mass polynomial,
$\delta m_u^2 + \delta m_d^2 + \delta m_s^2$,
so this polynomial is allowed to appear in equations for singlet
and $27$-plet physical quantities, but not in equations for
any other $SU(3)$ representation. This polynomial is recorded
in Table~\ref{cubic} with the symmetry representations $A_1$ 
and $1$ or $27$.

We can continue and use the methods of this subsection to classify all
polynomials up to cubic order, the results are recorded in
Table~\ref{cubic}.


\subsection{Matrix representations of $\mathbf{SU(3)}$}
\label{grp_anal_mass}


To construct hadron mass matrices for octet and decuplet
hadrons we need to analyse $8 \times 8$ and $10 \times 10$ 
matrices by their $S_3$ and $SU(3)$ properties. 

To get started we need to construct $8 \times 8$ and $10 \times 10$ 
representations of the $SU(3)$ generators. We can do this by
considering the known behaviour of the hadron multiplets under
the $SU(2)$ subgroups, isospin $I$, $U$-spin and $V$-spin,
and the hypercharge, $Y$, \cite{gasiorowicz66}:
\begin{eqnarray} 
   \lambda_1 = 2 I_1 & & \lambda_2 = 2 I_2 
                         \qquad \lambda_3 = 2 I_3 
                                                            \nonumber   \\
   \lambda_4 = 2 V_1 & & \lambda_5 = 2 V_2 
                                                                        \\
   \lambda_6 = 2 U_1 & & \lambda_7 = 2 U_2 
                         \qquad \lambda_8 = \sqrt{3}\; Y \,.
                                                            \nonumber 
\end{eqnarray} 
These $8 \times 8$ or $10 \times 10$ $\lambda$ matrices 
have the same commutation relations as the usual $3 \times 3$ matrices
\begin{equation} 
   [\lambda_i , \lambda_j] = 2 i f_{ijk} \lambda_k \,.
\end{equation} 
Once we have the eight $\lambda$ matrices for our hadron multiplet 
we can use eq.~(\ref{CasiMat}) to classify any other matrices. 
For the hadron mass matrices, we need all the flavour-conserving
matrices. For the decuplet mass matrix these are all diagonal 
matrices; but for the octet mass matrix they can include some 
off-diagonal elements, because the $\Sigma^0$ and $\Lambda$ have
the same flavour quantum numbers. Our results for the decuplet and octet 
matrices are given in Tables~\ref{mat10} and \ref{mat8}. 
 
We have other methods of constructing the matrix representations
of $SU(3)$. In addition to the Casimir projection method sketched here, 
we can start with one matrix which belongs to a known $SU(3)$ 
representation, and then build all the other matrices in that
representation by repeatedly acting with raising and lowering 
operators. For example, in the decuplet case we know that the 
$10 \times 10$ matrix 
\begin{equation} 
   \left( \begin{array}{cccccccccc}
             0&0&0&0&0&0&0&0&0&1 \\
             0&0&0&0&0&0&0&0&0&0 \\
             0&0&0&0&0&0&0&0&0&0 \\
             0&0&0&0&0&0&0&0&0&0 \\
             0&0&0&0&0&0&0&0&0&0 \\
             0&0&0&0&0&0&0&0&0&0 \\
             0&0&0&0&0&0&0&0&0&0 \\
             0&0&0&0&0&0&0&0&0&0 \\
             0&0&0&0&0&0&0&0&0&0 \\
             1&0&0&0&0&0&0&0&0&0 \\
          \end{array}
   \right) \,,
\label{m64} 
\end{equation} 
must be a pure $64$-plet, because it interchanges the $\Delta^-$ and 
the $\Omega^-$, which changes strangeness by $\pm 3$. From 
Fig.~\ref{decup_weight} we see that the $64$-plet is the only representation 
in $10 \otimes \overline{10}$ that can change strangeness by $3$ units. 
Starting from the matrix in eq.~(\ref{m64}) we can construct a 
set of $64$ matrices which transform amongst themselves under all
group operations eq.~(\ref{group_ops}).

Once we have classified all the $10 \times 10$
and $8 \times 8$ matrices according to their 
$SU(3)$ and $S_3$ behaviour, we can read off the
rows of Tables~\ref{mat10} and \ref{mat8}.
For example, knowing that the following 
$8 \times 8$ matrix is an octet with symmetry $E^-$ 
gives the fifth row of Table~\ref{mat8},
\begin{equation} 
   \left( \begin{array}{cccccccc}
             -1 & 0 & 0 & 0 & 0 & 0 & 0 & 0 \\
              0 & 1 & 0 & 0 & 0 & 0 & 0 & 0 \\
              0 & 0 & 0 & 0 & 0 & 0 & 0 & 0 \\
              0 & 0 & 0 & 0 & {2 \over \sqrt{3}} & 0 & 0 & 0 \\
              0 & 0 & 0 & {2 \over \sqrt{3}} & 0 & 0 & 0 & 0 \\
              0 & 0 & 0 & 0 & 0 & 0 & 0 & 0 \\
              0 & 0 & 0 & 0 & 0 & 0 & 1 & 0 \\
              0 & 0 & 0 & 0 & 0 & 0 & 0 & -1\\
          \end{array}
   \right) \,.
 \end{equation}


\subsection{Hadron mass matrices}
\label{had_mass_matrices}


We describe the hadron masses via a hadron mass matrix ${\cal H}$, 
a $10 \times 10$ matrix for the decuplet baryons, an $8 \times 8$ 
matrix for octet baryons or mesons. 

If the different quark flavours have different masses, a global 
$SU(3)$ rotation of the quark mass matrix, eq.~(\ref{mtrans}),
leads to a change in the quark mass matrix, ${\cal M} \to {\cal M}^\prime$,
but does not change the eigenvalues of the matrix,
or the essential physics of the situation. 

What will be the effect of a flavour rotation of the quark 
Lagrangian on a hadronic mass matrix ${\cal H}$? 

If we are considering a Taylor expansion for hadronic masses, 
all the elements of ${\cal H}$ will be polynomials of the
elements in the quark mass matrix, ${\cal M}_{ij}$.
Flavour blindness requires that we still get equivalent physics
when we change ${\cal M}_{ij} \to {\cal M}_{ij}^\prime$,
i.e.\ that the eigenvalues of ${\cal H}$ are unchanged,  
and the eigenvectors of ${\cal H}$ rotate according
to eq.~(\ref{psi_transf}). Writing this as an equation, 
\begin{equation}
   {\cal H}^\prime \equiv {\cal H}( {\cal M}_{ij}^\prime ) 
                    = U  {\cal H}( {\cal M}_{ij} ) U^\dagger \,. 
\end{equation} 
Using the unitarity of $U$ we can rewrite this as an
invariance condition, 
\begin{equation}
   U^\dagger {\cal H}( {\cal M}_{ij}^\prime ) U 
      =  {\cal H}( {\cal M}_{ij} ) \,.
\label{CMinvar} 
\end{equation}
The effect of changing ${\cal M}$ to ${\cal M}^\prime$ can be 
exactly cancelled by the effect of an $SU(3)$ rotation on ${\cal H}$. 

To construct an invariant matrix satisfying eq.~(\ref{CMinvar}) 
we have to pair up matrices of known symmetry, constructed
as described in Appendix~\ref{grp_anal_mass}, with ${\cal M}$
polynomials of known symmetry, constructed using the methods
of Appendix~\ref{group_class}. This gives us a hadron mass
matrix of the form
\begin{eqnarray}
   {\cal H}
     &=& \sum (\mbox{singlet mass polynomial}) \times
               (\mbox{singlet matrix})
                                                     \nonumber  \\
     & & + \sum (\mbox{octet mass polynomial}) \times
                (\mbox{octet matrix}) 
                                               \label{schema}   \\
     & & + \sum (\mbox{27-plet mass polynomial}) \times
                (\mbox{27-plet matrix}) 
                                                     \nonumber  \\
     & & + \, \cdots
                                                     \nonumber
\end{eqnarray}
To give an $SU(2)$ analogy, we can make a rotationally invariant
system (i.e.\ a system with total spin zero), by coupling together 
two particles with the same $J$, but not by coupling together 
two particles with different $J$. Similarly, to give a hadron mass matrix
under the $SU(3)$ operation, eq.~(\ref{CMinvar}), we must give every
matrix a coefficient of the same symmetry, as shown schematically
in eq.~(\ref{schema}). 

Once we have (with the help of Mathematica), constructed the 
most general matrix satisfying eq.~(\ref{CMinvar}), we make the 
substitutions ${\cal M}_{11} \to m_u$, ${\cal M}_{22} \to m_d$,
${\cal M}_{33} \to m_s$, and ${\cal M}_{ij} \to 0$ if $i \ne j$
to get mass formulae for all the hadrons.  

We now consider an example. In Table~\ref{mat8mes}
we list the $6$ matrices which can occur in the
octet meson mass matrix in the $1+1+1$ flavour case. 
In the $2+1$ flavour case ($m_u = m_d$) the two $E^-$ matrices
drop out, because their coefficients must be odd under 
the exchange $u \leftrightarrow d$, leaving just $4$ matrices
which can contribute. We calculate the most general form of
the meson mass matrix, by demanding that it is invariant under
$SU(3)$ rotations, eq.~(\ref{CMinvar}), and find that in the
$2+1$ case with $\overline{m} = \mbox{constant}$ we get
\begin{eqnarray} 
   {\cal H}
     &=&  (M_0^2 + b_1 \delta m_l^2) 
          \left( \begin{array}{cccccccccc}
                    1 & 0 & 0 & 0 & 0 & 0 & 0 & 0 \\
                    0 & 1 & 0 & 0 & 0 & 0 & 0 & 0 \\
                    0 & 0 & 1 & 0 & 0 & 0 & 0 & 0 \\
                    0 & 0 & 0 & 1 & 0 & 0 & 0 & 0 \\
                    0 & 0 & 0 & 0 & 1 & 0 & 0 & 0 \\
                    0 & 0 & 0 & 0 & 0 & 1 & 0 & 0 \\
                    0 & 0 & 0 & 0 & 0 & 0 & 1 & 0 \\
                    0 & 0 & 0 & 0 & 0 & 0 & 0 & 1 \\
                 \end{array} \right)
                                                      \nonumber \\[0.8em]
     & &  + ( a_8 \delta m_l + b_8 \delta m_l^2 ) 
          \left( \begin{array}{cccccccccc}
                    1 & 0 & 0 & 0 & 0 & 0 & 0 & 0 \\
                    0 & 1 & 0 & 0 & 0 & 0 & 0 & 0 \\
                    0 & 0 &-2 & 0 & 0 & 0 & 0 & 0 \\
                    0 & 0 & 0 &-2 & 0 & 0 & 0 & 0 \\
                    0 & 0 & 0 & 0 & 2 & 0 & 0 & 0 \\
                    0 & 0 & 0 & 0 & 0 &-2 & 0 & 0 \\
                    0 & 0 & 0 & 0 & 0 & 0 & 1 & 0 \\
                    0 & 0 & 0 & 0 & 0 & 0 & 0 & 1 \\
                 \end{array} \right)
                                                      \nonumber \\[0.8em]
     & &  + 5 b_{27} \delta m_l^2 
          \left( \begin{array}{cccccccccc}
                    1 & 0 & 0 & 0 & 0 & 0 & 0 & 0 \\ 
                    0 & 1 & 0 & 0 & 0 & 0 & 0 & 0 \\
                    0 & 0 & 1 & 0 & 0 & 0 & 0 & 0 \\
                    0 & 0 & 0 &-3 & 0 & 0 & 0 & 0 \\
                    0 & 0 & 0 & 0 & -3& 0 & 0 & 0 \\
                    0 & 0 & 0 & 0 & 0 & 1 & 0 & 0 \\
                    0 & 0 & 0 & 0 & 0 & 0 & 1 & 0 \\
                    0 & 0 & 0 & 0 & 0 & 0 & 0 & 1 \\
                 \end{array} \right)
                                                      \nonumber \\[0.8em]
     & &  + 4 b_{27} \delta m_l^2 
          \left( \begin{array}{cccccccccc}
                    1 & 0 & 0 & 0 & 0 & 0 & 0 & 0 \\ 
                    0 & 1 & 0 & 0 & 0 & 0 & 0 & 0 \\
                    0 & 0 & -2& 0 & 0 & 0 & 0 & 0 \\
                    0 & 0 & 0 & 3 & 0 & 0 & 0 & 0 \\
                    0 & 0 & 0 & 0 & -3& 0 & 0 & 0 \\
                    0 & 0 & 0 & 0 & 0 &-2 & 0 & 0 \\
                    0 & 0 & 0 & 0 & 0 & 0 & 1 & 0 \\
                    0 & 0 & 0 & 0 & 0 & 0 & 0 & 1 \\
                 \end{array} \right)
\end{eqnarray} 
keeping terms up to quadratic order. From this we read off
\begin{eqnarray} 
   M_\pi^2 &=& M_0^2 - 2 a_8 \delta m_l + (b_1 - 2 b_8 - 3 b_{27}) \delta m_l^2 
                                                      \nonumber    \\
   M_K^2  &=& M_0^2 +  a_8 \delta m_l + (b_1 +   b_8 + 9 b_{27}) \delta m_l^2 
                                                      \nonumber    \\
   M_{\eta_8}^2 
         &=& M_0^2 + 2 a_8 \delta m_l + (b_1 + 2 b_8 -27 b_{27}) \delta m_l^2 \,.
\label{etamult}
\end{eqnarray} 
We can check that these equations are consistent with Table~\ref{meso2p1}:
\begin{eqnarray} 
   3 M_\pi^2 + 4 M_K^2 + M_{\eta_8}^2 &=& 8 (M_0^2 + b_1 \delta m_l^2)
                                                      \nonumber    \\
  -3 M_\pi^2 + 2 M_K^2 + M_{\eta_8}^2 &=& 10 (a_8 \delta m_l + b_8 \delta m_l^2) 
                                                      \nonumber    \\
   - M_\pi^2 + 4 M_K^2 -3M_{\eta_8}^2 &=& 120 b_{27} \delta m_l^2 \,.
\label{syst9}
\end{eqnarray} 
An alternative method, as discussed in section~\ref{theory_2p1},
would be to start from the simultaneous equations in eq.~(\ref{syst9})
and solve the system to derive eq.~(\ref{etamult}). Note that in
eq.~(\ref{fit_mpsO}) we have also re-written the results in a form
to agree with the notation of the partially quenched results,
so $a_8 = - \alpha$ and
\begin{eqnarray}
   b_1    &=& \beta_0 +4\beta_1 + 6\beta_2 + \eighth\beta_3
                                                      \nonumber    \\
   b_8    &=& \beta_1 + 3\beta_2 + \tenth\beta_3
                                                      \nonumber    \\
   b_{27} &=& - \fortieth\beta_3 \,.
\end{eqnarray}


\section{Coordinate choice for partially quenched formulae}
\label{coordinate_choice}


It is often convenient to plot quantities against the pseudoscalar meson
mass squared, because then we know better the location of the physical
point and the chiral limit. If we do want to use pseudoscalar mesons,
the best choice is to replace the light sea quark by the full
(non partially quenched) pion, the light valence quark by the
partially quenched pion, and to replace the valence strange
quark mass by the partially quenched  $\overline{s}_{val} s_{val}$
meson (the `strange pion'), which we call the $\eta_s$.
This is a particle that doesn't exist in the real world, but which
we can measure in the partially quenched channel. Determining the
valence $s$ quark mass from the kaon has disadvantages,
as we shall shortly see.

We introduce the mesonic variables
\begin{eqnarray} 
   x \equiv M^2_{\pi^{full}} - M^2_{\pi}|_0
      &=& 2 \alpha \delta m_l + \beta_0  \delta m_l^2+ 2 \beta_1 \delta m_l^2
                                                             \nonumber \\
   y \equiv M^2_{\pi^{PQ}} - M^2_{\pi}|_0
      &=& 2 \alpha \delta\mu_l + \beta_0 \delta m_l^2 + 2 \beta_1 \delta\mu_l^2
                                                             \nonumber \\
   z \equiv M^2_{\eta_s}  - M^2_{\pi}|_0
      &=& 2 \alpha \delta\mu_s + \beta_0 \delta m_l^2 + 2 \beta_1 \delta\mu_s^2
          \,,
\label{Szdef}
\end{eqnarray} 
keeping terms up to second order in the quark masses. In terms of these
variables the decuplet mass formulae eq.~(\ref{PQOmega}) become 
\begin{eqnarray} 
   M_\Delta &=& M_0 + 3  \tilde{A} y  +  \tilde{B}_0 x^2 
                   + 3 \tilde{B}_1 y^2
                                                             \nonumber \\
   M_{\Sigma^*} 
            &=& M_0 +  \tilde{A} (2y +z) +  \tilde{B}_0 x^2 
                    + \tilde{B}_1 (2 y^2 +z^2) + \tilde{B}_2 (z-y)^2
                                                             \nonumber \\
   M_{\Xi^*}&=& M_0 +  \tilde{A} (y + 2z) +  \tilde{B}_0 x^2 
                   + \tilde{B}_1 (y^2 + 2z^2) + \tilde{B}_2 (z-y)^2
                                                             \nonumber \\
   M_\Omega &=& M_0 + 3  \tilde{A} z  +  \tilde{B}_0 x^2 
                   + 3 \tilde{B}_1 z^2 \,,
\label{mesOmega} 
\end{eqnarray}  
with 
\begin{eqnarray} 
   \tilde{A}   &\equiv & {A \over 2 \alpha}
                                                             \nonumber \\
   \tilde{B}_0 &\equiv & {2 \alpha B_0 - 3 A \beta_0 \over 8 \alpha^3}
                                                             \nonumber \\
   \tilde{B}_1 &\equiv & {\alpha B_1 - A \beta_1 \over 4 \alpha^3}
                                                             \nonumber \\
   \tilde{B}_2 &\equiv & {B_2 \over 4 \alpha^2} \,.
\label{Btil} 
\end{eqnarray}  
The form of eq.~(\ref{mesOmega}) exactly repeats the 
form of eq.~(\ref{PQOmega}), but the new constants involve
a combination of curvature terms from the pion mass equation and from the 
baryon mass equation. 
     
Suppose we use the PQ kaon mass (instead of the strange pion)
to represent the strange quark mass, i.e.\ we replace $z$ defined in 
eq.~(\ref{Szdef}) by 
\begin{eqnarray}
   w &\equiv& 2 M^2_{K^{\PQ}} - M^2_{\pi^{\PQ}} - M^2_{\pi^{\full}} 
                                                             \nonumber \\
     &=& 2 \alpha \delta\mu_s + \beta_0 \delta m_l^2 + 2 \beta_1 \delta\mu_s^2
                             + 2 \beta_2 (\delta\mu_s - \delta\mu_l)^2 \,. 
\end{eqnarray}
At first order, $w$ is just as good as $z$, but if we are interested
in curvature, it is less suitable, because at second order it involves
both the valence $s$ and the valence $l$, unlike eq.~(\ref{Szdef}). 
Using $w$ instead of $z$, the decuplet mass formulae become 
\begin{eqnarray}
   M_\Delta &=& M_0 + 3 \tilde{A} y + \tilde{B}_0 x^2
                    + 3 \tilde{B}_1 y^2
                                                             \nonumber \\
   M_{\Sigma^*}
            &=& M_0 + \tilde{A}(2y +w) + \tilde{B}_0 x^2
                    + \tilde{B}_1 (2 y^2 +w^2) + \tilde{B}_2 (w-y)^2
                    + \tilde{B}_X (w-y)^2
                                                             \nonumber \\
    M_{\Xi^*}&=& M_0 + \tilde{A}(y + 2w) + \tilde{B}_0 x^2
                    + \tilde{B}_1 (y^2 + 2w^2) + \tilde{B}_2 (w-y)^2
                    + 2\tilde{B}_X (w-y)^2
                                                             \nonumber \\
    M_\Omega &=& M_0 + 3  \tilde{A} w  +  \tilde{B}_0 x^2
                     + 3 \tilde{B}_1 w^2 + 3 \tilde{B}_X  (w-y)^2 \,, 
\end{eqnarray}
with $ \tilde{A}, \tilde{B}_0, \tilde{B}_1, \tilde{B}_2 $ defined as
in eq.~(\ref{Btil}), but with an extra curvature coefficient 
\begin{eqnarray}
   \tilde{B}_X = - {A \beta_2 \over 4 \alpha^3}  \,,
\end{eqnarray}
so one fit constraint is lost (or deeply hidden) if we use the 
kaon mass to represent the strange mass. 

Finally, we want to relate the partially quenched fit
to the unitary results, on our trajectory $\third(2 m_l + m_s) = m_0$. 
If we use bare quark masses as our coordinates, we do this by using 
the substitutions 
\begin{equation}
   \delta\mu_l \to \delta m_l \,, \qquad \delta\mu_s \to -2 \delta m_l \,,
\end{equation}
giving
\begin{eqnarray} 
   M_\Delta  &=& M_0 + 3 A \delta m_l + [B_0 +3 B_1]\delta m_l^2 
                                                             \nonumber \\
   M_{\Sigma^*}
             &=& M_0 \phantom{+ 3 A \delta m_l\  }
                     + [B_0 + 6 B_1 + 9 B_2]\delta m_l^2
                                                             \nonumber \\
  M_{\Xi^*}  &=& M_0 - 3 A \delta m_l
                    + [ B_0 + 9 B_1 + 9 B_2]\delta m_l^2
                                                             \nonumber \\
   M_\Omega  &=& M_0 - 6 A \delta m_l + [B_0 +12 B_1]\delta m_l^2 \,.
\label{fullOmega} 
\end{eqnarray} 
However, if we use meson-based coordinates, such as 
eq.~(\ref{Szdef}), the mapping back to the 
unitary result is more complicated,
\begin{eqnarray}
   y & \to & x
                                                             \nonumber \\
   z & \to & -2 x + {3(\beta_0 + 4 \beta_1) \over 4 \alpha^2} x^2  \,.
\end{eqnarray} 
The mapping from $z$, our measure of the strange quark mass, 
back to $x$ is complicated by a second order term. 
The reason is clear. On our trajectory, the relation
$2 \delta m_l + \delta m_s = 0$ is made exactly true
for bare lattice quark masses, while the meson mass relations 
$2 M_\pi^2 + M_{\eta_s}^2 \approx \mbox{constant}$ or
$2 M_K^2 + M_\pi^2 \approx \mbox{constant}$ are only true to leading order. 
Thus in conclusion if we are considering the curvature terms 
it is definitely better to use (bare) lattice quark masses
as the coordinates.


\section{The action}
\label{slinc_action}


The particular clover action used here has a single iterated mild
stout smearing, \cite{morningstar03a} for the hopping terms together
with thin links for the clover term (this ensures that the fermion
matrix does not become too extended). Together with
the (tree level) Symanzik improved gluon action this gives
\begin{equation}
   S = S_G + S_{Fu} + S_{Fd} + S_{Fs} \,,
\end{equation}
with the gluon action
\begin{equation}
   S_G = {6 \over g_0^2} \, \left\{
          c_0 \sum_{\plaquette} {1 \over 3} \mbox{Re\,Tr}
                                                ( 1 - U_{\plaquette}) +
          c_1 \sum_{\rectangle} {1 \over 3} \mbox{Re\,Tr}
                                                ( 1 - U_{\rectangle})
                            \right\} \,,
\end{equation}
and
\begin{equation}
   \beta 
     = {6c_0\over g_0^2} = {10 \over g_0^2} \qquad 
       \mbox{and} \quad c_0 = {20 \over 12} \,, \,\, c_1 = - {1 \over 12} \,.
\end{equation}
For each flavour the Wilson--Dirac fermion action is
\begin{eqnarray}
   \lefteqn{S_{Fq} =}
                                                \nonumber              \\
   && \sum_x \left\{{1 \over 2} 
                        \sum_{\mu}[\overline{q}(x)(\gamma_\mu - 1)
                          \tilde{U}_\mu(x) q(x+a\hat{\mu}) -
                        \overline{q}(x)(\gamma_\mu + 1) 
                          \tilde{U}^\dagger_\mu(x-a\hat{\mu}) q(x-a\hat{\mu})]
                  \right.
                                                \nonumber              \\
   && \hspace*{0.75in} \left.
            + {1 \over 2\kappa_q} \overline{q}(x)q(x) -
             {1 \over 4} a c_{sw} \sum_{\mu\nu}
                 \overline{q}(x)\sigma_{\mu\nu}F_{\mu\nu}(x)q(x)
                            \right\} \,,
\label{action}
\end{eqnarray}
where $F$ is the `clover' field strength, necessary for $O(a)$-improvement.
As the up and down quarks are always taken here as mass degenerate
we have $\kappa_u = \kappa_d \equiv \kappa_l$.

To keep the action highly local, the hopping terms use a stout smeared link
(`fat link') with $\alpha = 0.1$ `mild smearing' for the Dirac kinetic
term and Wilson mass term,
\begin{eqnarray}
   \tilde{U}_\mu(x)
            &=& \exp\{iQ_\mu(x)\}\, U_\mu(x)
                                                            \nonumber \\
   Q_\mu    &=&
       {\alpha \over 2i} \left[ V_\mu U_\mu^\dagger - U_\mu V_\mu^\dagger 
       - \third \mbox{Tr} (V_\mu U_\mu^\dagger - U_\mu V_\mu^\dagger) \right] \,,
\end{eqnarray}
where $V_\mu(x)$ is the sum of all staples around $U_\mu(x)$.
The clover term is built from thin links as it is already of length
$4a$ and, as previously mentioned, we do not want the fermion matrix
to become too extended. Stout smearing is analytic and so a derivative
can be taken (so the HMC force is well defined) and also allows
for perturbative expansions \cite{horsley08a}.

The clover coefficient, $c_{sw}$, has recently been
non-perturbatively fixed, \cite{cundy09a}, by requiring
that the axial Ward identity (WI) quark mass determined in several
different ways is the same. A sensitive way of achieving this
is the Schr\"odinger functional formalism.
Further details of our results may be found in \cite{cundy09a}.
$c_{sw}$ is determined for $3$ mass degenerate or $SU(3)$
flavour symmetric quarks (where $\kappa_l = \kappa_s \equiv \kappa_0$)
in the chiral limit. A $5$th order polynomial in $g_0^2$ interpolating
between the numerically determined $c_{sw}(g_0)$ points was found to be
\cite{cundy09a}
\begin{equation}
   c_{sw}^*(g_0) 
      = 1 + 0.269041\, g_0^2 + 0.29910\, g_0^4 
          - 0.11491\, g_0^6 - 0.20003\, g_0^8 + 0.15359\, g_0^{10} \,.
\label{cswstar_poly}
\end{equation}
(This interpolation function is constrained to reproduce the
$O(g_0^2)$ perturbative results, \cite{horsley08a}, in the $\beta \to \infty$
limit and therefore has four free fit parameters.) We take this result
to define $c_{sw}$ for a given $\beta$.

Improving one on-shell quantity to $O(a^2)$ (here the axial WI quark mass)
fixes $c_{sw}(g_0^2)$ and then all masses are automatically improved to
$O(a^2)$,
\begin{equation}
   {M_H \over M_{H^\prime}}(a) 
      = {M_H \over M_{H^\prime}}(0) + O(a^2) \,,
                                                \nonumber
\end{equation}
rather than just to $O(a)$. Operators in general require
further $O(a)$ operators together with associated improvement
coefficients to ensure $O(a)$--improvement for physical on-shell
quantities.

This determination of $c_{sw}$ via the Schr\"odinger functional
formalism also provides an estimate for the critical $\kappa_0$,
\cite{cundy09a}, of
\begin{eqnarray}
   \kappa_{0;c}(g_0) 
     &=& {1 \over 8} \,
        \left[ 1  + 0.002391\, g_0^2 + 0.0122470\, g_0^4  - 0.0525676\, g_0^6
        \right.
                                                \nonumber  \\
     & & \hspace*{1.35in} \left.
                  + 0.0668197\, g_0^8  - 0.0242800\, g_0^{10} \right] \,.
\label{kapcstar_poly}
\end{eqnarray}
(Again this interpolation function is constrained to reproduce the
$O(g_0^2)$ perturbative results, \cite{horsley08a}, in the $\beta \to \infty$
limit. The errors for $c_{sw}^*$ from the fit are estimated to be
about $0.4\%$ while for $\kappa_c^*$ we have $0.02\%$
at $\beta = 14.0$ rising to $0.15\%$ at $\beta = 5.10$.)

The simulations only need knowledge of $c_{sw}$ to proceed;
however it is useful to check consistency between different
determinations of $\kappa_{0;c}$ (via the Schr\"odinger functional
or the pseudoscalar mass). For $\beta = 5.50$ then using
eq.~(\ref{kapcstar_poly}) we find $\kappa_{0;c} = 0.120996$ (the direct
simulation result is $\kappa_{0;c} = 0.121125(330)$, \cite{cundy09a}).
This is to be compared with the estimation in section~\ref{kappasymc}
which is quite close. (It should also be noted that different
determinations should only agree up to $O(a^2)$ effects.)


\section{Hadron masses}
\label{hadron_masses_numbers}


We collect here in Tables~\ref{table_run_M_sym} -- 
\ref{table_run_M_BD_ksym12092} values of the pseudoscalar octet,
vector octet, baryon octet and baryon decuplet masses.
In Table~\ref{table_run_M_sym} we give values along the flavour
symmetric line ($\kappa_l = \kappa_s = \kappa_0$),
while in Tables~\ref{table_run_M_MO} -- \ref{table_run_M_BD}
and in Tables~\ref{table_run_M_MO_ksym12092} -- 
\ref{table_run_M_BD_ksym12092} we give results for
$\kappa_0 = 0.12090$ and $\kappa_0 = 0.12092$ respectively,
while keeping $\overline{m} = \mbox{constant}$,
eq.~(\ref{kappas_mbar_const}).

In Tables \ref{table_run_M_MVOoX} -- \ref{table_run_M_BDoX}
we give the ratios (i.e.\ hadron octet or decuplet masses normalised
with their centre of mass).

The data sets are roughly $\sim O(2000)$ trajectories for the
$24^3\times 48$ lattices and $O(1500)$ -- $O(2000)$ trajectories
for the $32^3\times 64$ lattices (with the exception for the
$\kappa_0 = 0.12095$ results which are $\sim O(500)$ trajectories).
The errors are all taken from a bootstrap analysis of the ratio
(which often enables a smaller error to be given for the ratios than
simply using error propagation).



\begin{table}[p]
   \begin{center}
      \begin{tabular}{llllll}
\multicolumn{1}{c}{$\kappa_0$}                               &
\multicolumn{1}{c}{$N_S^3\times N_T$}                         &
\multicolumn{1}{c}{$aM_\pi$}                                  &
\multicolumn{1}{c}{$aM_\rho$}                                 &
\multicolumn{1}{c}{$aM_N$}                                    &
\multicolumn{1}{c}{$aM_\Delta$}                               \\
         \hline
  0.12000 & $16^3\times 32$ 
          & 0.4908(17) & 0.6427(23) & 0.9612(42) & 1.048(6)    \\
  0.12030 & $16^3\times 32$ 
          & 0.4026(19) & 0.5635(38) & 0.8374(74) & 0.9414(107) \\
  0.12050 & $24^3\times 48$ 
          & 0.3375(24) & 0.4953(47) & 0.7201(83) & 0.8216(89)  \\
  0.12080 & $24^3\times 48$ 
          & 0.2260(10) & 0.3903(55) & 0.5417(68) & 0.6415(99)  \\
  0.12090 & $16^3\times 32$ 
          & 0.2209(49) & 0.4192(97) & 0.6298(251)& 0.7811(274) \\
  0.12090 & $24^3\times 48$ & 
    \multicolumn{4}{c}
 {see Tables~\protect\ref{table_run_M_MO} -- \protect\ref{table_run_M_BD}} \\
  0.12090 & $32^3\times 64$ &
    \multicolumn{4}{c}
 {see Tables~\protect\ref{table_run_M_MO} -- \protect\ref{table_run_M_BD}} \\
  0.12092 & $24^3\times 48$ &
    \multicolumn{4}{c}
 {see Tables~\protect\ref{table_run_M_MO_ksym12092} --
                                  \protect\ref{table_run_M_BD_ksym12092}}  \\
  0.12095 & $32^3\times 64$ 
          & 0.1508(4)       & 0.3209(27) & 0.4329(49) & 0.5541(80)  \\
  0.12099 & $32^3\times 64$ 
          & 0.1297(10)      & 0.3154(67) & 0.4127(117)& 0.5476(168) \\
         \hline
      \end{tabular}
   \end{center}
\caption{The results for the hadrons on the symmetric line,
         $aM_\pi$, $aM_\rho$, $aM_N$ and $aM_\Delta$ for
         $(\beta, c_{sw}, \alpha) = (5.50, 2.65, 0.1)$.}
\label{table_run_M_sym}
\end{table}


\begin{table}[p]
   \begin{center}
      \begin{tabular}{llll}
\multicolumn{1}{c}{$(\kappa_l,\kappa_s)$}                     &
\multicolumn{1}{c}{$aM_\pi$}                                  &
\multicolumn{1}{c}{$aM_K$}                                    &
\multicolumn{1}{c}{$aM_{\eta_s}$}                               \\
         \hline
\multicolumn{4}{c}{$16^3\times 32$}                           \\
         \hline
  (0.121040, 0.120620) & 0.1962(74) & 0.2447(49) & 0.2773(37) \\
         \hline
\multicolumn{4}{c}{$24^3\times 48$}                           \\
         \hline
  (0.120830, 0.121040) & 0.1933(6)  & 0.1688(7)  & 0.1391(11) \\
  (0.120900, 0.120900) & 0.1779(6)  & 0.1779(6)  & 0.1779(6)  \\
  (0.120950, 0.120800) & 0.1661(8)  & 0.1845(7)  & 0.2011(7)  \\
  (0.121000, 0.120700) & 0.1515(10) & 0.1898(8)  & 0.2209(6)  \\
  (0.121040, 0.120620) & 0.1406(8)  & 0.1949(6)  & 0.2361(5)  \\
         \hline
\multicolumn{4}{c}{$32^3\times 64$}                           \\
         \hline
  (0.120900, 0.120900) & 0.1747(5)  & 0.1747(5)  & 0.1747(5)  \\
  (0.121040, 0.120620) & 0.1349(5)  & 0.1897(4)  & 0.2321(3)  \\
  (0.121095, 0.120512) & 0.1162(8)  & 0.1956(5)  & 0.2512(3)  \\
  (0.121145, 0.120413) & 0.09694(88)& 0.2016(4)  & 0.2683(3)  \\
         \hline
      \end{tabular}
   \end{center}
\caption{The results for the pseudoscalar octet mesons:
         $aM_\pi$, $aM_K$ and $aM_{\eta_s}$ for
         $(\beta, c_{sw}, \alpha) = (5.50, 2.65, 0.1)$
         where $\kappa_0 = 0.12090$.}
\label{table_run_M_MO}
\end{table}


\begin{table}[htb]
   \begin{center}
      \begin{tabular}{llll}
\multicolumn{1}{c}{$(\kappa_l,\kappa_s)$}                    &
\multicolumn{1}{c}{$aM_\rho$}                                 &
\multicolumn{1}{c}{$aM_{K^*}$}                                &
\multicolumn{1}{c}{$aM_{\phi_s}$}                             \\
         \hline
\multicolumn{4}{c}{$16^3\times 32$}                           \\
         \hline
  (0.121040, 0.120620) & 0.4353(123)& 0.4331(84) & 0.4380(60) \\
         \hline
\multicolumn{4}{c}{$24^3\times 48$}                           \\
         \hline
  (0.120830, 0.121040) & 0.3460(22) & 0.3335(30) & 0.3198(48) \\
  (0.120900, 0.120900) & 0.3494(25) & 0.3494(25) & 0.3494(25) \\
  (0.120950, 0.120800) & 0.3400(40) & 0.3473(32) & 0.3546(27) \\
  (0.121000, 0.120700) & 0.3364(43) & 0.3517(30) & 0.3663(20) \\
  (0.121040, 0.120620) & 0.3270(50) & 0.3484(28) & 0.3701(18) \\
         \hline
\multicolumn{4}{c}{$32^3\times 64$}                           \\
         \hline
  (0.120900, 0.120900) & 0.3341(34) & 0.3341(34) & 0.3341(34) \\
  (0.121040, 0.120620) & 0.3127(38) & 0.3380(21) & 0.3632(14) \\
  (0.121095, 0.120512) & 0.3123(43) & 0.3426(20) & 0.3738(11) \\
  (0.121145, 0.120413) & 0.3210(63) & 0.3500(24) & 0.3880(11) \\
         \hline
      \end{tabular}
   \end{center}
\caption{The results for the vector octet mesons:
         $aM_\rho$, $aM_{K^*}$ and $aM_{\phi_s}$ for
         $(\beta, c_{sw}, \alpha) = (5.50, 2.65, 0.1)$
         where $\kappa_0 = 0.12090$.}
\label{table_run_M_MVO}
\end{table}


\begin{table}[htb]
   \begin{center}
      \begin{tabular}{lllll}
\multicolumn{1}{c}{$(\kappa_l,\kappa_s)$}                     &
\multicolumn{1}{c}{$aM_N$}                                    &
\multicolumn{1}{c}{$aM_\Lambda$}                               &
\multicolumn{1}{c}{$aM_\Sigma$}                                &
\multicolumn{1}{c}{$aM_\Xi$}                                  \\
         \hline
\multicolumn{5}{c}{$16^3\times 32$}                           \\
         \hline
  (0.121040, 0.120620) & 0.5817(214)& 0.5941(182)& 0.6311(128)& 0.6353(121)\\
         \hline
\multicolumn{5}{c}{$24^3\times 48$}                           \\
         \hline
  (0.120830, 0.121040) & 0.4976(25) & 0.4859(43) & 0.4791(31) & 0.4679(39) \\
  (0.120900, 0.120900) & 0.4811(33) & 0.4811(33) & 0.4811(33) & 0.4811(33) \\
  (0.120950, 0.120800) & 0.4737(68) & 0.4794(58) & 0.4871(55) & 0.4938(48) \\
  (0.121000, 0.120700) & 0.4648(46) & 0.4815(49) & 0.4910(36) & 0.5055(28) \\
  (0.121040, 0.120620) & 0.4466(66) & 0.4810(57) & 0.4843(42) & 0.5068(32) \\
         \hline
\multicolumn{5}{c}{$32^3\times 64$}                           \\
         \hline
  (0.120900, 0.120900) & 0.4673(27) & 0.4673(27) & 0.4673(27) & 0.4673(27) \\
  (0.121040, 0.120620) & 0.4267(50) & 0.4547(43) & 0.4697(33) & 0.4907(21) \\
  (0.121095, 0.120512) & 0.4140(61) & 0.4510(58) & 0.4690(37) & 0.4971(21) \\
  (0.121145, 0.120413) & 0.4016(89) & 0.4507(65) & 0.4761(39) & 0.5092(19) \\
         \hline
      \end{tabular}
   \end{center}
\caption{The results for the octet baryons: $aM_N$, $aM_\Lambda$,
         $aM_\Sigma$ and $aM_\Xi$ for
         $(\beta, c_{sw}, \alpha) = (5.50, 2.65, 0.1)$
         where $\kappa_0 = 0.12090$.}
\label{table_run_M_BO}
\end{table}


\begin{table}[htb]
   \begin{center}
      \begin{tabular}{lllll}
\multicolumn{1}{c}{$(\kappa_l,\kappa_s)$}                     &
\multicolumn{1}{c}{$aM_\Delta$}                                &
\multicolumn{1}{c}{$aM_{\Sigma^*}$}                             &
\multicolumn{1}{c}{$aM_{\Xi^*}$}                               &
\multicolumn{1}{c}{$aM_\Omega$}                               \\
         \hline
\multicolumn{5}{c}{$16^3\times 32$}                          \\
         \hline
  (0.121040, 0.120620) & 0.7437(227)& 0.7490(184)& 0.7537(146)& 0.7595(114)\\
         \hline
\multicolumn{5}{c}{$24^3\times 48$}                          \\
         \hline
  (0.120830, 0.121040) & 0.5906(73) & 0.5801(89) & 0.5685(114)& 0.5548(151)\\
  (0.120900, 0.120900) & 0.5933(88) & 0.5933(88) & 0.5933(88) & 0.5933(88) \\
  (0.120950, 0.120800) & 0.5817(55) & 0.5895(48) & 0.5973(43) & 0.6050(38) \\
  (0.121000, 0.120700) & 0.5883(101)& 0.6006(77) & 0.6133(61) & 0.6262(51) \\
  (0.121040, 0.120620) & 0.5483(137)& 0.5679(90) & 0.5902(64) & 0.6108(48) \\
         \hline
\multicolumn{5}{c}{$32^3\times 64$}                           \\
         \hline
  (0.120900, 0.120900) & 0.5675(64))& 0.5675(64) & 0.5675(64) & 0.5675(64) \\
  (0.121040, 0.120620) & 0.5520(79) & 0.5744(48) & 0.5968(34) & 0.6194(28) \\
  (0.121095, 0.120512) & 0.5161(185)& 0.5541(98) & 0.5812(52) & 0.6104(33) \\
  (0.121145, 0.120413) & 0.5071(211)& 0.5576(105)& 0.6018(51) & 0.6420(29) \\
         \hline
      \end{tabular}
   \end{center}
\caption{The results for the decuplet baryons: $aM_\Delta$,
         $aM_{\Sigma^*}$, $aM_{\Xi^*}$ and $aM_\Omega$ for
         $(\beta, c_{sw}, \alpha) = (5.50, 2.65, 0.1)$
         where $\kappa_0 = 0.12090$.}
\label{table_run_M_BD}
\end{table}



\begin{table}[htb]
   \begin{center}
      \begin{tabular}{llll}
\multicolumn{1}{c}{$(\kappa_l,\kappa_s)$}                     &
\multicolumn{1}{c}{$aM_\pi$}                                  &
\multicolumn{1}{c}{$aM_K$}                                    &
\multicolumn{1}{c}{$aM_{\eta_s}$}                             \\
         \hline
\multicolumn{4}{c}{$24^3\times 48$}                           \\
         \hline
  (0.120920, 0.120920) & 0.1694(9)  & 0.1694(9)  & 0.1694(9)  \\
         \hline
\multicolumn{4}{c}{$32^3\times 64$}                           \\
         \hline
  (0.121050, 0.120661) & 0.1280(6)  & 0.1813(5)  & 0.2221(4)  \\
         \hline
      \end{tabular}
   \end{center}
\caption{The results for the pseudoscalar octet mesons:
         $aM_\pi$, $aM_K$ and $aM_{\eta_s}$ for
         $(\beta, c_{sw}, \alpha) = (5.50, 2.65, 0.1)$
         where $\kappa_0 = 0.12092$.}
\label{table_run_M_MO_ksym12092}
\end{table}


\begin{table}[htb]
   \begin{center}
      \begin{tabular}{llll}
\multicolumn{1}{c}{$(\kappa_l,\kappa_s)$}                     &
\multicolumn{1}{c}{$aM_\rho$}                                 &
\multicolumn{1}{c}{$aM_{K^*}$}                                &
\multicolumn{1}{c}{$aM_{\phi_s}$}                             \\
         \hline
\multicolumn{4}{c}{$24^3\times 48$}                           \\
         \hline
  (0.120920, 0.120920) & 0.3404(44) & 0.3404(44) & 0.3404(44) \\
         \hline
\multicolumn{4}{c}{$32^3\times 64$}                           \\
         \hline
  (0.121050, 0.120661) & 0.3161(38) & 0.3354(22) & 0.3564(16) \\
         \hline
      \end{tabular}
   \end{center}
\caption{The results for the vector octet mesons:
         $aM_\rho$, $aM_{K^*}$ and $aM_{\phi_s}$ for
         $(\beta, c_{sw}, \alpha) = (5.50, 2.65, 0.1)$
         where $\kappa_0 = 0.12092$.}
\label{table_run_M_MVO_ksym12092}
\end{table}


\begin{table}[htb]
   \begin{center}
      \begin{tabular}{lllll}
\multicolumn{1}{c}{$(\kappa_l,\kappa_s)$}                     &
\multicolumn{1}{c}{$aM_N$}                                    &
\multicolumn{1}{c}{$aM_\Lambda$}                               &
\multicolumn{1}{c}{$aM_\Sigma$}                                &
\multicolumn{1}{c}{$aM_\Xi$}                                  \\
         \hline
\multicolumn{5}{c}{$24^3\times 48$}                           \\
         \hline
  (0.120920, 0.120920) & 0.4725(39) & 0.4725(39) & 0.4725(39) & 0.4725(39) \\
         \hline
\multicolumn{5}{c}{$32^3\times 64$}                           \\
         \hline
  (0.121050, 0.120661) & 0.4127(42) & 0.4444(35) & 0.4580(31) &0.4798(22)  \\
         \hline
      \end{tabular}
   \end{center}
\caption{The results for the octet baryons: $aM_N$, $aM_\Lambda$,
         $aM_\Sigma$ and $aM_\Xi$ for
         $(\beta, c_{sw}, \alpha) = (5.50, 2.65, 0.1)$
         where $\kappa_0 = 0.12092$.}
\label{table_run_M_BO_ksym12092}
\end{table}


\begin{table}[htb]
   \begin{center}
      \begin{tabular}{lllll}
\multicolumn{1}{c}{$(\kappa_l,\kappa_s)$}                      &
\multicolumn{1}{c}{$aM_\Delta$}                                &
\multicolumn{1}{c}{$aM_{\Sigma^*}$}                            &
\multicolumn{1}{c}{$aM_{\Xi^*}$}                               &
\multicolumn{1}{c}{$aM_\Omega$}                               \\
         \hline
\multicolumn{5}{c}{$24^3\times 48$}                           \\
         \hline
  (0.120920, 0.120920) & 0.5790(97) & 0.5790(97) & 0.5790(97) & 0.5790(97) \\
         \hline
\multicolumn{5}{c}{$32^3\times 64$}                           \\
         \hline
  (0.121050, 0.120661) & 0.5457(108)& 0.5607(72) & 0.5800(51) & 0.6005(40) \\
         \hline
      \end{tabular}
   \end{center}
\caption{The results for the decuplet baryons: $aM_\Delta$,
         $aM_{\Sigma^*}$, $aM_{\Xi^*}$ and $aM_\Omega$ for
         $(\beta, c_{sw}, \alpha) = (5.50, 2.65, 0.1)$
         where $\kappa_0 = 0.12092$.}
\label{table_run_M_BD_ksym12092}
\end{table}


\begin{table}[htb]
   \begin{center}
      \begin{tabular}{llll}
\multicolumn{1}{c}{$(\kappa_l,\kappa_s)$}                     &
\multicolumn{1}{c}{$M_\rho/X_\rho$}                            &
\multicolumn{1}{c}{$M_{K^*}/X_\rho$}                           &
\multicolumn{1}{c}{$M_{\phi_s}/X_\rho$}                        \\
         \hline
\multicolumn{4}{c}{$24^3\times 48$}                           \\
         \hline
  (0.120830, 0.121040) & 1.025(2)   & 0.9877(12) & 0.9470(155)\\
  (0.120900, 0.120900) & 1.0        & 1.0        & 1.0        \\
  (0.120950, 0.120800) & 0.9859(22) & 1.007(1)   & 1.028(6)   \\
  (0.121000, 0.120700) & 0.9706(34) & 1.015(2)   & 1.057(12)  \\
  (0.121040, 0.120620) & 0.9581(60) & 1.021(3)   & 1.086(12)  \\
         \hline
\multicolumn{4}{c}{$32^3\times 64$}                           \\
         \hline
  (0.120900, 0.120900) & 1.0        & 1.0        & 1.0        \\
  (0.121040, 0.120620) & 0.9488(50) & 1.026(3)   & 1.102(6)   \\
  (0.121095, 0.120512) & 0.9392(63) & 1.030(3)   & 1.124(7)   \\
  (0.121145, 0.120413) & 0.9431(109)& 1.028(5)   & 1.140(9)   \\
         \hline
      \end{tabular}
   \end{center}
\caption{Ratio results for the vector octet mesons:
         $M_\rho/X_\rho$, $M_{K^*}/X_\rho$ and $M_{\phi_s}/X_\rho$ for
         $(\beta, c_{sw}, \alpha) = (5.50, 2.65, 0.1)$
         where $\kappa_0 = 0.12090$.}
\label{table_run_M_MVOoX}
\end{table}


\begin{table}[htb]
   \begin{center}
      \begin{tabular}{lllll}
\multicolumn{1}{c}{$(\kappa_l,\kappa_s)$}                     &
\multicolumn{1}{c}{$M_N/X_N$}                                 &
\multicolumn{1}{c}{$M_\Lambda/X_N$}                           &
\multicolumn{1}{c}{$M_\Sigma/X_N$}                            &
\multicolumn{1}{c}{$M_\Xi/X_N$}                               \\
         \hline
\multicolumn{5}{c}{$24^3\times 48$}                           \\
         \hline
  (0.120830, 0.121040) & 1.033(2)   & 1.009(6)   & 0.9949(13) & 0.9717(26) \\
  (0.120900, 0.120900) & 1.0        & 1.0        & 1.0        & 1.0        \\
  (0.120950, 0.120800) & 0.9769(33) & 0.9887(84) & 1.005(1)   & 1.018(3)   \\
  (0.121000, 0.120700) & 0.9543(32) & 0.9885(77) & 1.008(2)   & 1.038(3)   \\
  (0.121040, 0.120620) & 0.9319(56) & 1.004(7)   & 1.011(2)   & 1.058(4)   \\
         \hline
\multicolumn{5}{c}{$32^3\times 64$}                           \\
         \hline
  (0.120900, 0.120900) & 1.0        & 1.0        & 1.0        & 1.0        \\
  (0.121040, 0.120620) & 0.9229(47) & 0.9833(58) & 1.016(2)   & 1.061(4)   \\
  (0.121095, 0.120512) & 0.8999(77) & 0.9804(111)& 1.019(4)   & 1.081(8)   \\
  (0.121145, 0.120413) & 0.8688(118)& 0.9949(130)& 1.030(5)   &1.101(8)    \\
         \hline
      \end{tabular}
   \end{center}
\caption{Ratio results for the octet baryons: $M_N/X_N$, $M_\Lambda/X_N$,
         $M_\Sigma/X_N$ and $M_\Xi/X_N$ for
         $(\beta, c_{sw}, \alpha) = (5.50, 2.65, 0.1)$
         where $\kappa_0 = 0.12090$.}
\label{table_run_M_BOoX}
\end{table}



\begin{table}[htb]
   \begin{center}
      \begin{tabular}{lllll}
\multicolumn{1}{c}{$(\kappa_l,\kappa_s)$}                      &
\multicolumn{1}{c}{$M_\Delta/X_\Delta$}                         &
\multicolumn{1}{c}{$M_{\Sigma^*}/X_\Delta$}                     &
\multicolumn{1}{c}{$M_{\Xi^*}/X_\Delta$}                        &
\multicolumn{1}{c}{$M_\Omega/X_\Delta$}                        \\
         \hline
\multicolumn{5}{c}{$24^3\times 48$}                           \\
         \hline
  (0.120830, 0.121040) & 1.021(6)   & 1.003(2)   & 0.9824(44) & 0.9588(121)\\
  (0.120900, 0.120900) & 1.0        & 1.0        & 1.0        & 1.0        \\
  (0.120950, 0.120800) & 0.9868(14) & 1.0002(3)  & 1.013(2)   & 1.026(3)   \\
  (0.121000, 0.120700) & 0.9790(42) & 0.9993(24) & 1.020(6)   & 1.042(8)   \\
  (0.121040, 0.120620) & 0.9634(72) & 0.9978(53) & 1.037(11)  & 1.073(14)  \\
         \hline
\multicolumn{5}{c}{$32^3\times 64$}                           \\
         \hline
  (0.120900, 0.120900) & 1.0        & 1.0        & 1.0        & 1.0        \\
  (0.121040, 0.120620) & 0.9609(44) & 0.9999(28) & 1.039(7)   & 1.078(9)   \\
  (0.121095, 0.120512) & 0.9426(120)& 1.012(10)  & 1.062(19)  & 1.115(24)  \\
  (0.121145, 0.120413) & 0.9185(145)& 1.010(13)  & 1.090(23)  & 1.163(29)  \\
         \hline
      \end{tabular}
   \end{center}
\caption{Ratio results for the decuplet baryons: $M_\Delta/X_\Delta$,
         $M_{\Sigma^*}/X_\Delta$, $M_{\Xi^*}/X_\Delta$ and $M_\Omega/X_\Delta$ for
         $(\beta, c_{sw}, \alpha) = (5.50, 2.65, 0.1)$
         where $\kappa_0 = 0.12090$.}
\label{table_run_M_BDoX}
\end{table}

\clearpage



\end{document}